\documentclass[12pt]{article}
\usepackage{amsfonts}
\usepackage{amsfonts}
\usepackage[dvips]{graphicx}
\usepackage{latexsym,amsmath,amssymb,stmaryrd}
\usepackage{cite}
\usepackage{array}
\usepackage{subfigure}
\usepackage{multirow}
\usepackage{epsfig}
\usepackage[dvips]{color}
\usepackage{pstricks}
\usepackage{pst-node}
\usepackage{pst-plot}
\usepackage{dsfont}
\usepackage{amsthm}
\usepackage{graphics}
\usepackage{fancyhdr}
\usepackage{hhline} 
\usepackage{longtable}
\usepackage{feynmp-eps}
\usepackage{booktabs}
\usepackage{calc}
\usepackage{rotating}

\allowdisplaybreaks[1] 

\newcommand{\covder}{D}
\newcommand{\bcovder}{\bar{D}}

\newcommand{\basisop}[2]{\mathcal{O}_{#1,#2}}

\DeclareMathOperator{\dperm}{\mathbf{P}}
\newcommand{\dchi}[1]{\boldsymbol{\chi}(#1)}

\newcommand{\dpthree}[3]{{}\boldsymbol{\{#1,#2,#3\}}{}}

\newcommand{\swrapthree}[2]{W_{0}^{(3)}}
\newcommand{\bswrapthree}[2]{\bar{W}_{0}^{(3)}}

\newcommand{\sint}[2]{I_{#1}^{(#2)}}

\newcommand{\cfact}[2]{C_{#2}^{(#1)}}
\newcommand{\kint}[2]{K_{#2}^{(#1)}}
\newcommand{\pint}[1]{P^{(#1)}}

\newcommand{\swrap}[2]{W_{#2}^{(#1)}}
\newcommand{\bswrap}[2]{\bar{W}_{#2}^{(#1)}}

\newcommand{\subfigspace}{\phantom{**}} 

\newcommand{\nwgraph}[3]{G_{(#3)}^{(#1)\ifx\am #2\am \else{,#2}\fi}}
\newcommand{\wgraph}[3]{W_{(#3)}^{(#1)\ifx\am #2\am \else{,#2}\fi}}
\newcommand{\scgraph}[3]{s_{(#3)}^{(#1)\ifx\am #2\am \else{,#2}\fi}}
\newcommand{\swgraph}[3]{W_{(#3)}^{(#1)\ifx\am #2\am \else{,#2}\fi}}
\newcommand{\sscgraph}[3]{s_{(#3)}^{(#1)\ifx\am #2\am \else{,#2}\fi}}

\newcommand{\cf}[1]{}
\newcommand{\intf}[1]{}

\newcommand{\sub}[1]{S^{(#1)}}
\newcommand{\bsub}[1]{\bar{S}^{(#1)}}

\definecolor{light-gray}{gray}{0.5}

\fancypagestyle{plain}{
\fancyhf{}
\newcommand{\fpage}{\iffloatpage{}{\thepage}}
\fancyfoot[C]{\fpage}

}

\newcommand{\col}{~,}
\newcommand{\pnt}{~.}
\newcommand{\AdS}{\text{AdS}}
\newcommand{\CFT}{\text{CFT}}

\newcommand{\N}{\mathcal{N}}

\newcommand{\unitmatrix}{\mathds{1}}
\newcommand{\de}{\operatorname{d}\!}

\newcommand{\pfour}[4]{{}\{#1,#2,#3,#4\}{}}
\newcommand{\pthree}[3]{{}\{#1,#2,#3\}{}}
\newcommand{\ptwo}[2]{{}\{#1,#2\}{}}
\newcommand{\pone}[1]{{}\{#1\}{}}
\newcommand{\pid}{{}\{ \}{}}

\newlength{\neglength}
\newlength{\diameter}

\newcommand{\svertex}[2]{%
\fmfiequ{#1}{point length(#2)/2 of #2}
}
\newcommand{\dvertex}[3]{%
\fmfiequ{#1}{point length(#3)/3 of #3}
\fmfiequ{#2}{point 2length(#3)/3 of #3}
}
\newcommand{\vvertex}[3]{%
\fmfipath{px}
\fmfiequ{px}{(0,ypart(#2))..(100,ypart(#2))}
\fmfiequ{#1}{point xpart(#3 intersectiontimes px) of #3}
}

\newcommand{\plainwrap}[4]{%
\fmfipath{pi[]}
\fmfiset{pi1}{vloc(__#1) ..controls (-0.175w,ypart(vloc(__#1))) and (-0.175w,-0.15w) .. (xpart(vloc(__#2)),-0.15w)}
\fmfiset{pi2}{(xpart(vloc(__#2)),-0.15w) ..(xpart(vloc(__#3)),-0.15w)}

\fmfiset{pi3}{(xpart(vloc(__#3)),-0.15w) ..controls (1.175w,-0.15w) and (1.175w,ypart(vloc(__#4))) .. vloc(__#4)}
\fmfi{plain}{pi1 ..pi2 ..pi3}
}
\newcommand{\wigglywrap}[4]{%
\fmfipath{pi[]}
\fmfiset{pi1}{#1 ..controls (-0.175w,ypart(#1)) and (-0.175w,-0.15w) .. (xpart(vloc(__#2)),-0.15w)}
\fmfiset{pi2}{(xpart(vloc(__#2)),-0.15w) ..(xpart(vloc(__#3)),-0.15w)}

\fmfiset{pi3}{(xpart(vloc(__#3)),-0.15w) ..controls (1.175w,-0.15w) and (1.175w,ypart(#4)) .. #4}
\fmfi{wiggly}{pi1 ..pi2 ..pi3}
}

\newcommand{\WoneplainB}{%
\fmftop{v1}
\fmfbottom{v5}
\fmfforce{(0.125w,h)}{v1}
\fmfforce{(0.125w,0)}{v5}
\fmffixed{(0.25w,0)}{v1,v2}
\fmffixed{(0.25w,0)}{v2,v3}
\fmffixed{(0.25w,0)}{v3,v4}
\fmffixed{(0.25w,0)}{v5,v6}
\fmffixed{(0.25w,0)}{v6,v7}
\fmffixed{(0.25w,0)}{v7,v8}
\fmf{plain,tension=0.5,right=0.25}{v2,vc1}
\fmf{plain,tension=0.5,left=0.25}{v3,vc1}
  \fmf{plain}{vc1,vc2}
\fmf{plain,tension=0.5,left=0.125}{vc3,vc2}
\fmf{plain,tension=0.5,left=0.25}{v6,vc2}
\fmf{plain,tension=0.5,right=0.25}{v7,vc2}
\fmfposition
\fmfipath{p[]}
\fmfiset{p1}{vpath(__v1,__v5)}
\fmfiset{p2}{vpath(__v2,__vc1)}
\fmfiset{p3}{vpath(__v3,__vc1)}
\fmfiset{p4}{vpath(__vc1,__vc2)}
\fmfiset{p5}{vpath(__v6,__vc2)}
\fmfiset{p6}{vpath(__v7,__vc2)}
\fmfiset{p7}{vpath(__v4,__v8)}
}

\newcommand{\Wthreeplain}{%
\fmftop{v1}
\fmfbottom{v5}
\fmfforce{(0.125w,h)}{v1}
\fmfforce{(0.125w,0)}{v5}
\fmffixed{(0.25w,0)}{v1,v2}
\fmffixed{(0.25w,0)}{v2,v3}
\fmffixed{(0.25w,0)}{v3,v4}
\fmffixed{(0.25w,0)}{v5,v6}
\fmffixed{(0.25w,0)}{v6,v7}
\fmffixed{(0.25w,0)}{v7,v8}
\fmffixed{(0,whatever)}{vc1,vc2}
\fmffixed{(0,whatever)}{vc3,vc4}
\fmf{plain,tension=0.5,right=0.25}{v2,vc1}
\fmf{plain,tension=0.5,left=0.25}{v3,vc1}
\fmf{plain,right=0.25}{v1,vc3}
\fmf{plain,tension=0.5,left=0.25}{v5,vc4}
\fmf{plain,tension=0.5,right=0.25}{v6,vc4}
\fmf{plain,right=0.25}{v7,vc2}
\fmf{plain}{v8,v4}
  \fmf{plain,tension=1}{vc1,vc2}
  \fmf{plain,tension=0.5}{vc2,vc3}
  \fmf{plain,tension=1}{vc3,vc4}
\fmf{plain,tension=0.5,right=0,width=1mm}{v5,v8}
\fmfposition
\fmfipath{p[]}
\fmfiset{p1}{vpath(__v1,__vc3)}
\fmfiset{p2}{vpath(__vc3,__vc4)}
\fmfiset{p3}{vpath(__v5,__vc4)}
\fmfiset{p4}{vpath(__v3,__vc1)}
\fmfiset{p5}{vpath(__vc1,__vc2)}
\fmfiset{p6}{vpath(__v7,__vc2)}
\fmfiset{p7}{vpath(__v4,__v8)}
}

\DeclareMathOperator{\tr}{tr}

\DeclareMathOperator{\perm}{P}

\numberwithin{equation}{section}
\newlength{\eqoff}
\newlength{\eqofftwo}

\newlength{\unit}
\psset{xunit=\unit,yunit=\unit,runit=\unit}
\newlength{\linew}
\psset{linewidth=\linew}
\begin{document}
\begin{fmffile}{fullgraphs}

\newlength{\mwidth}
\settowidth{\mwidth}{\widthof{$_m$}}
\newlength{\lwidth}
\settowidth{\lwidth}{\widthof{$_l$}}
\newlength{\mldiff}
\setlength{\mldiff}{\mwidth-\lwidth}
\newlength{\fwidth}
\settowidth{\fwidth}{\widthof{$_f$}}
\newlength{\ewidth}
\settowidth{\ewidth}{\widthof{$_e$}}
\newlength{\fediff}
\setlength{\fediff}{\fwidth-\ewidth}

\fmfcmd{
wiggly_len := 2mm;
vardef wiggly expr p_arg =
 save wpp,len;
 numeric wpp,alen;
 wpp = ceiling (pixlen (p_arg, 10) / wiggly_len);
 len=length p_arg;
 for k=0 upto wpp - 1:
  point arctime k/(wpp-1)*arclength(p_arg) of p_arg of p_arg
    {direction arctime k/(wpp-1)*arclength(p_arg) of p_arg of p_arg rotated wiggly_slope} ..
  point  arctime (k+.5)/(wpp-1)*arclength(p_arg) of p_arg of p_arg
 {direction arctime (k+.5)/(wpp-1)*arclength(p_arg) of p_arg of p_arg rotated - wiggly_slope} ..
 endfor
 if cycle p_arg: cycle else: point infinity of p_arg fi
enddef;
}
\fmfcmd{%
marksize=2mm;
def draw_mark(expr p,a) =
  begingroup
    save t,tip,dma,dmb; pair tip,dma,dmb;
    t=arctime a of p;
    tip =marksize*unitvector direction t of p;
    dma =marksize*unitvector direction t of p rotated -45;
    dmb =marksize*unitvector direction t of p rotated 45;
    linejoin:=beveled;
    draw (-.5dma.. .5tip-- -.5dmb) shifted point t of p;
  endgroup
enddef;
style_def derplain expr p =
    save amid;
    amid=.5*arclength p;
    draw_mark(p, amid);
    draw p;
enddef;
def draw_point(expr p,a) =
  begingroup
    save t,tip,dma,dmb,dmo; pair tip,dma,dmb,dmo;
    t=arctime a of p;
    tip =marksize*unitvector direction t of p;
    dma =marksize*unitvector direction t of p rotated -45;
    dmb =marksize*unitvector direction t of p rotated 45;
    dmo =marksize*unitvector direction t of p rotated 90;
    linejoin:=beveled;
    draw (-.5dma.. .5tip-- -.5dmb) shifted point t of p withcolor 0white;
    draw (-.5dmo.. .5dmo) shifted point t of p;
  endgroup
enddef;
style_def derplainpt expr p =
    save amid;
    amid=.5*arclength p;
    draw_point(p, amid);
    draw p;
enddef;
style_def dblderplain expr p =
    save amidm;
    save amidp;
    amidm=.5*arclength p-0.75mm;
    amidp=.5*arclength p+0.75mm;
    draw_mark(p, amidm);
    draw_point(p, amidp);
    draw p;
enddef;
}

\begin{titlepage}
\begin{flushright}
IFUM-954-FT \\
\end{flushright}
\mbox{ }
\vspace{7ex}

\Large
\begin {center}     
{\bf
Superspace methods for the computation of wrapping effects in the standard and $\beta$-deformed ${\cal{N}}=4$ SYM}
\end {center}

\renewcommand{\thefootnote}{\fnsymbol{footnote}}

\large
\vspace{1cm}
\centerline{F.\ Fiamberti ${}^{a,b}$, A.\ Santambrogio ${}^b$, 
C.\ Sieg ${}^c$
\footnote[1]{\noindent \tt
francesco.fiamberti@mi.infn.it \\
\hspace*{6.3mm}alberto.santambrogio@mi.infn.it \\ 
\hspace*{6.3mm}csieg@nbi.dk}}
\vspace{4ex}
\normalsize
\begin{center}
\emph{$^a$  Dipartimento di Fisica, Universit\`a degli Studi di Milano, \\
Via Celoria 16, 20133 Milano, Italy}\\
\vspace{0.2cm}
\emph{$^b$ INFN--Sezione di Milano,\\
Via Celoria 16, 20133 Milano, Italy}\\
\vspace{0.2cm}
\emph{$^c$ The Niels Bohr International Academy,\\
The Niels Bohr Institute,\\
Blegdamsvej 17, DK-2100 Copenhagen, Denmark\\
}
\end{center}
\vspace{0.5cm}
\rm
\abstract
We review the general procedure for the field-theoretical computation of wrapping effects in standard and $\beta$-deformed $\N=4$ super Yang-Mills by means of $\N=1$ superspace techniques. In the undeformed theory, these methods allowed to find explicit results at four and five loops for two-impurity operators. In the deformed case, a general expression for the finite-size correction to the anomalous dimension of single-impurity operators at the critical order was obtained.
\normalsize 
\noindent 

\vspace{1cm}
\vfill
\end{titlepage} 

\section{Introduction}
A great progress has been made in the last years towards a better understanding of the original formulation of the AdS/CFT correspondence~\cite{Maldacena:1998re}, which conjectures the equivalence of the $\N=4$ super Yang-Mills (SYM) theory in four dimensions and type II B superstrings on $\AdS_5\times S^5$, with the discovery of several strong hints of integrability on both sides of the duality. The powerful techniques available for integrable systems allow now to compute the planar spectra of the two theories in a particular limit, and several tests on the validity of the conjecture have become possible.

The first discovery of integrability in $\N=4$ SYM was the demonstration, based on the analogy with spin chains, that the restriction of the theory to the $SU(2)$ sector, containing operators built using only two out of the three available complex scalar fields, is integrable at one loop~\cite{Minahan:2002ve,Minahan:2006sk}. In this spin-chain picture, it is natural to see the operators as excited states obtained by adding fields of one type (\emph{impurities}) to a ground state built using only fields of the other kind.
Afterwards, it was first shown that integrability is valid for the whole theory at one loop~\cite{Beisert:2003jj,Beisert:2003yb,Beisert:2004ag,SchaferNameki:2004ik,Beisert:2005di}, and then strong evidence for its extension to two and three loops was found~\cite{Beisert:2003tq,Serban:2004jf,Kotikov:2004er,Beisert:2004hm,Eden:2004ua}. These results led to the believe that integrability may be an exact property to all orders, and a great amount of work was dedicated to the subject, which culminated with the formulation of a proposal for an all-order Bethe ansatz~\cite{Beisert:2003xu,Beisert:2003jb,Beisert:2003ys,Staudacher:2004tk,Beisert:2005fw,Beisert:2005tm}, from which the dilatation operator can be computed. This extension to arbitrary order was made possible by the discovery that the S-matrix of $\N=4$ SYM is fixed (up to a phase factor) by the symmetries of the theory~\cite{Beisert:2005tm}.

During the same years, a great effort was dedicated also to the study of integrability on the string dual of $\N=4$ SYM. The starting point was the realization that classical strings on the $\AdS_5\times S^5$ background are integrable~\cite{Mandal:2002fs,Bena:2003wd,Kazakov:2004qf,Arutyunov:2004yx,Alday:2005gi}, followed by the proposal of a Bethe ansatz to describe strings on the $\mathds{R} \times S^3$ subspace of $\AdS_5\times S^5$~\cite{Arutyunov:2003za,Arutyunov:2004vx,Gromov:2006cq}, with the discovery that the S-matrices of the gauge and string theory must be related by a global dressing phase~\cite{Callan:2003xr,Serban:2004jf,Callan:2004uv,Hernandez:2006tk,Beisert:2006ib,Beisert:2006ez,Arutyunov:2004vx,Eden:2006rx,Beisert:2007hz}.
As on the field theory side, the result for the Bethe ansatz and the S-matrix have been later extended to the full theory~\cite{Beisert:2003ea,Arutyunov:2003rg,Arutyunov:2004xy,Kazakov:2004nh,Swanson:2004qa,Beisert:2005bm,Beisert:2005cw,SchaferNameki:2005is,Staudacher:2004tk,Freyhult:2006vr}.

The powerful integrability techniques allow to compute easily the spectrum of $\N=4$ SYM in the limit of long operators. For any given operator, however, they are forced to fail when the loop order becomes high enough~\cite{Beisert:2004hm}.The definition of the S-matrix and of the Bethe ansatz, in fact, requires the existence of an asymptotic regime, i.e.\ the interaction 
must not involve all fields at the same time.
But since the range of the interaction grows with the loop order $\ell$ as $(\ell+1)$, for an operator of length (number of elementary fields) $L$, the interaction range exceeds the length $L$ at loop orders 
$\ell\ge L$.
Thus, the Bethe ansatz breaks down, and it no longer produces the correct components of the anomalous dimension. In terms of Feynman diagrams, such failure of the asymptotic tools is caused by the appearance of wrapping diagrams, in which
the interactions wrap around the composite operator.

The finite-size effects generated by wrapping interactions must be studied in order to compute the exact spectrum of $\N=4$ SYM. The first analysis of these corrections in terms of Feynman diagrams was performed in~\cite{Sieg:2005kd}, and in~\cite{Ambjorn:2005wa,Janik:2007wt} some proposals were considered for their general description, among which the use of the thermodynamic Bethe ansatz. The first exact computation of a field-theory quantity affected by wrapping corrections appeared in~\cite{us,uslong}, where the four-loop anomalous dimension of the Konishi operator was calculated by means of superspace techniques. The result was then confirmed first by an independent analysis, based on the L\"uscher approach~\cite{Luscher:1985dn}, performed on the string side~\cite{Bajnok:2008bm}, which thus constitutes a highly non-trivial check of the AdS/CFT correspondence, and later by a computer-made direct computation based on the component-field approach~\cite{Velizhanin:2009zz}. The superspace computation has been recently extended to consider length-five operators at five loops~\cite{usfive}. 

At the same time, finite-size effects were studied also on the string side~\cite{SchaferNameki:2006gk,Arutyunov:2006gs,SchaferNameki:2006ey,Astolfi:2007uz,Ambjorn:2005wa,Minahan:2008re,Gromov:2008ie,Heller:2008at,Hatsuda:2008gd,Ramadanovic:2008qd,Hatsuda:2008na,Janik:2007wt,Sax:2008in}. In most cases, however, the calculations were carried out in the limit of large 't~Hooft coupling $\lambda$ from the very beginning, so that the final results cannot be compared directly with the ones coming from the gauge side, where $\lambda\ll1$ must be considered in a perturbative approach.

In parallel with the explicit computations based on diagrammatic techniques, also the method based on the L\"uscher approach has been generalized to deal with different classes of operators~\cite{Penedones:2008rv,Bajnok:2008qj,Beccaria:2009eq,Lukowski:2009ce,Velizhanin:2010cm}, and a proposal for the five-loop component of the Konishi anomalous dimension has been obtained~\cite{Bajnok:2009vm}.
Despite these successful applications, however, it is very difficult to extend the L\"uscher technique to the most general cases,
and hence also different approaches have been taken into consideration. Currently, the most promising one is based on the thermodynamic Bethe ansatz~\cite{Zamolodchikov:1989cf,Arutyunov:2007tc,deLeeuw:2008ye,Arutyunov:2009zu,deLeeuw:2009hn,Arutyunov:2009ga,Arutyunov:2009mi,Arutyunov:2009ce,Arutyunov:2009ux,Arutyunov:2009ax,Arutyunov:2010gb,Balog:2010xa,Balog:2010vf}, which led to the proposal that the full, exact spectrum of $\N=4$ SYM is captured
in terms of a so-called Y-system~\cite{Gromov:2009tv,Gromov:2009bc,Bombardelli:2009ns,Arutyunov:2009ur,Hegedus:2009ky,Cavaglia:2010nm}.
This method has already been shown to reproduce the known field theory results at four and five loops with wrapping effects included~\cite{us,uslong,Gromov:2009tv,usfive,Beccaria:2009eq}, the five-loop result for the Konishi operator from the L\"uscher approach \cite{Bajnok:2009vm,Balog:2010xa,Arutyunov:2010gb}, and it has also been used to compute 
finite-size corrections 
at strong coupling~\cite{Gromov:2009zb,Roiban:2009aa,Gromov:2009tq,Gromov:2010vb}.

A similar agreement was found at four loops also in the context of the $\AdS_4/\CFT_3$ correspondence~\cite{Gromov:2009tv,Minahan:2009aq,Minahan:2009wg}.
In the future it will be very important to further check the validity of the Y-system against additional direct perturbative results.

Besides $\N=4$ SYM, also its $\beta$-deformed version, preserving only $\N=1$ supersymmetry, offers an interesting environment for the analysis of wrapping effects. The study of finite-size corrections in this case started in~\cite{betadef} and was later extended in~\cite{Fiamberti:2008sn}, whereas in~\cite{Bykov:2008bj} wrapping corrections were discussed on the deformed string background. The most interesting feature of the deformed theory is the possibility to study a simpler class of operators, so that perturbative computations at higher orders become feasible. 

In this paper, we review all the perturbative computations of wrapping effects, based on $\N=1$ superspace techniques, that have been performed both in the standard and in the $\beta$-deformed $\N=4$ SYM. First of all, in Section~\ref{superspace}, we present the main technical results that make superspace techniques so useful for this kind of calculations. Then, in Section~\ref{undeformed}, we describe the general procedure for the analysis of finite-size corrections in the standard $\N=4$ SYM and its applications to four and five loops. Section~\ref{betadef} is dedicated to the topic of wrapping effects in the $\beta$-deformed $\N=4$ SYM. We conclude with some comments in Section~\ref{comments}.

\section{Superspace techniques}
\label{superspace}
Perturbative computations in gauge field theories at high loop orders, based on the standard component-field approach, are usually very complicated. Hence, when supersymmetry is present, it is very useful to exploit it by making use of superspace techniques. For both standard and $\beta$-deformed $\N=4$ SYM, it is convenient to use the $\N=1$ superspace description~\cite{Gates:1983nr}, where all the field content is encoded into one vector $V$ and three scalar superfields $\Phi^i$, which we denote as $\phi$, $Z$ and $\psi$. Therefore, fermionic matter fields never appear explicitly, which considerably simplifies all the computations. Moreover, every supergraph combines the information on a large number of standard component-field diagrams, so that the total number of relevant contributions is reduced, too.

\subsection{The action of $\N=4$ SYM}
The action for undeformed $\N=4$ SYM reads, in the notation of~\cite{Gates:1983nr}, 
\begin{equation}
\label{action}
\begin{aligned}
S &= \int\de^4 x\de^4 \theta \, \tr \left(e^{-gV} \bar \Phi_i e^{gV}
\Phi^i\right) + \frac 1{2g^2} \int \de^4 x \de^2 \theta
\,\tr \left(W^\alpha W_\alpha\right)\\
&\phantom{{}={}}
+i \frac{g}{3!}  \int\de^4 x\de^2 \theta \,\epsilon^{ijk}\,\tr \left(\Phi_i
\left[\Phi_j , \Phi_k\right]\right) + \text{h.c.}\pnt
\end{aligned}
\end{equation}
Here, $W_\alpha = i\bar \covder^2 \left(e^{-gV} \covder_\alpha\,e^{gV}\right)$,
$V=V^aT_a$, $\Phi^i=\Phi_i^aT_a$, {\small $i=1,2,3$}, and the $T_a$ are matrices that are normalized as 
\begin{equation}
\tr(T_a T_b) =\delta_{ab}
\col
\end{equation}
and that obey the $SU(N)$ algebra
\begin{equation}
\label{Tcomm}
[T_a, T_b] = i f_{abc} T_c \col
\end{equation}
where the $f_{abc}$ are the $SU(N)$ structure constants. The latter can be written in terms of the $T_a$ through to the inverse of~\eqref{Tcomm}
\begin{equation}
f_{abc}=-i\,\mathrm{tr}([T_a,T_b]T_c) \pnt
\end{equation}
This relation and the identity
\begin{equation}
T_a^{ij}T_a^{kl}=\left(\delta_{il}\delta_{jk}-\frac{1}{N}\delta_{ij}\delta_{kl}\right) \col
\end{equation}
allow to determine the colour structures of Feynman diagrams.
Since we will make all our computations in the planar limit, in addition to the gauge coupling $g$, it will be useful to define also the rescaled 't~Hooft coupling
\begin{equation}
\lambda=\frac{g^2N}{(4\pi)^2} \pnt
\end{equation}
The Feynman rules for propagators and vertices in supergraphs can be derived from the action~\eqref{action}. In momentum space, the propagators are
\begin{equation}
\label{propagators}
\langle V^a V^b\rangle=-\frac{\delta^{ab}}{p^2} \col \qquad\langle\Phi^a_i\bar{\Phi}^b_j\rangle=\delta_{ij}\frac{\delta^{ab}}{p^2} \pnt
\end{equation}
As for the vertex factors, the last term in~\eqref{action} describes the interactions among three scalar chiral or anti-chiral superfields, whose contributions are
\begin{equation}
\label{vertices-scalar}
V_C=-\frac{g}{3!}\epsilon^{ijk}f_{abc}\Phi^a_i\Phi^b_j\Phi^c_k\ ,\quad V_A=-\frac{g}{3!}\epsilon^{ijk}f_{abc}\bar{\Phi}^a_i\bar{\Phi}^b_j\bar{\Phi}^c_k\col
\end{equation}
whereas the first term in the action generates vertices with one chiral and one anti-chiral scalar, plus a maximum number of vectors growing with the perturbative order under consideration. In this paper, we will encounter only the cases with one or two vectors, whose factors are respectively
\begin{equation}
\label{vertices-vector}
V_V^{(1)}=i g f_{abc}\delta^{ij}\bar{\Phi}^a_i V^b \Phi^c_j\ ,\quad V_V^{(2)}=\frac{g^2}{2}\delta^{ij}f_{adm}f_{bcm}V^a V^b \bar{\Phi}^c_i\Phi^d_j \pnt
\end{equation}
Finally, the second term in the action~\eqref{action} produces vertices where only vector superfields interact, and which will not be needed for our computations. In the same way, ghost fields will never be relevant in this work.

In addition to such vertex factors, one must remember to add a $\bar{D}^2$ or a $D^2$ covariant derivative to each chiral or antichiral line, respectively, to restore the full $\de^4\theta$ integration measure on the Grassmann variables. In the case of three-scalar vertices, only two out of the three lines carry derivatives. Once a supergraph has been built, we apply the D-algebra procedure~\cite{Gates:1983nr}, which consists in a sequence of integration by parts of the covariant derivatives resulting in a reduction of the diagram to a standard momentum integral.

\subsection{The $\beta$-deformed theory}
The $\beta$-deformed $\N=4$ SYM theory is obtained from the standard one through the following modification of the superpotential for chiral and anti-chiral superfields
\begin{equation}
ig\,\tr\left(\phi\,\psi\,Z -  \phi\,Z\,\psi\right)~\longrightarrow ~ih\,\tr\left(e^{i\pi\beta} \phi\,\psi\,Z - e^{-i\pi\beta} \phi\,Z\,\psi\right)\col
\end{equation}
where $h$ and $\beta$ are complex constants, and we recall that $\phi$, $Z$ and $\psi$ are the three scalar chiral superfields. Such a deformation is marginal, and thus the theory remains conformally invariant, at all orders~\cite{Leigh:1995ep,Mauri:2005pa} if
\begin{equation}
h\bar{h}=g^2 \col
\end{equation}
where $g$ is the Yang-Mills coupling constant. In this paper we consider only the case of real $\beta$. In fact in this case the deformed theory is also believed to be integrable~\cite{Roiban:2003dw,Berenstein:2004ys,Beisert:2005if}, together with its string dual, namely superstring theory on the Lunin-Maldacena background~\cite{Lunin:2005jy,Frolov:2005dj,Frolov:2005ty}, and an all-loop Bethe ansatz similar to the standard one has been formulated~\cite{Beisert:2005if}.

As far as the Feynman rules are concerned, only the vertex coefficients for the three-scalar interactions will be modified by the deformation, with the appearance of a factor of $q\equiv e^{i\pi\beta}$ or $\bar{q}=e^{-i\pi\beta}$, depending on the order of the fields
\begin{equation}
\begin{aligned}
V_C&= - h f_{abc} (e^{i\pi\beta} \Phi_1^a\,\Phi_2^b\,\Phi_3^c - e^{-i\pi\beta} \Phi_1^a\,\Phi_3^b\,\Phi_2^c) \col \\
V_A&= - \bar{h} f_{abc} (e^{-i\pi\beta}\bar{\Phi}_1^a\, \bar{\Phi}_2^b\, \bar{\Phi}_3^c-e^{i\pi\beta} \bar{\Phi}_1^a\, \bar{\Phi}_3^b\, \bar{\Phi}_2^c) \pnt
\end{aligned}
\end{equation}
All the other vertex factors and the propagators remain the same as in the undeformed case.

The $\beta$-deformed theory is particularly interesting because some classes of simple operators that were protected by supersymmetry in the undeformed $\N=4$ SYM now acquire a non-trivial anomalous dimension. As we will explain later, this fact will allow us to perform perturbative computations to orders beyond four or five loops.

\subsection{Anomalous dimensions}
The analysis of wrapping effects that we are going to present is based on the computation of anomalous dimensions of composite operators. Given a set of bare operators $\{\mathcal{O}_1,\ldots,\mathcal{O}_n\}$ with the same classical dimension, in general they will be mixed by quantum corrections, and their renormalized versions will be given by
\begin{equation}
\mathcal{O}_i^{\mathrm{ren}}=\mathcal{Z}_i^j\mathcal{O}_j^{\mathrm{bare}} \col
\end{equation}
where $\mathcal{Z}_i^j$ is the one-point function matrix, which gets contributions from all the Feynman diagrams with one insertion of one of the composite operators $\mathcal{O}_i$. In order to find the linear combinations of the $\mathcal{O}_i$ with well-defined anomalous dimensions, we must diagonalize this matrix. Working in dimensional regularization, with $d=4-2\varepsilon$ spacetime dimensions, the anomalous dimensions will be the eigenvalues of the mixing matrix $\mathcal{M}$ defined as
\begin{equation}
\label{anomalous}
\gamma_k=\mathrm{eig}(\mathcal{M})_k\ ,\qquad\mathcal{M}_i^j=\lim_{\varepsilon\rightarrow0}\left[\varepsilon g\frac{\de}{\de g}\mathrm{log}\mathcal{Z}_i^j(g,\varepsilon)\right] \pnt
\end{equation}
It is worth emphasizing here that the computation of anomalous dimensions requires only the knowledge of the \emph{divergent} part of the expansion of every graph in powers of $1/\varepsilon$. As we will explain later, this fact leads to a great simplification in the calculation of the integrals. 

It is useful to present now a different approach~\cite{Beisert:2003tq} to the computation of anomalous dimensions. In a conformal theory, the possible quantum dimensions of operators are the eigenvalues of the dilatation operator $\mathcal{D}$, which represents the generator of dilatation transformations on the operator algebra, whereas the corresponding eigenvectors are the composite operators that renormalize multiplicatively. Thus,
\begin{equation}
\mathcal{D}\,\mathcal{O}=\Delta(\lambda)\mathcal{O}\ ,\qquad\Delta(\lambda)=\Delta_0+\gamma(\lambda) \col 
\end{equation}
where $\Delta_0$ is the classical dimension and $\gamma(\lambda)$ is the anomalous one.
We will focus on the perturbative expansion of $\mathcal{D}$ in powers of the 't~Hooft coupling
\begin{equation}
\mathcal{D}(\lambda)=\sum_{k=0}^\infty\lambda^k \mathcal{D}_k \pnt
\end{equation}
The importance of the dilatation operator in $\N=4$ SYM (and in its deformed version) comes from the fact that the hypothesis of integrability allows to compute its components in the planar limit without the need for explicit diagrammatic computations. This simplification allowed to find the explicit form of $\mathcal{D}$ in a particular sector up to five loops.

From a simple analysis of the general properties of planar Feynman diagrams, it follows that 
the range of interaction, that is, the maximum number of fields of the composite operator that interact with each other, 
grows with the perturbative order. In fact, at $\ell$ loops the dilatation operator will get contributions from graphs with range up to $\ell+1$. Note that the range of a Feynman diagram is a meaningful quantity only in the planar limit.

\subsection{A theorem about supergraphs}
We now demonstrate a very useful result~\cite{uslong}, which allows to simplify considerably the D-algebra procedure for a Feynman supergraph by identifying operations that lead to finite contributions and hence are irrelevant for our analysis of anomalous dimensions. Let us consider a planar graph with a single insertion of a composite operator made of chiral superfields, and where the final operator contains the same fields. In Figure~\ref{example-01} an example is shown, where the thick line represents the composite operator, straight thin lines are scalar propagators and wiggly lines are vector fields. Moreover, chiral scalar vertices are marked with a circle, and for the sake of simplicity we did not show explicitly the covariant derivatives. 

We want to show that divergent contributions can be obtained after D-algebra only if all the covariant derivatives of chiral type $D$ are kept inside the diagram during the sequence of integrations by parts (with the exception, as will be clear in the following, of derivatives on lines not belonging to any loop). 

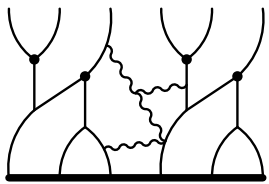
\begin{table}[t]
\begin{tabular}{ll}
\toprule
$V_C$ & number of chiral vertices \\
$V_A$ & number of antichiral vertices \\
$V_V^{(n)}$ & number of vertices with a chiral, an antichiral and $n$ vector lines \\
$\tilde{V}_V^{(n)}$ & number of vertices with $n$ vector lines \\
$p_S$ & number of scalar propagators belonging to at least one loop \\
$p_V$ & number of vector propagators \\
$p$ & total number of propagators belonging to at least one loop \\
$p_E$ & number of scalar propagators not belonging to any loop \\
$V_{EC}$ & number of chiral vertices not belonging to any loop \\
$N_\ell$ & number of loops \\
$N_\covder$, $N_{\bcovder}$ & numbers of $\covder$ and $\bcovder$ derivatives \\
\bottomrule
\end{tabular}
\caption{Useful definitions}
\label{tab:proof}
\end{table}

\begin{figure}[!h]
\vspace{0.5cm}
\begin{minipage}[c]{0.5\linewidth}
\centering
\unitlength=0.75mm
\settoheight{\eqoff}{$\times$}%
\setlength{\eqoff}{0.5\eqoff}%
\addtolength{\eqoff}{-12.5\unitlength}%
\settoheight{\eqofftwo}{$\times$}%
\setlength{\eqofftwo}{0.5\eqofftwo}%
\addtolength{\eqofftwo}{-7.5\unitlength}%
\raisebox{\eqoff}{%
\fmfframe(3,1)(1,4){%
\begin{fmfchar*}(45,30)
\fmftop{v1}
\fmfbottom{v7}
\fmfforce{(0w,h)}{v1}
\fmfforce{(0w,0)}{v7}
\fmffixed{(0.2w,0)}{v1,v2}
\fmffixed{(0.2w,0)}{v2,v3}
\fmffixed{(0.2w,0)}{v3,v4}
\fmffixed{(0.2w,0)}{v4,v5}
\fmffixed{(0.2w,0)}{v5,v6}
\fmffixed{(0.2w,0)}{v7,v8}
\fmffixed{(0.2w,0)}{v8,v9}
\fmffixed{(0.2w,0)}{v9,v10}
\fmffixed{(0.2w,0)}{v10,v11}
\fmffixed{(0.2w,0)}{v11,v12}
\fmffixed{(0.1w,0)}{v3,vb3}
\fmffixed{(0.1w,0)}{v9,vb9}
\fmf{plain,tension=0.5,right=0.25}{v1,vc1}
\fmf{plain,tension=0.5,left=0.25}{v2,vc1}
\fmf{plain,left=0.25}{v3,vc3}
\fmf{plain,tension=0.5,left=0.25}{v8,vc4}
\fmf{plain,tension=0.5,right=0.25}{v9,vc4}
\fmf{plain,left=0.25}{v7,vc2}
\fmf{plain,tension=1}{vc1,vc2}
\fmf{plain,tension=0.5}{vc2,vc3}
\fmf{plain,tension=1}{vc3,vc4}
\fmf{plain,tension=0.5,right=0.25}{v4,vc7}
\fmf{plain,tension=0.5,left=0.25}{v5,vc7}
\fmf{plain,left=0.25}{v6,vc9}
\fmf{plain,tension=0.5,left=0.25}{v11,vc10}
\fmf{plain,tension=0.5,right=0.25}{v12,vc10}
\fmf{plain,left=0.25}{v10,vc8}
\fmf{plain,tension=1}{vc7,vc8}
\fmf{plain,tension=0.5}{vc8,vc9}
\fmf{plain,tension=1}{vc9,vc10}
\fmf{phantom,tension=1}{vb3,vb9}
\fmffreeze
\fmfposition
\fmf{plain,tension=1,right=0,width=1mm}{v7,v12}
\fmfipath{p[]}
\fmfiset{p1}{vpath(__vc3,__vc4)}
\fmfiset{p2}{vpath(__v10,__vc8)}
\fmfiset{p3}{vpath(__v9,__vc4)}
\fmfiset{p4}{vpath(__v3,__vc3)}
\fmfiset{p5}{vpath(__vc7,__vc8)}
\fmfiset{p6}{vpath(__vb3,__vb9)}
\fmfiset{p7}{vpath(__vc1,__vc2)}
\fmfiset{p8}{vpath(__v6,__vc9)}
\fmfipair{w[]}
\fmfipair{wa[]}
\svertex{w1}{p1}
\svertex{w2}{p2}
\svertex{w3}{p3}
\svertex{w4}{p4}
\svertex{w5}{p5}
\svertex{w6}{p6}
\fmfi{wiggly}{w2..w3}
\fmfi{wiggly}{w4..w6}
\fmfi{wiggly}{w6..w5}
\fmfi{wiggly}{w6..w2}
\fmfiequ{wa1}{point 0*length(p7) of p7}
\fmfiequ{wa2}{point 1*length(p4) of p4}
\fmfiequ{wa3}{point 0*length(p5) of p5}
\fmfiequ{wa4}{point 1*length(p8) of p8}
\fmfiv{d.shape=circle,d.size=3}{wa1}
\fmfiv{d.shape=circle,d.size=3}{wa2}
\fmfiv{d.shape=circle,d.size=3}{wa3}
\fmfiv{d.shape=circle,d.size=3}{wa4}
\fmffreeze
\end{fmfchar*}}}
\end{minipage}
\begin{minipage}[c]{0.5\linewidth}
\footnotesize
\begin{tabular}{l}
$N_\ell=7$ \\
$V_C=V_A=4$ \\
$V_V^{(1)}=3\col\quad V_V^{(2)}=1\col\quad V_V^{(n>2)}=0$ \\
$\tilde{V}_V^{(3)}=1\col\quad \tilde{V}_V^{(m>3)}=0$ \\
$p_S=14\col\quad p_V=4\col\quad p=p_S+p_V=18$ \\
$p_E=2\col\quad V_{EC}=2$ \\
\end{tabular}
\normalsize
\end{minipage}
\caption{Example of Feynman supergraph}
\label{example-01}
\end{figure}

To demonstrate our assertion, it is useful to define the quantities of Table~\ref{tab:proof}. We can find several equations involving them, based on the possible types of vertices in the graph. In particular,
\begin{itemize}
\item from each chiral (anti-chiral) vertex, three scalar propagators start. On two of them, a $\bar{D}^2$ ($D^2$) double covariant derivative acts.
\item From each vertex of type $V_V^{(n)}$, two scalar (one chiral and one anti-chiral) and $n$ vector lines start. On the chiral and anti-chiral lines, a $\bar{D}^2$ and a $D^2$ act respectively.
\item From each vertex of type $\tilde{V}_V^{(n)}$, $n$ vector propagators start. A complicated derivative structure will act on the lines, but in all the cases two $\bar{D}$ and two $D$ derivatives will be involved. 
\end{itemize}
In this way, every propagator is counted twice, since by hypothesis the outgoing fields are the same as the ones in the composite operator. So for the propagator numbers we find
\begin{equation}
\label{proof:propagators}
\begin{aligned}
p_S&=\frac{1}{2}\left[3(V_C+V_A)+2\sum_{n\geq1} V_V^{(n)}\right]-p_E \col \\
p_V&=\frac{1}{2}\left[\sum_{n\geq1} nV_V^{(n)}+\sum_{m\geq3}m\tilde{V}_V^{(m)}\right] \col
\end{aligned}
\end{equation}
whereas the numbers of covariant derivatives fulfill
\begin{equation}
\label{proof:ND}
\begin{aligned}
N_\covder&=4V_C+2\sum_{n\geq1}V_V^{(n)}+2\sum_{m\geq3}\tilde{V}_V^{(m)} \col \\
N_{\bcovder}&=4V_A+2\sum_{n\geq1}V_V^{(n)}+2\sum_{m\geq3}\tilde{V}_V^{(m)} \pnt
\end{aligned}
\end{equation}
For the hypothesis on the outgoing fields, the numbers of chiral and anti-chiral vertices must be equal
\begin{equation}
V_C=V_A \col
\end{equation}
and hence we have $N_\covder=N_{\bcovder}$. Moreover, the number of scalar propagators that do not belong to any loop and the number of chiral vertices with the same feature must be equal: $p_E=V_{EC}$. 
We can now combine the equations~\eqref{proof:propagators} and~\eqref{proof:ND} into
\begin{equation}
\label{proof:p}
p=\frac{1}{2}\left[N_\covder+V_C+V_A+\sum_{n\geq1}n V_V^{(n)}+\sum_{m\geq3}(m-2)\tilde{V}_V^{(m)}\right]-p_E \pnt
\end{equation}
From a simple power counting, it follows that the final integral will be at least logarithmically divergent only if the D-algebra generates at least $(2p-4N_\ell)$ momenta in the numerator. The construction of each momentum requires a $D$ derivative (and a $\bar{D}$). Moreover, for every loop a $D^2$ and a $\bar{D}^2$ are required to complete the superspace integration. Thus, performing the D-algebra we need to keep inside the diagram at least $(2p-2N_\ell)$ derivatives of type $D$.

Now we must get rid of propagators not belonging to any loop. It is easy to see that on any one of them, either a $D^2$ is already there from the beginning, as it appears when the propagator is connected to a vertex of kind $V_V^{(n)}$, or it can be moved there by an integration by parts at the vertex at which the propagator is attached to the rest of the diagram. So we can always assume that every such propagator has a $D^2$, which is hence effectively \emph{outside} the graph. Thus, the actual number of $D$ derivatives that can be effectively used is just $(N_D-2p_E)$. We will be allowed to move one more $D$ out of the diagram only if this number exceeds the required minimum, i.e.\ if
\begin{equation}
\label{proof:ineq}
N_\covder-2p_E>2p-2N_\ell \col
\end{equation}
which we can rewrite using~\eqref{proof:p} as
\begin{equation}
\label{proof:ineq2}
N_\ell>\frac{1}{2}\left[V_C+V_A+\sum_{n\geq1}n V_V^{(n)}+\sum_{m\geq3}(m-2)\tilde{V}_V^{(m)}\right] \pnt
\end{equation}
As the last step of the proof, it is enough to show that this inequality can never be fulfilled. In fact, let us make use of Euler's formula for planar connected graphs
\begin{equation}
\label{proof:Euler}
\mathcal{V}-\mathcal{E}+\mathcal{F}=2 \col
\end{equation}
where $\mathcal{V}$ is the total number of vertices, $\mathcal{E}$ is the number of edges and $\mathcal{F}$ is the number of faces, including the external unbounded region. For a Feynman supergraph, as the operator insertion behaves as an additional vertex, we have
\begin{equation}
\mathcal{V}=V_C+V_A+\sum_{n\geq1}V_V^{(n)}+\sum_{m\geq3}\tilde{V}_V^{(m)}+1\ ,\qquad \mathcal{E}=p+p_E\ ,\qquad \mathcal{F}=N_\ell+1\pnt
\end{equation}
So we find
\begin{equation}
N_\ell=\frac{1}{2}\left[V_C+V_A+\sum_{n\geq1}n V_V^{(n)}+\sum_{m\geq3}(m-2)\tilde{V}_V^{(m)}\right] \col
\end{equation}
which is not compatible with~\eqref{proof:ineq}. We have therefore shown that none of the derivatives of type $D$ can be moved out of the diagram during the D-algebra.

\subsection{Cancellation identities for supergraphs}
The property that we have demonstrated above can be exploited to show that large classes of supergraphs entering the computation of anomalous dimensions sum up to finite expressions. In particular, we are interested in diagrams with the maximum interaction range allowed by the corresponding number of loops.

Consider a maximum-range diagram that contains a single-vector vertex breaking an outgoing scalar propagator, as in the example of Figure~\ref{example-02}. 

\begin{figure}[!h]
\vspace{0.5cm}
\centering
\unitlength=0.75mm
\settoheight{\eqoff}{$\times$}%
\setlength{\eqoff}{0.5\eqoff}%
\addtolength{\eqoff}{-12.5\unitlength}%
\settoheight{\eqofftwo}{$\times$}%
\setlength{\eqofftwo}{0.5\eqofftwo}%
\addtolength{\eqofftwo}{-7.5\unitlength}%
\raisebox{\eqoff}{%
\fmfframe(3,1)(1,4){%
\begin{fmfchar*}(45,30)
\fmftop{v1}
\fmfbottom{v7}
\fmfforce{(0w,h)}{v1}
\fmfforce{(0w,0)}{v7}
\fmffixed{(0.2w,0)}{v1,v2}
\fmffixed{(0.2w,0)}{v2,v3}
\fmffixed{(0.2w,0)}{v3,v4}
\fmffixed{(0.2w,0)}{v4,v5}
\fmffixed{(0.2w,0)}{v5,v6}
\fmffixed{(0.2w,0)}{v7,v8}
\fmffixed{(0.2w,0)}{v8,v9}
\fmffixed{(0.2w,0)}{v9,v10}
\fmffixed{(0.2w,0)}{v10,v11}
\fmffixed{(0.2w,0)}{v11,v12}
\fmffixed{(0.1w,0)}{v3,vb3}
\fmffixed{(0.1w,0)}{v9,vb9}
\fmf{plain,tension=0.5,right=0.25}{v1,vc1}
\fmf{plain,tension=0.5,left=0.25}{v2,vc1}
\fmf{plain,left=0.25}{v3,vc3}
\fmf{plain,tension=0.5,left=0.25}{v8,vc4}
\fmf{plain,tension=0.5,right=0.25}{v9,vc4}
\fmf{plain,left=0.25}{v7,vc2}
\fmf{plain,tension=1}{vc1,vc2}
\fmf{plain,tension=0.5}{vc2,vc3}
\fmf{plain,tension=1}{vc3,vc4}
\fmf{plain,tension=0.5}{v4,v10}
\fmf{plain,tension=0.5,right=0.25}{v5,vc9}
\fmf{plain,tension=0.5,left=0.25}{v6,vc9}
\fmf{plain,tension=0.5,left=0.25}{v11,vc10}
\fmf{plain,tension=0.5,right=0.25}{v12,vc10}
\fmf{plain,tension=1}{vc9,vc10}
\fmffreeze
\fmfposition
\fmf{plain,tension=1,right=0,width=1mm}{v7,v12}
\fmfipath{p[]}
\fmfiset{p1}{vpath(__vc3,__vc4)}
\fmfiset{p2}{vpath(__v10,__v4)}
\fmfiset{p3}{vpath(__v9,__vc4)}
\fmfiset{p4}{vpath(__v3,__vc3)}
\fmfiset{p5}{vpath(__vc9,__vc10)}
\fmfipair{w[]}
\svertex{w1}{p1}
\svertex{w2}{p2}
\svertex{w3}{p3}
\svertex{w4}{p4}
\vvertex{w5}{w3}{p2}
\dvertex{w6}{w7}{p2}
\svertex{w8}{p5}
\fmfi{wiggly}{w3..w7}
\fmfi{wiggly}{w6..w8}
\fmffreeze
\end{fmfchar*}}}
\caption{Example of maximum-range supergraph}
\label{example-02}
\end{figure}

\begin{figure}[t]
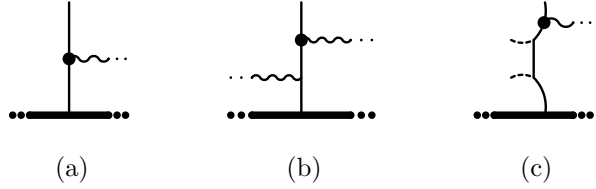

\unitlength=0.75mm
\settoheight{\eqoff}{$\times$}%
\setlength{\eqoff}{0.5\eqoff}%
\addtolength{\eqoff}{-12.5\unitlength}%
\settoheight{\eqofftwo}{$\times$}%
\setlength{\eqofftwo}{0.5\eqofftwo}%
\addtolength{\eqofftwo}{-7.5\unitlength}%
\centering
\raisebox{\eqoff}{%
\subfigure[\hspace{-0.5cm}]{
\label{onevector}
\fmfframe(3,1)(1,4){%
\begin{fmfchar*}(20,20)
\fmftop{v1}
\fmfbottom{v6}
\fmfforce{(0w,h)}{v1}
\fmfforce{(0w,0)}{v6}
\fmffixed{(0.15w,0)}{v1,v2}
\fmffixed{(0.35w,0)}{v2,v3}
\fmffixed{(0.35w,0)}{v3,v4}
\fmffixed{(0.15w,0)}{v4,v5}
\fmffixed{(0.15w,0)}{v6,v7}
\fmffixed{(0.35w,0)}{v7,v8}
\fmffixed{(0.35w,0)}{v8,v9}
\fmffixed{(0.15w,0)}{v9,v10}
\fmf{plain}{v3,v8}
\fmf{phantom}{v2,v7}
\fmf{phantom}{v1,v6}
\fmf{phantom}{v4,v9}
\fmf{phantom}{v5,v10}
\fmf{plain,tension=0.5,right=0,width=1mm}{v7,v9}
\fmf{dots,tension=0.5,right=0,width=1mm}{v6,v7}
\fmf{dots,tension=0.5,right=0,width=1mm}{v9,v10}
\fmffreeze
\fmfposition
\fmfipath{p[]}
\fmfipair{w[]}
\fmfiset{p1}{vpath(__v3,__v8)}
\fmfiset{p2}{vpath(__v2,__v7)}
\fmfiset{p3}{vpath(__v4,__v9)}
\fmfiset{p4}{vpath(__v1,__v6)}
\fmfiset{p5}{vpath(__v5,__v10)}
\fmfiequ{w1}{point length(p1)/2 of p1}
\fmfiequ{w2}{point 2length(p1)/3 of p1}
\vvertex{w4}{w1}{p3}
\vvertex{w6}{w1}{p5}
\fmfi{wiggly}{w1..w4}
\fmfi{dots}{w4..w6}
\fmfiv{d.sh=circle,d.f=1,d.size=4}{w1}
\fmfposition
\end{fmfchar*}}}
\qquad
\subfigure[\hspace{-0.5cm}]{
\label{twovectors}
\fmfframe(3,1)(1,4){%
\begin{fmfchar*}(25,20)
\fmftop{v1}
\fmfbottom{v6}
\fmfforce{(0w,h)}{v1}
\fmfforce{(0w,0)}{v6}
\fmffixed{(0.15w,0)}{v1,v2}
\fmffixed{(0.35w,0)}{v2,v3}
\fmffixed{(0.35w,0)}{v3,v4}
\fmffixed{(0.15w,0)}{v4,v5}
\fmffixed{(0.15w,0)}{v6,v7}
\fmffixed{(0.35w,0)}{v7,v8}
\fmffixed{(0.35w,0)}{v8,v9}
\fmffixed{(0.15w,0)}{v9,v10}
\fmf{plain}{v3,v8}
\fmf{phantom}{v2,v7}
\fmf{phantom}{v1,v6}
\fmf{phantom}{v4,v9}
\fmf{phantom}{v5,v10}
\fmf{plain,tension=0.5,right=0,width=1mm}{v7,v9}
\fmf{dots,tension=0.5,right=0,width=1mm}{v6,v7}
\fmf{dots,tension=0.5,right=0,width=1mm}{v9,v10}
\fmffreeze
\fmfposition
\fmfipath{p[]}
\fmfipair{w[]}
\fmfiset{p1}{vpath(__v3,__v8)}
\fmfiset{p2}{vpath(__v2,__v7)}
\fmfiset{p3}{vpath(__v4,__v9)}
\fmfiset{p4}{vpath(__v1,__v6)}
\fmfiset{p5}{vpath(__v5,__v10)}
\fmfiequ{w1}{point length(p1)/3 of p1}
\fmfiequ{w2}{point 2length(p1)/3 of p1}
\vvertex{w3}{w2}{p2}
\vvertex{w4}{w1}{p3}
\vvertex{w5}{w2}{p4}
\vvertex{w6}{w1}{p5}
\fmfi{wiggly}{w3..w2}
\fmfi{wiggly}{w1..w4}
\fmfi{dots}{w5..w3}
\fmfi{dots}{w4..w6}
\fmfiv{d.sh=circle,d.f=1,d.size=4}{w1}
\fmfposition
\end{fmfchar*}}}
\qquad
\subfigure[\hspace{-0.5cm}]{
\fmfframe(3,1)(1,4){%
\begin{fmfchar*}(20,20)
\fmftop{v1}
\fmfbottom{v7}
\fmfforce{(0w,h)}{v1}
\fmfforce{(0w,0)}{v7}
\fmffixed{(0.15w,0)}{v1,v2}
\fmffixed{(0.25w,0)}{v2,v3}
\fmffixed{(0.2w,0)}{v3,v4}
\fmffixed{(0.25w,0)}{v4,v5}
\fmffixed{(0.15w,0)}{v5,v6}
\fmffixed{(0.15w,0)}{v7,v8}
\fmffixed{(0.25w,0)}{v8,v9}
\fmffixed{(0.2w,0)}{v9,v10}
\fmffixed{(0.25w,0)}{v10,v11}
\fmffixed{(0.15w,0)}{v11,v12}
\fmf{phantom}{v1,v7}
\fmf{phantom}{v2,v8}
\fmf{phantom}{v5,v11}
\fmf{phantom}{v6,v12}
\fmf{plain,tension=0.5,right=0,width=1mm}{v8,v11}
\fmf{dots,tension=0.5,right=0,width=1mm}{v7,v8}
\fmf{dots,tension=0.5,right=0,width=1mm}{v11,v12}
\fmf{phantom,tension=0.25,right=0.25}{v3,vc1}
\fmfset{dash_len}{1.5mm}
\fmf{plain,tension=0.25,left=0.25}{v4,vc1}
\fmf{phantom,tension=0.25,left=0.25}{v9,vc2}
\fmf{plain,tension=0.25,right=0.25}{v10,vc2}
\fmf{plain,tension=0.5}{vc1,vc2}
\fmffreeze
\fmffixed{(0.2w,0)}{vc3,vc1}
\fmffixed{(0.2w,0)}{vc4,vc2}
\fmf{dashes,tension=0.25,right=0.25}{vc3,vc1}
\fmf{dashes,tension=0.25,left=0.25}{vc4,vc2}
\fmfposition
\fmfipath{p[]}
\fmfipair{w[]}
\fmfiset{p1}{vpath(__vc1,__vc2)}
\fmfiset{p2}{vpath(__v5,__v11)}
\fmfiset{p3}{vpath(__v6,__v12)}
\fmfiset{p4}{vpath(__vc1,__v4)}
\svertex{w1}{p1}
\vvertex{w2}{w1}{p2}
\vvertex{w3}{w1}{p3}
\fmfiequ{w4}{point length(p4)/2 of p4}
\vvertex{w5}{w4}{p2}
\vvertex{w6}{w4}{p3}
\fmfi{wiggly}{w4..w5}
\fmfi{dots}{w5..w6}
\fmfiv{d.sh=circle,d.f=1,d.size=4}{w4}
\fmfposition
\end{fmfchar*}}}
}
\caption{Scalar line with one or two distinct single-vector vertices}
\label{startstruct}
\end{figure}

\begin{figure}[t]
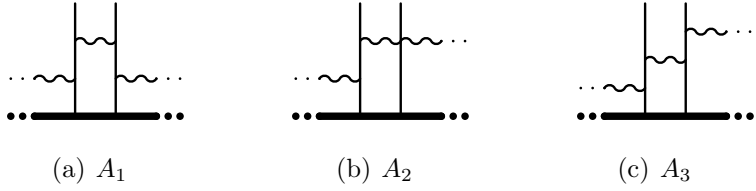

\unitlength=0.75mm
\settoheight{\eqoff}{$\times$}%
\setlength{\eqoff}{0.5\eqoff}%
\addtolength{\eqoff}{-12.5\unitlength}%
\settoheight{\eqofftwo}{$\times$}%
\setlength{\eqofftwo}{0.5\eqofftwo}%
\addtolength{\eqofftwo}{-7.5\unitlength}%
\centering
\raisebox{\eqoff}{%
\subfigure[$A_1$]{
\fmfframe(3,1)(1,4){%
\begin{fmfchar*}(30,20)
\fmftop{v1}
\fmfbottom{v7}
\fmfforce{(0w,h)}{v1}
\fmfforce{(0w,0)}{v7}
\fmffixed{(0.14w,0)}{v1,v2}
\fmffixed{(0.24w,0)}{v2,v3}
\fmffixed{(0.24w,0)}{v3,v4}
\fmffixed{(0.24w,0)}{v4,v5}
\fmffixed{(0.14w,0)}{v5,v6}
\fmffixed{(0.14w,0)}{v7,v8}
\fmffixed{(0.24w,0)}{v8,v9}
\fmffixed{(0.24w,0)}{v9,v10}
\fmffixed{(0.24w,0)}{v10,v11}
\fmffixed{(0.14w,0)}{v11,v12}
\fmf{phantom}{v1,v7}
\fmf{phantom}{v2,v8}
\fmf{plain}{v3,v9}
\fmf{plain}{v4,v10}
\fmf{phantom}{v5,v11}
\fmf{phantom}{v6,v12}
\fmf{plain,tension=0.5,right=0,width=1mm}{v8,v11}
\fmf{dots,tension=0.5,right=0,width=1mm}{v7,v8}
\fmf{dots,tension=0.5,right=0,width=1mm}{v11,v12}
\fmffreeze
\fmfposition
\fmfipath{p[]}
\fmfipair{w[]}
\fmfiset{p1}{vpath(__v1,__v7)}
\fmfiset{p2}{vpath(__v2,__v8)}
\fmfiset{p3}{vpath(__v3,__v9)}
\fmfiset{p4}{vpath(__v4,__v10)}
\fmfiset{p5}{vpath(__v5,__v11)}
\fmfiset{p6}{vpath(__v6,__v12)}
\fmfiequ{w1}{point length(p3)/3 of p3}
\fmfiequ{w2}{point 2length(p3)/3 of p3}
\vvertex{w3}{w2}{p1}
\vvertex{w4}{w2}{p2}
\vvertex{w5}{w1}{p4}
\vvertex{w6}{w2}{p5}
\vvertex{w7}{w2}{p6}
\vvertex{w8}{w2}{p4}
\fmfi{wiggly}{w1..w5}
\fmfi{wiggly}{w4..w2}
\fmfi{wiggly}{w8..w6}
\fmfi{dots}{w3..w4}
\fmfi{dots}{w6..w7}
\fmfposition
\end{fmfchar*}}
}
\qquad
\subfigure[$A_2$]{
\fmfframe(3,1)(1,4){%
\begin{fmfchar*}(30,20)
\fmftop{v1}
\fmfbottom{v7}
\fmfforce{(0w,h)}{v1}
\fmfforce{(0w,0)}{v7}
\fmffixed{(0.14w,0)}{v1,v2}
\fmffixed{(0.24w,0)}{v2,v3}
\fmffixed{(0.24w,0)}{v3,v4}
\fmffixed{(0.24w,0)}{v4,v5}
\fmffixed{(0.14w,0)}{v5,v6}
\fmffixed{(0.14w,0)}{v7,v8}
\fmffixed{(0.24w,0)}{v8,v9}
\fmffixed{(0.24w,0)}{v9,v10}
\fmffixed{(0.24w,0)}{v10,v11}
\fmffixed{(0.14w,0)}{v11,v12}
\fmf{phantom}{v1,v7}
\fmf{phantom}{v2,v8}
\fmf{plain}{v3,v9}
\fmf{plain}{v4,v10}
\fmf{phantom}{v5,v11}
\fmf{phantom}{v6,v12}
\fmf{plain,tension=0.5,right=0,width=1mm}{v8,v11}
\fmf{dots,tension=0.5,right=0,width=1mm}{v7,v8}
\fmf{dots,tension=0.5,right=0,width=1mm}{v11,v12}
\fmffreeze
\fmfposition
\fmfipath{p[]}
\fmfipair{w[]}
\fmfiset{p1}{vpath(__v1,__v7)}
\fmfiset{p2}{vpath(__v2,__v8)}
\fmfiset{p3}{vpath(__v3,__v9)}
\fmfiset{p4}{vpath(__v4,__v10)}
\fmfiset{p5}{vpath(__v5,__v11)}
\fmfiset{p6}{vpath(__v6,__v12)}
\fmfiequ{w1}{point length(p3)/3 of p3}
\fmfiequ{w2}{point 2length(p3)/3 of p3}
\vvertex{w3}{w2}{p1}
\vvertex{w4}{w2}{p2}
\vvertex{w5}{w1}{p4}
\vvertex{w6}{w1}{p5}
\vvertex{w7}{w1}{p6}
\vvertex{w8}{w1}{p4}
\fmfi{wiggly}{w1..w5}
\fmfi{wiggly}{w4..w2}
\fmfi{wiggly}{w8..w6}
\fmfi{dots}{w3..w4}
\fmfi{dots}{w6..w7}
\fmfposition
\end{fmfchar*}}
}
\qquad
\subfigure[$A_3$]{
\fmfframe(3,1)(1,4){%
\begin{fmfchar*}(30,20)
\fmftop{v1}
\fmfbottom{v7}
\fmfforce{(0w,h)}{v1}
\fmfforce{(0w,0)}{v7}
\fmffixed{(0.14w,0)}{v1,v2}
\fmffixed{(0.24w,0)}{v2,v3}
\fmffixed{(0.24w,0)}{v3,v4}
\fmffixed{(0.24w,0)}{v4,v5}
\fmffixed{(0.14w,0)}{v5,v6}
\fmffixed{(0.14w,0)}{v7,v8}
\fmffixed{(0.24w,0)}{v8,v9}
\fmffixed{(0.24w,0)}{v9,v10}
\fmffixed{(0.24w,0)}{v10,v11}
\fmffixed{(0.14w,0)}{v11,v12}
\fmf{phantom}{v1,v7}
\fmf{phantom}{v2,v8}
\fmf{plain}{v3,v9}
\fmf{plain}{v4,v10}
\fmf{phantom}{v5,v11}
\fmf{phantom}{v6,v12}
\fmf{plain,tension=0.5,right=0,width=1mm}{v8,v11}
\fmf{dots,tension=0.5,right=0,width=1mm}{v7,v8}
\fmf{dots,tension=0.5,right=0,width=1mm}{v11,v12}
\fmffreeze
\fmfposition
\fmfipath{p[]}
\fmfipair{w[]}
\fmfiset{p1}{vpath(__v1,__v7)}
\fmfiset{p2}{vpath(__v2,__v8)}
\fmfiset{p3}{vpath(__v3,__v9)}
\fmfiset{p4}{vpath(__v4,__v10)}
\fmfiset{p5}{vpath(__v5,__v11)}
\fmfiset{p6}{vpath(__v6,__v12)}
\fmfiequ{w1}{point length(p3)/4 of p3}
\fmfiequ{w2}{point length(p3)/2 of p3}
\fmfiequ{w3}{point 3length(p3)/4 of p3}
\vvertex{w4}{w3}{p1}
\vvertex{w5}{w3}{p2}
\vvertex{w6}{w2}{p4}
\vvertex{w7}{w1}{p4}
\vvertex{w8}{w1}{p5}
\vvertex{w9}{w1}{p6}
\fmfi{wiggly}{w5..w3}
\fmfi{wiggly}{w2..w6}
\fmfi{wiggly}{w7..w8}
\fmfi{dots}{w4..w5}
\fmfi{dots}{w8..w9}
\fmfposition
\end{fmfchar*}}}
}
\caption{Diagrams of class A}
\label{diagrams-A}
\end{figure}

\begin{figure}[p]
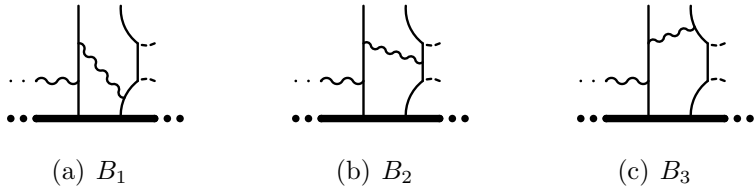

\unitlength=0.75mm
\settoheight{\eqoff}{$\times$}%
\setlength{\eqoff}{0.5\eqoff}%
\addtolength{\eqoff}{-12.5\unitlength}%
\settoheight{\eqofftwo}{$\times$}%
\setlength{\eqofftwo}{0.5\eqofftwo}%
\addtolength{\eqofftwo}{-7.5\unitlength}%
\centering
\raisebox{\eqoff}{%
\subfigure[$B_1$]{
\fmfframe(3,1)(1,4){%
\begin{fmfchar*}(30,20)
\fmftop{v1}
\fmfbottom{v7}
\fmfforce{(0w,h)}{v1}
\fmfforce{(0w,0)}{v7}
\fmffixed{(0.15w,0)}{v1,v2}
\fmffixed{(0.25w,0)}{v2,v3}
\fmffixed{(0.25w,0)}{v3,v4}
\fmffixed{(0.2w,0)}{v4,v5}
\fmffixed{(0.15w,0)}{v5,v6}
\fmffixed{(0.15w,0)}{v7,v8}
\fmffixed{(0.25w,0)}{v8,v9}
\fmffixed{(0.25w,0)}{v9,v10}
\fmffixed{(0.2w,0)}{v10,v11}
\fmffixed{(0.15w,0)}{v11,v12}
\fmf{phantom}{v1,v7}
\fmf{phantom}{v2,v8}
\fmf{plain}{v3,v9}
\fmfset{dash_len}{1.5mm}
\fmf{plain,tension=0.25,right=0.25}{v4,vc1}
\fmf{phantom,tension=0.25,left=0.25}{v5,vc1}
\fmf{plain,tension=0.25,left=0.25}{v10,vc2}
\fmf{phantom,tension=0.25,right=0.25}{v11,vc2}
\fmf{plain,tension=0.5}{vc1,vc2}
\fmf{plain,tension=0.5,right=0,width=1mm}{v8,v11}
\fmf{dots,tension=0.5,right=0,width=1mm}{v7,v8}
\fmf{dots,tension=0.5,right=0,width=1mm}{v11,v12}
\fmffreeze
\fmfposition
\fmffixed{(0.1w,0)}{vc1,vc3}
\fmffixed{(0.1w,0)}{vc2,vc4}
\fmf{dashes,tension=0.25,left=0.25}{vc3,vc1}
\fmf{dashes,tension=0.25,right=0.25}{vc4,vc2}
\fmfipath{p[]}
\fmfipair{w[]}
\fmfiset{p1}{vpath(__v1,__v7)}
\fmfiset{p2}{vpath(__v2,__v8)}
\fmfiset{p3}{vpath(__v3,__v9)}
\fmfiset{p4}{vpath(__vc2,__v10)}
\fmfiequ{w1}{point length(p3)/3 of p3}
\fmfiequ{w2}{point 2length(p3)/3 of p3}
\vvertex{w3}{w2}{p1}
\vvertex{w4}{w2}{p2}
\svertex{w5}{p4}
\fmfi{wiggly}{w4..w2}
\fmfi{wiggly}{w1..w5}
\fmfi{dots}{w3..w4}
\fmfposition
\end{fmfchar*}}
}
\qquad
\subfigure[$B_2$]{
\fmfframe(3,1)(1,4){%
\begin{fmfchar*}(30,20)
\fmftop{v1}
\fmfbottom{v7}
\fmfforce{(0w,h)}{v1}
\fmfforce{(0w,0)}{v7}
\fmffixed{(0.15w,0)}{v1,v2}
\fmffixed{(0.25w,0)}{v2,v3}
\fmffixed{(0.25w,0)}{v3,v4}
\fmffixed{(0.2w,0)}{v4,v5}
\fmffixed{(0.15w,0)}{v5,v6}
\fmffixed{(0.15w,0)}{v7,v8}
\fmffixed{(0.25w,0)}{v8,v9}
\fmffixed{(0.25w,0)}{v9,v10}
\fmffixed{(0.2w,0)}{v10,v11}
\fmffixed{(0.15w,0)}{v11,v12}
\fmf{phantom}{v1,v7}
\fmf{phantom}{v2,v8}
\fmf{plain}{v3,v9}
\fmfset{dash_len}{1.5mm}
\fmf{plain,tension=0.25,right=0.25}{v4,vc1}
\fmf{phantom,tension=0.25,left=0.25}{v5,vc1}
\fmf{plain,tension=0.25,left=0.25}{v10,vc2}
\fmf{phantom,tension=0.25,right=0.25}{v11,vc2}
\fmf{plain,tension=0.5}{vc1,vc2}
\fmf{plain,tension=0.5,right=0,width=1mm}{v8,v11}
\fmf{dots,tension=0.5,right=0,width=1mm}{v7,v8}
\fmf{dots,tension=0.5,right=0,width=1mm}{v11,v12}
\fmffreeze
\fmfposition
\fmffixed{(0.1w,0)}{vc1,vc3}
\fmffixed{(0.1w,0)}{vc2,vc4}
\fmf{dashes,tension=0.25,left=0.25}{vc3,vc1}
\fmf{dashes,tension=0.25,right=0.25}{vc4,vc2}
\fmfipath{p[]}
\fmfipair{w[]}
\fmfiset{p1}{vpath(__v1,__v7)}
\fmfiset{p2}{vpath(__v2,__v8)}
\fmfiset{p3}{vpath(__v3,__v9)}
\fmfiset{p4}{vpath(__vc2,__vc1)}
\fmfiequ{w1}{point length(p3)/3 of p3}
\fmfiequ{w2}{point 2length(p3)/3 of p3}
\vvertex{w3}{w2}{p1}
\vvertex{w4}{w2}{p2}
\svertex{w5}{p4}
\fmfi{wiggly}{w4..w2}
\fmfi{wiggly}{w1..w5}
\fmfi{dots}{w3..w4}
\fmfposition
\end{fmfchar*}}
}
\qquad
\subfigure[$B_3$]{
\fmfframe(3,1)(1,4){%
\begin{fmfchar*}(30,20)
\fmftop{v1}
\fmfbottom{v7}
\fmfforce{(0w,h)}{v1}
\fmfforce{(0w,0)}{v7}
\fmffixed{(0.15w,0)}{v1,v2}
\fmffixed{(0.25w,0)}{v2,v3}
\fmffixed{(0.25w,0)}{v3,v4}
\fmffixed{(0.2w,0)}{v4,v5}
\fmffixed{(0.15w,0)}{v5,v6}
\fmffixed{(0.15w,0)}{v7,v8}
\fmffixed{(0.25w,0)}{v8,v9}
\fmffixed{(0.25w,0)}{v9,v10}
\fmffixed{(0.2w,0)}{v10,v11}
\fmffixed{(0.15w,0)}{v11,v12}
\fmf{phantom}{v1,v7}
\fmf{phantom}{v2,v8}
\fmf{plain}{v3,v9}
\fmfset{dash_len}{1.5mm}
\fmf{plain,tension=0.25,right=0.25}{v4,vc1}
\fmf{phantom,tension=0.25,left=0.25}{v5,vc1}
\fmf{plain,tension=0.25,left=0.25}{v10,vc2}
\fmf{phantom,tension=0.25,right=0.25}{v11,vc2}
\fmf{plain,tension=0.5}{vc1,vc2}
\fmf{plain,tension=0.5,right=0,width=1mm}{v8,v11}
\fmf{dots,tension=0.5,right=0,width=1mm}{v7,v8}
\fmf{dots,tension=0.5,right=0,width=1mm}{v11,v12}
\fmffreeze
\fmfposition
\fmffixed{(0.1w,0)}{vc1,vc3}
\fmffixed{(0.1w,0)}{vc2,vc4}
\fmf{dashes,tension=0.25,left=0.25}{vc3,vc1}
\fmf{dashes,tension=0.25,right=0.25}{vc4,vc2}
\fmfipath{p[]}
\fmfipair{w[]}
\fmfiset{p1}{vpath(__v1,__v7)}
\fmfiset{p2}{vpath(__v2,__v8)}
\fmfiset{p3}{vpath(__v3,__v9)}
\fmfiset{p4}{vpath(__v4,__vc1)}
\fmfiequ{w1}{point length(p3)/3 of p3}
\fmfiequ{w2}{point 2length(p3)/3 of p3}
\vvertex{w3}{w2}{p1}
\vvertex{w4}{w2}{p2}
\svertex{w5}{p4}
\fmfi{wiggly}{w4..w2}
\fmfi{wiggly}{w1..w5}
\fmfi{dots}{w3..w4}
\fmfposition
\end{fmfchar*}}}
}
\caption{Diagrams of class B}
\label{diagrams-B}
\end{figure}

\begin{figure}[p]
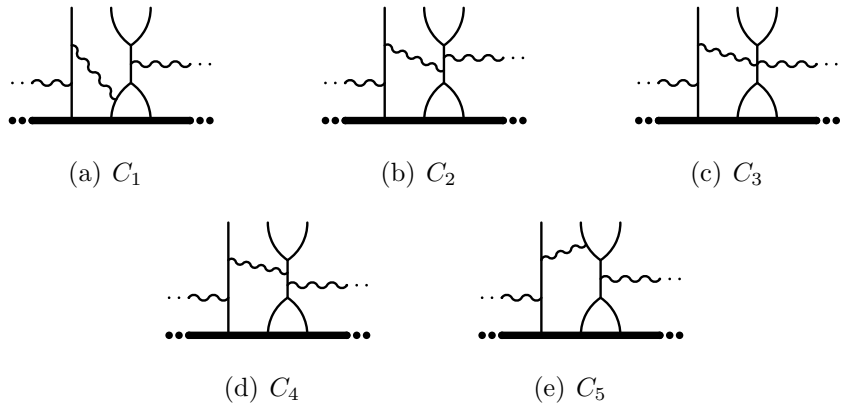

\unitlength=0.75mm
\settoheight{\eqoff}{$\times$}%
\setlength{\eqoff}{0.5\eqoff}%
\addtolength{\eqoff}{-12.5\unitlength}%
\settoheight{\eqofftwo}{$\times$}%
\setlength{\eqofftwo}{0.5\eqofftwo}%
\addtolength{\eqofftwo}{-7.5\unitlength}%
\centering
\raisebox{\eqoff}{%
\subfigure[$C_1$]{
\fmfframe(3,1)(1,4){%
\begin{fmfchar*}(35,20)
\fmftop{v1}
\fmfbottom{v8}
\fmfforce{(0w,h)}{v1}
\fmfforce{(0w,0)}{v8}
\fmffixed{(0.1w,0)}{v1,v2}
\fmffixed{(0.2w,0)}{v2,v3}
\fmffixed{(0.2w,0)}{v3,v4}
\fmffixed{(0.2w,0)}{v4,v5}
\fmffixed{(0.2w,0)}{v5,v6}
\fmffixed{(0.1w,0)}{v6,v7}
\fmffixed{(0.1w,0)}{v8,v9}
\fmffixed{(0.2w,0)}{v9,v10}
\fmffixed{(0.2w,0)}{v10,v11}
\fmffixed{(0.2w,0)}{v11,v12}
\fmffixed{(0.2w,0)}{v12,v13}
\fmffixed{(0.1w,0)}{v13,v14}
\fmf{phantom}{v1,v8}
\fmf{phantom}{v2,v9}
\fmf{plain}{v3,v10}
\fmf{phantom}{v6,v13}
\fmf{phantom}{v7,v14}
\fmf{plain,tension=0.25,right=0.25}{v4,vc1}
\fmf{plain,tension=0.25,left=0.25}{v5,vc1}
\fmf{plain,tension=0.25,left=0.25}{v11,vc2}
\fmf{plain,tension=0.25,right=0.25}{v12,vc2}
\fmf{plain,tension=0.5}{vc1,vc2}
\fmf{plain,tension=0.5,right=0,width=1mm}{v9,v13}
\fmf{dots,tension=0.5,right=0,width=1mm}{v8,v9}
\fmf{dots,tension=0.5,right=0,width=1mm}{v13,v14}
\fmffreeze
\fmfposition
\fmfipath{p[]}
\fmfipair{w[]}
\fmfiset{p1}{vpath(__v1,__v8)}
\fmfiset{p2}{vpath(__v2,__v9)}
\fmfiset{p3}{vpath(__v3,__v10)}
\fmfiset{p4}{vpath(__v11,__vc2)}
\fmfiset{p5}{vpath(__v6,__v13)}
\fmfiset{p6}{vpath(__v7,__v14)}
\fmfiset{p7}{vpath(__vc1,__vc2)}
\fmfiequ{w1}{point length(p3)/3 of p3}
\fmfiequ{w2}{point 2length(p3)/3 of p3}
\vvertex{w3}{w2}{p1}
\vvertex{w4}{w2}{p2}
\svertex{w5}{p4}
\svertex{w6}{p7}
\vvertex{w7}{w6}{p5}
\vvertex{w8}{w6}{p6}
\fmfi{wiggly}{w4..w2}
\fmfi{wiggly}{w1..w5}
\fmfi{dots}{w3..w4}
\fmfi{wiggly}{w6..w7}
\fmfi{dots}{w7..w8}
\fmfposition
\end{fmfchar*}}
}
\qquad
\subfigure[$C_2$]{
\fmfframe(3,1)(1,4){%
\begin{fmfchar*}(35,20)
\fmftop{v1}
\fmfbottom{v8}
\fmfforce{(0w,h)}{v1}
\fmfforce{(0w,0)}{v8}
\fmffixed{(0.1w,0)}{v1,v2}
\fmffixed{(0.2w,0)}{v2,v3}
\fmffixed{(0.2w,0)}{v3,v4}
\fmffixed{(0.2w,0)}{v4,v5}
\fmffixed{(0.2w,0)}{v5,v6}
\fmffixed{(0.1w,0)}{v6,v7}
\fmffixed{(0.1w,0)}{v8,v9}
\fmffixed{(0.2w,0)}{v9,v10}
\fmffixed{(0.2w,0)}{v10,v11}
\fmffixed{(0.2w,0)}{v11,v12}
\fmffixed{(0.2w,0)}{v12,v13}
\fmffixed{(0.1w,0)}{v13,v14}
\fmf{phantom}{v1,v8}
\fmf{phantom}{v2,v9}
\fmf{plain}{v3,v10}
\fmf{phantom}{v6,v13}
\fmf{phantom}{v7,v14}
\fmf{plain,tension=0.25,right=0.25}{v4,vc1}
\fmf{plain,tension=0.25,left=0.25}{v5,vc1}
\fmf{plain,tension=0.25,left=0.25}{v11,vc2}
\fmf{plain,tension=0.25,right=0.25}{v12,vc2}
\fmf{plain,tension=0.5}{vc1,vc2}
\fmf{plain,tension=0.5,right=0,width=1mm}{v9,v13}
\fmf{dots,tension=0.5,right=0,width=1mm}{v8,v9}
\fmf{dots,tension=0.5,right=0,width=1mm}{v13,v14}
\fmffreeze
\fmfposition
\fmfipath{p[]}
\fmfipair{w[]}
\fmfiset{p1}{vpath(__v1,__v8)}
\fmfiset{p2}{vpath(__v2,__v9)}
\fmfiset{p3}{vpath(__v3,__v10)}
\fmfiset{p4}{vpath(__v11,__vc2)}
\fmfiset{p5}{vpath(__v6,__v13)}
\fmfiset{p6}{vpath(__v7,__v14)}
\fmfiset{p7}{vpath(__vc1,__vc2)}
\fmfiequ{w1}{point length(p3)/3 of p3}
\fmfiequ{w2}{point 2length(p3)/3 of p3}
\vvertex{w3}{w2}{p1}
\vvertex{w4}{w2}{p2}
\fmfiequ{w5}{point 2length(p3)/3 of p7}
\fmfiequ{w6}{point length(p3)/3 of p7}
\vvertex{w7}{w6}{p5}
\vvertex{w8}{w6}{p6}
\fmfi{wiggly}{w4..w2}
\fmfi{wiggly}{w1..w5}
\fmfi{dots}{w3..w4}
\fmfi{wiggly}{w6..w7}
\fmfi{dots}{w7..w8}
\fmfposition
\end{fmfchar*}}
}
\qquad
\subfigure[$C_3$]{
\fmfframe(3,1)(1,4){%
\begin{fmfchar*}(35,20)
\fmftop{v1}
\fmfbottom{v8}
\fmfforce{(0w,h)}{v1}
\fmfforce{(0w,0)}{v8}
\fmffixed{(0.1w,0)}{v1,v2}
\fmffixed{(0.2w,0)}{v2,v3}
\fmffixed{(0.2w,0)}{v3,v4}
\fmffixed{(0.2w,0)}{v4,v5}
\fmffixed{(0.2w,0)}{v5,v6}
\fmffixed{(0.1w,0)}{v6,v7}
\fmffixed{(0.1w,0)}{v8,v9}
\fmffixed{(0.2w,0)}{v9,v10}
\fmffixed{(0.2w,0)}{v10,v11}
\fmffixed{(0.2w,0)}{v11,v12}
\fmffixed{(0.2w,0)}{v12,v13}
\fmffixed{(0.1w,0)}{v13,v14}
\fmf{phantom}{v1,v8}
\fmf{phantom}{v2,v9}
\fmf{plain}{v3,v10}
\fmf{phantom}{v6,v13}
\fmf{phantom}{v7,v14}
\fmf{plain,tension=0.25,right=0.25}{v4,vc1}
\fmf{plain,tension=0.25,left=0.25}{v5,vc1}
\fmf{plain,tension=0.25,left=0.25}{v11,vc2}
\fmf{plain,tension=0.25,right=0.25}{v12,vc2}
\fmf{plain,tension=0.5}{vc1,vc2}
\fmf{plain,tension=0.5,right=0,width=1mm}{v9,v13}
\fmf{dots,tension=0.5,right=0,width=1mm}{v8,v9}
\fmf{dots,tension=0.5,right=0,width=1mm}{v13,v14}
\fmffreeze
\fmfposition
\fmfipath{p[]}
\fmfipair{w[]}
\fmfiset{p1}{vpath(__v1,__v8)}
\fmfiset{p2}{vpath(__v2,__v9)}
\fmfiset{p3}{vpath(__v3,__v10)}
\fmfiset{p4}{vpath(__v11,__vc2)}
\fmfiset{p5}{vpath(__v6,__v13)}
\fmfiset{p6}{vpath(__v7,__v14)}
\fmfiset{p7}{vpath(__vc1,__vc2)}
\fmfiequ{w1}{point length(p3)/3 of p3}
\fmfiequ{w2}{point 2length(p3)/3 of p3}
\vvertex{w3}{w2}{p1}
\vvertex{w4}{w2}{p2}
\fmfiequ{w5}{point length(p3)/2 of p7}
\fmfiequ{w6}{point length(p3)/2 of p7}
\vvertex{w7}{w6}{p5}
\vvertex{w8}{w6}{p6}
\fmfi{wiggly}{w4..w2}
\fmfi{wiggly}{w1..w5}
\fmfi{dots}{w3..w4}
\fmfi{wiggly}{w6..w7}
\fmfi{dots}{w7..w8}
\fmfposition
\end{fmfchar*}}}
}
\\
\raisebox{\eqoff}{%
\subfigure[$C_4$]{
\fmfframe(3,1)(1,4){%
\begin{fmfchar*}(35,20)
\fmftop{v1}
\fmfbottom{v8}
\fmfforce{(0w,h)}{v1}
\fmfforce{(0w,0)}{v8}
\fmffixed{(0.1w,0)}{v1,v2}
\fmffixed{(0.2w,0)}{v2,v3}
\fmffixed{(0.2w,0)}{v3,v4}
\fmffixed{(0.2w,0)}{v4,v5}
\fmffixed{(0.2w,0)}{v5,v6}
\fmffixed{(0.1w,0)}{v6,v7}
\fmffixed{(0.1w,0)}{v8,v9}
\fmffixed{(0.2w,0)}{v9,v10}
\fmffixed{(0.2w,0)}{v10,v11}
\fmffixed{(0.2w,0)}{v11,v12}
\fmffixed{(0.2w,0)}{v12,v13}
\fmffixed{(0.1w,0)}{v13,v14}
\fmf{phantom}{v1,v8}
\fmf{phantom}{v2,v9}
\fmf{plain}{v3,v10}
\fmf{phantom}{v6,v13}
\fmf{phantom}{v7,v14}
\fmf{plain,tension=0.25,right=0.25}{v4,vc1}
\fmf{plain,tension=0.25,left=0.25}{v5,vc1}
\fmf{plain,tension=0.25,left=0.25}{v11,vc2}
\fmf{plain,tension=0.25,right=0.25}{v12,vc2}
\fmf{plain,tension=0.5}{vc1,vc2}
\fmf{plain,tension=0.5,right=0,width=1mm}{v9,v13}
\fmf{dots,tension=0.5,right=0,width=1mm}{v8,v9}
\fmf{dots,tension=0.5,right=0,width=1mm}{v13,v14}
\fmffreeze
\fmfposition
\fmfipath{p[]}
\fmfipair{w[]}
\fmfiset{p1}{vpath(__v1,__v8)}
\fmfiset{p2}{vpath(__v2,__v9)}
\fmfiset{p3}{vpath(__v3,__v10)}
\fmfiset{p4}{vpath(__v11,__vc2)}
\fmfiset{p5}{vpath(__v6,__v13)}
\fmfiset{p6}{vpath(__v7,__v14)}
\fmfiset{p7}{vpath(__vc1,__vc2)}
\fmfiequ{w1}{point length(p3)/3 of p3}
\fmfiequ{w2}{point 2length(p3)/3 of p3}
\vvertex{w3}{w2}{p1}
\vvertex{w4}{w2}{p2}
\fmfiequ{w5}{point length(p3)/3 of p7}
\fmfiequ{w6}{point 2length(p3)/3 of p7}
\vvertex{w7}{w6}{p5}
\vvertex{w8}{w6}{p6}
\fmfi{wiggly}{w4..w2}
\fmfi{wiggly}{w1..w5}
\fmfi{dots}{w3..w4}
\fmfi{wiggly}{w6..w7}
\fmfi{dots}{w7..w8}
\fmfposition
\end{fmfchar*}}
}
\qquad
\subfigure[$C_5$]{
\fmfframe(3,1)(1,4){%
\begin{fmfchar*}(35,20)
\fmftop{v1}
\fmfbottom{v8}
\fmfforce{(0w,h)}{v1}
\fmfforce{(0w,0)}{v8}
\fmffixed{(0.1w,0)}{v1,v2}
\fmffixed{(0.2w,0)}{v2,v3}
\fmffixed{(0.2w,0)}{v3,v4}
\fmffixed{(0.2w,0)}{v4,v5}
\fmffixed{(0.2w,0)}{v5,v6}
\fmffixed{(0.1w,0)}{v6,v7}
\fmffixed{(0.1w,0)}{v8,v9}
\fmffixed{(0.2w,0)}{v9,v10}
\fmffixed{(0.2w,0)}{v10,v11}
\fmffixed{(0.2w,0)}{v11,v12}
\fmffixed{(0.2w,0)}{v12,v13}
\fmffixed{(0.1w,0)}{v13,v14}
\fmf{phantom}{v1,v8}
\fmf{phantom}{v2,v9}
\fmf{plain}{v3,v10}
\fmf{phantom}{v6,v13}
\fmf{phantom}{v7,v14}
\fmf{plain,tension=0.25,right=0.25}{v4,vc1}
\fmf{plain,tension=0.25,left=0.25}{v5,vc1}
\fmf{plain,tension=0.25,left=0.25}{v11,vc2}
\fmf{plain,tension=0.25,right=0.25}{v12,vc2}
\fmf{plain,tension=0.5}{vc1,vc2}
\fmf{plain,tension=0.5,right=0,width=1mm}{v9,v13}
\fmf{dots,tension=0.5,right=0,width=1mm}{v8,v9}
\fmf{dots,tension=0.5,right=0,width=1mm}{v13,v14}
\fmffreeze
\fmfposition
\fmfipath{p[]}
\fmfipair{w[]}
\fmfiset{p1}{vpath(__v1,__v8)}
\fmfiset{p2}{vpath(__v2,__v9)}
\fmfiset{p3}{vpath(__v3,__v10)}
\fmfiset{p4}{vpath(__v11,__vc2)}
\fmfiset{p5}{vpath(__v6,__v13)}
\fmfiset{p6}{vpath(__v7,__v14)}
\fmfiset{p7}{vpath(__vc1,__vc2)}
\fmfiset{p8}{vpath(__v4,__vc1)}
\fmfiequ{w1}{point length(p3)/3 of p3}
\fmfiequ{w2}{point 2length(p3)/3 of p3}
\vvertex{w3}{w2}{p1}
\vvertex{w4}{w2}{p2}
\svertex{w5}{p8}
\fmfiequ{w6}{point length(p3)/2 of p7}
\vvertex{w7}{w6}{p5}
\vvertex{w8}{w6}{p6}
\fmfi{wiggly}{w4..w2}
\fmfi{wiggly}{w1..w5}
\fmfi{dots}{w3..w4}
\fmfi{wiggly}{w6..w7}
\fmfi{dots}{w7..w8}
\fmfposition
\end{fmfchar*}}}
}
\caption{Diagrams of class C}
\label{diagrams-C}
\end{figure}

Because of the assumption on the interaction range, the only possible configurations are shown in Figure~\ref{startstruct}.
The three possibilities are exactly equivalent for our present purpose, so we will focus on the second one in the following.
\begin{figure}[t]
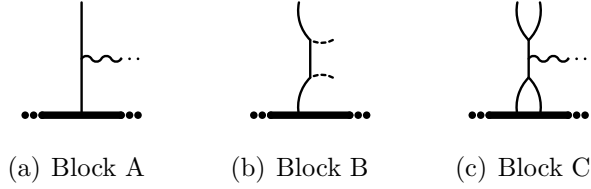

\unitlength=0.75mm
\settoheight{\eqoff}{$\times$}%
\setlength{\eqoff}{0.5\eqoff}%
\addtolength{\eqoff}{-12.5\unitlength}%
\settoheight{\eqofftwo}{$\times$}%
\setlength{\eqofftwo}{0.5\eqofftwo}%
\addtolength{\eqofftwo}{-7.5\unitlength}%
\centering
\raisebox{\eqoff}{%
\subfigure[Block A]{
\label{block-A}
\fmfframe(3,1)(1,4){%
\begin{fmfchar*}(20,20)
\fmftop{v1}
\fmfbottom{v6}
\fmfforce{(0w,h)}{v1}
\fmfforce{(0w,0)}{v6}
\fmffixed{(0.15w,0)}{v1,v2}
\fmffixed{(0.35w,0)}{v2,v3}
\fmffixed{(0.35w,0)}{v3,v4}
\fmffixed{(0.15w,0)}{v4,v5}
\fmffixed{(0.15w,0)}{v6,v7}
\fmffixed{(0.35w,0)}{v7,v8}
\fmffixed{(0.35w,0)}{v8,v9}
\fmffixed{(0.15w,0)}{v9,v10}
\fmf{plain}{v3,v8}
\fmf{phantom}{v2,v7}
\fmf{phantom}{v1,v6}
\fmf{phantom}{v4,v9}
\fmf{phantom}{v5,v10}
\fmf{plain,tension=0.5,right=0,width=1mm}{v7,v9}
\fmf{dots,tension=0.5,right=0,width=1mm}{v6,v7}
\fmf{dots,tension=0.5,right=0,width=1mm}{v9,v10}
\fmffreeze
\fmfposition
\fmfipath{p[]}
\fmfipair{w[]}
\fmfiset{p1}{vpath(__v3,__v8)}
\fmfiset{p2}{vpath(__v2,__v7)}
\fmfiset{p3}{vpath(__v4,__v9)}
\fmfiset{p4}{vpath(__v1,__v6)}
\fmfiset{p5}{vpath(__v5,__v10)}
\fmfiequ{w1}{point length(p1)/2 of p1}
\fmfiequ{w2}{point 2length(p1)/3 of p1}
\vvertex{w4}{w1}{p3}
\vvertex{w6}{w1}{p5}
\fmfi{wiggly}{w1..w4}
\fmfi{dots}{w4..w6}
\fmfposition
\end{fmfchar*}}
}
\qquad
\subfigure[Block B]{
\label{block-B}
\fmfframe(3,1)(1,4){%
\begin{fmfchar*}(20,20)
\fmftop{v1}
\fmfbottom{v7}
\fmfforce{(0w,h)}{v1}
\fmfforce{(0w,0)}{v7}
\fmffixed{(0.15w,0)}{v1,v2}
\fmffixed{(0.25w,0)}{v2,v3}
\fmffixed{(0.2w,0)}{v3,v4}
\fmffixed{(0.25w,0)}{v4,v5}
\fmffixed{(0.15w,0)}{v5,v6}
\fmffixed{(0.15w,0)}{v7,v8}
\fmffixed{(0.25w,0)}{v8,v9}
\fmffixed{(0.2w,0)}{v9,v10}
\fmffixed{(0.25w,0)}{v10,v11}
\fmffixed{(0.15w,0)}{v11,v12}
\fmf{phantom}{v1,v7}
\fmf{phantom}{v2,v8}
\fmf{phantom}{v5,v11}
\fmf{phantom}{v6,v12}
\fmf{plain,tension=0.5,right=0,width=1mm}{v8,v11}
\fmf{dots,tension=0.5,right=0,width=1mm}{v7,v8}
\fmf{dots,tension=0.5,right=0,width=1mm}{v11,v12}
\fmf{plain,tension=0.25,right=0.25}{v3,vc1}
\fmfset{dash_len}{1.5mm}
\fmf{phantom,tension=0.25,left=0.25}{v4,vc1}
\fmf{plain,tension=0.25,left=0.25}{v9,vc2}
\fmf{phantom,tension=0.25,right=0.25}{v10,vc2}
\fmf{plain,tension=0.5}{vc1,vc2}
\fmffreeze
\fmffixed{(0.2w,0)}{vc1,vc3}
\fmffixed{(0.2w,0)}{vc2,vc4}
\fmf{dashes,tension=0.25,left=0.25}{vc3,vc1}
\fmf{dashes,tension=0.25,right=0.25}{vc4,vc2}
\fmfposition
\fmfipath{p[]}
\fmfipair{w[]}
\fmfiset{p1}{vpath(__vc1,__vc2)}
\fmfiset{p2}{vpath(__v5,__v11)}
\fmfiset{p3}{vpath(__v6,__v12)}
\svertex{w1}{p1}
\vvertex{w2}{w1}{p2}
\vvertex{w3}{w1}{p3}
\fmfposition
\end{fmfchar*}}}
\qquad
\subfigure[Block C]{
\label{block-C}
\fmfframe(3,1)(1,4){%
\begin{fmfchar*}(20,20)
\fmftop{v1}
\fmfbottom{v7}
\fmfforce{(0w,h)}{v1}
\fmfforce{(0w,0)}{v7}
\fmffixed{(0.15w,0)}{v1,v2}
\fmffixed{(0.25w,0)}{v2,v3}
\fmffixed{(0.2w,0)}{v3,v4}
\fmffixed{(0.25w,0)}{v4,v5}
\fmffixed{(0.15w,0)}{v5,v6}
\fmffixed{(0.15w,0)}{v7,v8}
\fmffixed{(0.25w,0)}{v8,v9}
\fmffixed{(0.2w,0)}{v9,v10}
\fmffixed{(0.25w,0)}{v10,v11}
\fmffixed{(0.15w,0)}{v11,v12}
\fmf{phantom}{v1,v7}
\fmf{phantom}{v2,v8}
\fmf{phantom}{v5,v11}
\fmf{phantom}{v6,v12}
\fmf{plain,tension=0.5,right=0,width=1mm}{v8,v11}
\fmf{dots,tension=0.5,right=0,width=1mm}{v7,v8}
\fmf{dots,tension=0.5,right=0,width=1mm}{v11,v12}
\fmfset{dash_len}{1.5mm}
\fmf{plain,tension=0.25,right=0.25}{v3,vc1}
\fmf{plain,tension=0.25,left=0.25}{v4,vc1}
\fmf{plain,tension=0.25,left=0.25}{v9,vc2}
\fmf{plain,tension=0.25,right=0.25}{v10,vc2}
\fmf{plain,tension=0.5}{vc1,vc2}
\fmffreeze
\fmfposition
\fmfipath{p[]}
\fmfipair{w[]}
\fmfiset{p1}{vpath(__vc1,__vc2)}
\fmfiset{p2}{vpath(__v5,__v11)}
\fmfiset{p3}{vpath(__v6,__v12)}
\svertex{w1}{p1}
\vvertex{w2}{w1}{p2}
\vvertex{w3}{w1}{p3}
\fmfi{wiggly}{w1..w2}
\fmfi{dots}{w2..w3}
\fmfposition
\end{fmfchar*}}
}
}
\caption{Building blocks for maximum-range diagrams with vectors}
\end{figure}
In a maximum-range graph, the propagator leaving from the single-vector vertex can be attached only to one out of three structures, shown in Figures~\ref{block-A}, \ref{block-B} and \ref{block-C}. All the possible respective combinations are listed in Figures~\ref{diagrams-A}, ~\ref{diagrams-B} and \ref{diagrams-C}.
We can now demonstrate that the divergent parts of the diagrams in each class sum up to zero.
The key point of the proof is the fact that when we integrate by parts the double covariant derivative $\covder^2$ at the single-vector vertex,
we are forced, by the property demonstrated previously, to keep it inside the graph, thus moving it on the vector line starting from the vertex.
We can then shift the $\covder^2$ to the opposite end of the vector propagator, where a second integration by parts is possible. Since once again we can neglect the terms where the derivatives would act on the external fields, the integration produces a single contribution, which can be simplified thanks to the standard D-algebra identities, and in the end we find a modification in the original superspace integral. Now, it turns out that in all the cases of interest the divergent part of the found integral is cancelled by the one obtained from a different original supergraph. We now summarize the details of such cancellations for the three classes:
\begin{itemize}
\item Class A (Fig.~\ref{diagrams-A}):\\
The diagram $A_1$ is finite, since we are forced to move the $\covder^2$ outside the diagram.
The first steps of D-algebra for $A_3$ transform the graph into the same structure as $A_2$, with an additional minus sign coming from the $\Box=-p^2$ that cancels a propagator. Since $A_2$ and $A_3$ have the same colour factor, their divergent parts cancel each other.
\item Class B (Fig.~\ref{diagrams-B}):\\
In this case, we begin the D-algebra for both $B_1$ and $B_2$, finding the same superspace integral. However, their colour factors are opposite, and hence also the divergent parts of $B_1$ and $B_2$ sum up to zero.
The diagram $B_3$ is finite, for the same reason as $A_1$.
\item Class C (Fig.~\ref{diagrams-C}):\\
For $C_1$ and $C_2$, we proceed as in the case of $B_1$ and $B_2$, and we conclude again that the divergent parts of the two diagrams cancel out.
For $C_3$ and $C_4$, we have the same situation as for $A_2$ and $A_3$.
Finally, $C_5$ is finite, similarly to $A_1$ and $B_3$.
\end{itemize}


We have thus demonstrated that none of the maximum-range diagrams that contain one of the structures of Figure~\ref{startstruct} is relevant for the calculation of anomalous dimensions. As we will see, this represents a very important simplification.
\clearpage
\section{Wrapping interactions in $\N=4$ SYM}
\label{undeformed}
We are now ready to begin the analysis of wrapping effects in $\N=4$ SYM by exploiting the results of the previous section. 
Since single-impurity operators of the $SU(2)$ sector are protected, we are forced to study two-impurity ones. The shortest non-protected operators of this class have length $L=4$, and a possible choice for two independent states in the length-four subsector is~\cite{us,uslong}
\begin{equation}
\label{Opbasis}
\basisop{4}{1}=\tr(\phi Z\phi Z)\col\qquad\basisop{4}{2}=\tr(\phi\phi ZZ)\col
\end{equation}
which will in general mix under renormalization. 
Note that the anti-symmetric combination of the basis~\eqref{Opbasis} is a descendant of the Konishi operator.
Because of the relationship between the perturbative order and the interaction range, wrapping effects can appear in this subsector only at four loops and beyond. 

\subsection{The spin-chain analogy}
Since we work in the planar limit, we can restrict our analysis to single-trace operators. We will focus on the $SU(2)$ sector, containing operators built using only two out of the three available superfields. A single-trace operator in this sector can be easily associated to a state of an $SU(2)$ spin chain, once a simple correspondence between the field flavour and the spin projection on a chosen axis is fixed~\cite{Minahan:2002ve}. 
Thanks to this correspondence, we will refer to composite operators also as states of the corresponding chain. Because of the relationship between the loop order and the interaction range, we will be forced to work with long-range chains.
An operator given by the product of $L$ fields of the same flavour $Z$ will be related to the ground state of a ferromagnetic chain, from which we can build excited states by replacing some of the $Z$ fields with impurities, i.e.\ fields of type $\phi$. An operator with $n$ $\phi$ fields will thus be equivalent to an $n$-magnon state of the chain.

The interactions among the spins of the chain are conveniently described in terms of permutations of neighbouring sites, from which the following operators can be constructed~\cite{Beisert:2003tq}
\begin{equation}\label{permstrucdef}
\pthree{a_1}{\dots}{a_n}=\sum_{r=0}^{L-1}\perm_{a_1+r,a_1+r+1}\cdots
\perm_{a_n+r,a_n+r+1}
\col
\end{equation}
where $\perm_{a,a+1}$ swaps the spins at sites $a$ and $a+1$. For a chain of length $L$, we must impose the cyclic identification $\perm_{a,a+1}\simeq\perm_{a+L,a+L+1}$.
The range of an operator of the form~\eqref{permstrucdef} can be computed from the list of arguments $a_1,\ldots a_n$ as
\begin{equation}\label{nneighbourint}
\kappa=2+\max\{a_1,\dots, a_n\}-\min\{a_1,\dots, a_n\}\pnt
\end{equation}

This analogy is very convenient because it allows to restate all the field-theory problems in terms of spin chains, which are the traditional and best-known environment for the study of integrability properties. In fact, the one-loop integrability of $\N=4$ SYM in the $SU(2)$ sector was recognized through the discovery that the one-loop dilatation operator, once it is translated into the spin chain language, coincides with the Hamiltonian of the Heisenberg $XXX_{1/2}$ chain, which was known to be integrable~\cite{Minahan:2002ve,Minahan:2006sk}. 

Using the basis of permutation operators~\eqref{permstrucdef}, we can write the components of the asymptotic dilatation operator, as obtained from the asymptotic Bethe ansatz~\cite{Beisert:2004ry,Beisert:2007hz,Beisert:2003tq}. The expressions up to three loops are very simple and they are given by
\begin{equation}
\label{Duptothree}
\begin{aligned}
\mathcal{D}_0&={}+{}\{\} \col \\
\mathcal{D}_1&=2\left(\{\}-\{1\}\right) \col \\
\mathcal{D}_2&=2\left(-4\{\}+6\{1\}-\left(\{1,2\}+\{2,1\}\right)\right) \col \\
\mathcal{D}_3&=60\{\}-104\{1\}+4\{1,3\}+24\left(\{1,2\}+\{2,1\}\right) \\
&\phantom{{}={}}-4i\epsilon_{2a}\{1,3,2\}+4i\epsilon_{2a}\{2,1,3\}-4\left(\{1,2,3\}+\{3,2,1\}\right) \pnt
\end{aligned}
\end{equation}
The meaning of the undetermined coefficient $\epsilon_{2a}$ will be discussed later.

\subsection{The general procedure}
The asymptotic Bethe ansatz will give the correct value of the Konishi anomalous dimension only up to three loops. However, we can still take advantage from the knowledge of the asymptotic result to extract the information on the most complicated diagrams that otherwise we should consider explicitly. We summarize here the main steps of our procedure for the calculation of the exact $\ell$-loop anomalous dimensions of length-$L$ operators, which has general validity and will be applied also to the five-loop case in $\N=4$ SYM and to even higher orders in the $\beta$-deformed theory:
\begin{enumerate}
\item compute the $\ell$-loop component $\mathcal{D}_\ell$ of the asymptotic dilatation operator from the hypothesis of all-loop integrability and the asymptotic Bethe equations~\cite{Beisert:2004ry} (see~\cite{usfive} for an explicit example of this step, realized in the five-loop case). We can divide the diagrams contributing to the dilatation operator into two classes: the first one contains graphs with range less than or equal to $L$, which are allowed both in the asymptotic and in the length-$L$ case, possibly with different combinatorial factors. The second class is made of higher-range graphs, which appear in the asymptotic computation but not in the finite-length case. 
\item Subtract the contribution of the second class, to obtain all the information on the first one. Such a trick works because when the dilatation operator is written in the basis of permutation operators~\eqref{permstrucdef}, its functional form as an operator on the states of the $SU(2)$ sector automatically accounts for the change in the combinatorial factors of a given Feynman diagram applied to operators of different lengths. Moreover, this approach reveals to be very useful because the number and complexity of diagrams typically grow when the range is reduced, so that the lower-range graphs of which we avoid the direct computation constitute in fact the most difficult classes to calculate.
\item Add the contribution of wrapping diagrams, which must be computed explicitly.
\end{enumerate}
This general procedure becomes particularly simple when we consider finite-size effects at the critical order, that is $L$ loops for length-$L$ operators. Moreover, both the tasks of subtraction and computation of wrapping graphs will be greatly further simplified by the application of $\N=1$ superspace techniques.

\begin{figure}[!h]
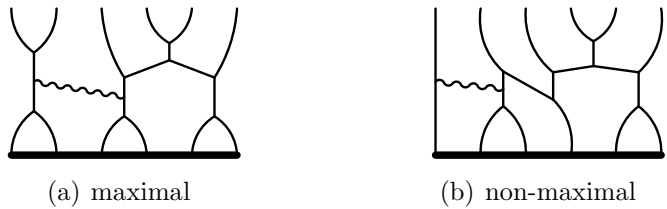

\vspace{0.5cm}
\centering
\unitlength=0.75mm
\settoheight{\eqoff}{$\times$}%
\setlength{\eqoff}{0.5\eqoff}%
\addtolength{\eqoff}{-12.5\unitlength}%
\settoheight{\eqofftwo}{$\times$}%
\setlength{\eqofftwo}{0.5\eqofftwo}%
\addtolength{\eqofftwo}{-7.5\unitlength}%
\subfigure[maximal]{
\raisebox{\eqoff}{%
\fmfframe(3,1)(1,4){%
\begin{fmfchar*}(40,26)
\fmftop{v1}
\fmfbottom{v7}
\fmfforce{(0w,h)}{v1}
\fmfforce{(0w,0)}{v7}
\fmffixed{(0.2w,0)}{v1,v2}
\fmffixed{(0.2w,0)}{v2,v3}
\fmffixed{(0.2w,0)}{v3,v4}
\fmffixed{(0.2w,0)}{v4,v5}
\fmffixed{(0.2w,0)}{v5,v6}
\fmffixed{(0.2w,0)}{v7,v8}
\fmffixed{(0.2w,0)}{v8,v9}
\fmffixed{(0.2w,0)}{v9,v10}
\fmffixed{(0.2w,0)}{v10,v11}
\fmffixed{(0.2w,0)}{v11,v12}
\fmf{plain,tension=1,left=0.25}{v9,vc6}
\fmf{plain,tension=1,right=0.25}{v10,vc6}
\fmf{plain,tension=1,left=0.25}{v11,vc10}
\fmf{plain,tension=1,right=0.25}{v12,vc10}
\fmf{plain,tension=1,right=0.125}{v3,vc5}
\fmf{plain,tension=0.25,right=0.25}{v4,vc7}
\fmf{plain,tension=0.25,left=0.25}{v5,vc7}
\fmf{plain,tension=1,left=0.125}{v6,vc9}
\fmf{plain,tension=2}{vc5,vc6}
\fmf{plain,tension=2}{vc9,vc10}
\fmf{plain,tension=0.5}{vc5,vc8}
\fmf{plain,tension=0.5}{vc8,vc9}
\fmf{plain,tension=1}{vc7,vc8}
\fmf{plain,tension=1,right=0.25}{v1,vc1}
\fmf{plain,tension=1,left=0.25}{v2,vc1}
\fmf{plain,tension=1,left=0.25}{v7,vc2}
\fmf{plain,tension=1,right=0.25}{v8,vc2}
\fmf{plain,tension=1.5}{vc1,vc2}
\fmffreeze
\fmfposition
\fmf{plain,tension=1,right=0,width=1mm}{v7,v12}
\fmfipath{p[]}
\fmfiset{p1}{vpath(__vc1,__vc2)}
\fmfiset{p2}{vpath(__vc5,__vc6)}
\fmfipair{w[]}
\svertex{w1}{p1}
\svertex{w2}{p2}
\fmfi{wiggly}{w1..w2}
\fmffreeze
\end{fmfchar*}}}
}
$\qquad\qquad$
\subfigure[non-maximal]{
\raisebox{\eqoff}{
\fmfframe(3,1)(1,4){
\begin{fmfchar*}(40,26)
\fmftop{va1}
\fmfbottom{va5}
\fmfforce{(0w,h)}{va1}
\fmfforce{(0w,0)}{va5}
\fmffixed{(0.2w,0)}{va1,v1}
\fmffixed{(0.2w,0)}{va5,v5}
\fmffixed{(0.2w,0)}{v1,v2}
\fmffixed{(0.2w,0)}{v2,v3}
\fmffixed{(0.2w,0)}{v3,v4}
\fmffixed{(0.2w,0)}{v4,v9}
\fmffixed{(0.2w,0)}{v5,v6}
\fmffixed{(0.2w,0)}{v6,v7}
\fmffixed{(0.2w,0)}{v7,v8}
\fmffixed{(0.2w,0)}{v8,v10}
\fmffixed{(0,whatever)}{vb1,vc2}
\fmffixed{(0,whatever)}{vc3,vb4}
\fmffixed{(0,whatever)}{vb7,vc8}
\fmf{plain,tension=0.25,left=0.25}{v5,vc2}
\fmf{plain,tension=0.25,right=0.25}{v6,vc2}
\fmf{phantom,tension=0.25,right=0.25}{v1,vb1}
\fmf{phantom,tension=0.25,left=0.25}{v2,vb1}
\fmf{phantom,tension=0.5}{vc2,vb1}
\fmf{plain,tension=0.25,left=0.25}{v8,vc8}
\fmf{plain,tension=0.25,right=0.25}{v10,vc8}
\fmf{phantom,tension=0.25,right=0.25}{v4,vb7}
\fmf{phantom,tension=0.25,left=0.25}{v9,vb7}
\fmf{phantom,tension=0.5}{vb7,vc8}
\fmf{plain,tension=0.25,right=0.25}{v3,vc3}
\fmf{plain,tension=0.25,left=0.25}{v4,vc3}
\fmf{phantom,tension=0.25,left=0.25}{v7,vb4}
\fmf{phantom,tension=0.25,right=0.25}{v8,vb4}
\fmf{phantom,tension=0.2}{vc3,vb4}
\fmf{plain}{va1,va5}
\fmffreeze
\fmffixed{(0,whatever)}{vc1,vc2}
\fmffixed{(0,whatever)}{vc3,vc4}
\fmffixed{(0,whatever)}{vc5,vc6}
\fmffixed{(0,whatever)}{vc7,vc8}
\fmffixed{(whatever,0)}{vc7,vc5}
\fmf{plain,tension=0.5,right=0.25}{v1,vc1}
\fmf{plain,tension=0.15,right=0.25}{v2,vc5}
\fmf{plain,tension=0.5,left=0.25}{v9,vc7}
\fmf{plain,tension=0.25}{vc3,vc4}
\fmf{plain,tension=0.5}{vc1,vc2}
\fmf{plain,tension=0.5}{vc6,vc1}
\fmf{plain,tension=0.5}{vc7,vc4}
\fmf{plain,tension=0.5}{vc4,vc5}
\fmf{plain,tension=0.5}{vc5,vc6}
\fmf{plain,tension=0.5,left=0.25}{vc6,v7}
\fmf{plain,tension=1}{vc7,vc8}
\fmf{plain,tension=0.5,right=0,width=1mm}{va5,v10}
\fmfposition
\fmfipath{p[]}
\fmfiset{p0}{vpath(__v5,__vc2)}
\fmfiset{p1}{vpath(__vc1,__vc2)}
\fmfiset{p2}{vpath(__v1,__vc1)}
\fmfiset{p3}{vpath(__v10,__vc8)}
\fmfiset{p4}{vpath(__vc8,__vc7)}
\fmfiset{p5}{vpath(__v9,__vc7)}
\fmfiset{p6}{vpath(__va1,__va5)}
\fmfipair{w[]}
\svertex{w1}{p1}
\svertex{w2}{p6}
\fmfi{wiggly}{w1..w2}
\fmffreeze
\end{fmfchar*}}}
}
\caption{Examples of range-$(L+1)$, $L$-loop diagrams with $L=5$}
\label{example-03}
\end{figure}

Before focusing on the actual four-loop case, we present here the general discussion for the subtraction step in the special case where the order is critical. We must subtract all the $L$-loop diagrams of range $(L+1)$, which is the maximum one allowed at this order. Now, let us divide all such graphs into two groups, with the first one made of diagrams (that we define maximal) whose chiral structure alone (i.e.\ the structure of scalar propagators and vertices) already has range $(L+1)$, whereas the second group contains graphs (referred to as non-maximal) with a lower-range chiral structure, whose range is increased by vector interactions. An example from each class is shown in Figure~\ref{example-03}. The second class fulfills all the assumptions we made to demonstrate the cancellation identities for supergraphs in Section~\ref{superspace}, and we conclude that the corresponding diagrams are not relevant for the computation of anomalous dimensions. 

We are therefore left with the subtraction of the maximal diagrams. To deal with them, it is better to consider a new basis, directly related to Feynman supergraphs, instead of the standard permutation one~\eqref{permstrucdef}. In order to find it, we must analyze the general properties of the possible chiral structures. 
First of all, we can have diagrams with only vector interactions. From the point of view of flavour permutations, their chiral structure is just the identity. 
Let us consider now the basic structure of Figure~\ref{buildingblock}. Since scalar propagator always connect a chiral and an anti-chiral vertex, any non-trivial chiral structure for an $\ell$-loop diagram with one insertion of a composite operator made of chiral superfields must be constructable by assembling up to $\ell$ copies of this building block. We can now use the standard permutation basis~\eqref{permstrucdef} to write down the explicit action of $\chi(1)$ on the flavours of the superfields. Starting from the structure of $\chi(1)$ and iterating it, the more complicated structures can be found. The results up to four loops are
\begin{figure}[t]
\vspace{1cm}
\unitlength=0.75mm
\settoheight{\eqoff}{$\times$}%
\setlength{\eqoff}{0.5\eqoff}%
\addtolength{\eqoff}{-12.5\unitlength}%
\settoheight{\eqofftwo}{$\times$}%
\setlength{\eqofftwo}{0.5\eqofftwo}%
\addtolength{\eqofftwo}{-7.5\unitlength}%
\begin{equation*}
\begin{aligned}
\chi(1):\quad-&
\raisebox{1.5\eqoff}{%
\fmfframe(1,1)(3,4){%
\begin{fmfchar*}(30,30)
\WoneplainB
\fmfipair{w[]}
\svertex{w1}{p4}
\fmfiv{l=\footnotesize{$\phi$},l.a=-90,l.d=5}{vloc(__v6)}
\fmfiv{l=\footnotesize{$Z$},l.a=-90,l.d=5}{vloc(__v7)}
\fmfiv{l=\footnotesize{$\phi$},l.a=90,l.d=5}{vloc(__v2)}
\fmfiv{l=\footnotesize{$Z$},l.a=90,l.d=5}{vloc(__v3)}
\fmfiv{l=\footnotesize{$\psi$},l.a=0,l.d=5}{w1}
\end{fmfchar*}}}
&+&
\raisebox{1.5\eqoff}{%
\fmfframe(3,1)(1,4){%
\begin{fmfchar*}(30,30)
\WoneplainB
\fmfipair{w[]}
\svertex{w1}{p4}
\fmfiv{l=\footnotesize{$\phi$},l.a=-90,l.d=5}{vloc(__v6)}
\fmfiv{l=\footnotesize{$Z$},l.a=-90,l.d=5}{vloc(__v7)}
\fmfiv{l=\footnotesize{$Z$},l.a=90,l.d=5}{vloc(__v2)}
\fmfiv{l=\footnotesize{$\phi$},l.a=90,l.d=5}{vloc(__v3)}
\fmfiv{l=\footnotesize{$\psi$},l.a=0,l.d=5}{w1}
\end{fmfchar*}}}
\end{aligned}
\end{equation*}
\caption{Building block for chiral structures}
\label{buildingblock}
\end{figure}
\begin{equation}
\label{chistruc}
\begin{aligned}
\chi(a,b,c,d)&=\pid-4\pone1
+\ptwo ab+\ptwo ac+\ptwo ad+\ptwo bc+\ptwo bd+\ptwo cd\\
&\phantom{{}={}}
-\pthree abc-\pthree abd-\pthree acd-\pthree bcd
+\pfour abcd\col\\
\chi(a,b,c)&=-\pid+3\pone1
-\ptwo ab-\ptwo ac-\ptwo bc+\pthree abc\col\\
\chi(a,b)&=\pid-2\pone1+\ptwo ab\col\\
\chi(1)&=-\pid+\pone1\col\\
\chi() &=\pid
\pnt
\end{aligned}
\normalsize
\end{equation}
We will refer to the $\chi(\ldots)$ functions as chiral functions. Since they can be written as linear combination of the old permutation basis, we can take them as new basis elements. The number $n$ of arguments in each chiral function is equal to the number of copies of $\chi(1)$ that are needed to assemble it. To obtain an $\ell$-loop graph we will need to add $(\ell-n)$ vector propagators. As anticipated, $\chi()$ is just the identity, corresponding to diagrams with only vector interactions. When the dilatation operator is written in the new basis, the coefficient of each chiral function will be equal to the coefficient of the $1/\varepsilon$ pole of the sum of all the relevant diagrams with the chosen chiral structure, multiplied by $(-2\ell)$ according to the definition of anomalous dimension~\eqref{anomalous}.

From the previous considerations, it follows that we can subtract the contribution of maximal range-$(L+1)$ diagrams by simply deleting the terms with the corresponding chiral functions from the expression of the asymptotic dilatation operator in the basis~\eqref{chistruc}. Note that this approach would not have been applicable to non-maximal graphs, whose contributions would mix with those of lower-range diagrams with the same chiral structure.

In summary, when the order is critical we can subtract all the contributions of range-$(L+1)$ supergraphs without the need for explicit diagrammatic computations. This result makes the subtraction step nearly trivial, reducing the original computation to the analysis of wrapping diagrams, and is a consequence of the use of $\N=1$ superspace techniques. Just to have an idea of the simplification deriving from it, the range-five diagrams that should be calculated explicitly in the four-loop case, without the results of Section~\ref{superspace}, are more than one hundred.

Now we can apply all this machinery to the four-loop case to find the exact four-loop anomalous dimension of the Konishi operator.

\subsection{The four-loop case}
We applied the general procedure that we have presented in the previous section to the four-loop case in~\cite{us,uslong}.
To do this, we started from the expression of the four-loop component of the asymptotic dilatation operator, which was given in the permutation basis~\eqref{permstrucdef} in~\cite{Beisert:2007hz}, and we wrote it in the chiral basis~\eqref{chistruc}. Then, we subtracted the range-five contributions by deleting all the terms with a range-five chiral function, i.e.\ all those with both 1 and 4 among the arguments. Afterwards, we added the contribution from wrapping supergraphs. We will not present here the details of the computation, which can be found in~\cite{us,uslong}. Instead, we prefer to stress the most interesting features of the procedure. 
\begin{figure}[t]
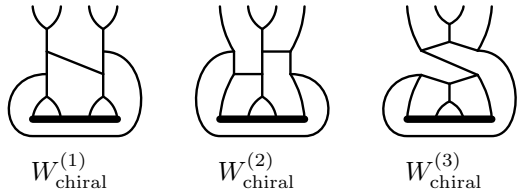

\centering
\renewcommand*{\thesubfigure}{}
\unitlength=0.75mm
\settoheight{\eqoff}{$\times$}%
\setlength{\eqoff}{0.5\eqoff}%
\addtolength{\eqoff}{-12.5\unitlength}%
\settoheight{\eqofftwo}{$\times$}%
\setlength{\eqofftwo}{0.5\eqofftwo}%
\addtolength{\eqofftwo}{-7.5\unitlength}%
\raisebox{\eqoff}{%
\subfigure[$W_{\mathrm{chiral}}^{(1)}$]{
\fmfframe(3,1)(1,4){%
\begin{fmfchar*}(20,20)
\fmftop{v1}
\fmfbottom{v5}
\fmfforce{(0.125w,h)}{v1}
\fmfforce{(0.125w,0)}{v5}
\fmffixed{(0.25w,0)}{v1,v2}
\fmffixed{(0.25w,0)}{v2,v3}
\fmffixed{(0.25w,0)}{v3,v4}
\fmffixed{(0.25w,0)}{v5,v6}
\fmffixed{(0.25w,0)}{v6,v7}
\fmffixed{(0.25w,0)}{v7,v8}
\fmffixed{(0,0.9w)}{v5,vh1}
\fmf{plain,tension=0.5,right=0.25}{v1,vc1}
\fmf{plain,tension=0.5,left=0.25}{v2,vc1}
\fmf{plain,tension=0.5,right=0.25}{v3,vc2}
\fmf{plain,tension=0.5,left=0.25}{v4,vc2}
\fmf{plain}{vc1,vc3}
\fmf{plain}{vc3,vc7}
\fmf{plain}{vc7,vc5}
\fmf{plain}{vc2,vc8}
\fmf{plain}{vc8,vc4}
\fmf{plain}{vc4,vc6}
\fmf{plain,tension=0}{vc3,vc4}
\fmf{plain,tension=0.5,left=0.25}{v5,vc5}
\fmf{plain,tension=0.5,right=0.25}{v6,vc5}
\fmf{plain,tension=0.5,left=0.25}{v7,vc6}
\fmf{plain,tension=0.5,right=0.25}{v8,vc6}
\fmf{plain,tension=0.5,right=0,width=1mm}{v5,v8}
\fmffreeze
\fmfposition
\plainwrap{vc7}{v5}{v8}{vc8}
\end{fmfchar*}}}
\subfigspace
\subfigure[$W_{\mathrm{chiral}}^{(2)}$]{
\fmfframe(3,1)(1,4){%
\begin{fmfchar*}(20,20)
\fmftop{v1}
\fmfforce{(0.125w,h)}{v1}
\fmfforce{(0.125w,0)}{v5}
\fmffixed{(0.25w,0)}{v1,v2}
\fmffixed{(0.25w,0)}{v2,v3}
\fmffixed{(0.25w,0)}{v3,v4}
\fmffixed{(0.25w,0)}{v5,v6}
\fmffixed{(0.25w,0)}{v6,v7}
\fmffixed{(0.25w,0)}{v7,v8}
\fmffixed{(whatever,0)}{vc1,vc3}
\fmffixed{(whatever,0)}{vc5,vc7}
\fmffixed{(whatever,0)}{vc3,vc4}
\fmffixed{(whatever,0)}{vc7,vc8}
\fmf{plain,tension=1,right=0.125}{v1,vc1}
\fmf{plain,tension=0.5,right=0.25}{v2,vc2}
\fmf{plain,tension=0.5,left=0.25}{v3,vc2}
\fmf{plain,tension=1,left=0.125}{v4,vc4}
\fmf{plain,tension=1,left=0.125}{v5,vc5}
\fmf{plain,tension=0.5,left=0.25}{v6,vc6}
\fmf{plain,tension=0.5,right=0.25}{v7,vc6}
\fmf{plain,tension=1,right=0.125}{v8,vc8}
\fmf{plain}{vc1,vc5}
\fmf{plain}{vc4,vc8}
\fmf{plain}{vc2,vc3}
\fmf{plain}{vc6,vc7}
\fmf{plain,tension=3}{vc3,vc7}
\fmf{plain,tension=0.5}{vc3,vc4}
\fmf{plain,tension=0.5}{vc5,vc7}
\fmf{phantom,tension=0.5}{vc7,vc8}
\fmf{phantom,tension=0.5}{vc1,vc3}
\fmf{plain,tension=0.5,right=0,width=1mm}{v5,v8}
\fmffreeze
\fmfposition
\plainwrap{vc1}{v5}{v8}{vc8}
\fmffreeze
\end{fmfchar*}}}
\subfigspace
\subfigure[$W_{\mathrm{chiral}}^{(3)}$]{
\fmfframe(3,1)(1,4){%
\begin{fmfchar*}(20,20)
\fmftop{v1}
\fmfbottom{v5}
\fmfforce{(0.125w,h)}{v1}
\fmfforce{(0.125w,0)}{v5}
\fmffixed{(0.25w,0)}{v1,v2}
\fmffixed{(0.25w,0)}{v2,v3}
\fmffixed{(0.25w,0)}{v3,v4}
\fmffixed{(0.25w,0)}{v5,v6}
\fmffixed{(0.25w,0)}{v6,v7}
\fmffixed{(0.25w,0)}{v7,v8}
\fmffixed{(0,whatever)}{vc1,vc5}
\fmffixed{(0,whatever)}{vc2,vc3}
\fmffixed{(0,whatever)}{vc3,vc6}
\fmffixed{(0,whatever)}{vc6,vc7}
\fmffixed{(0,whatever)}{vc4,vc8}
\fmffixed{(0.5w,0)}{vc1,vc4}
\fmffixed{(0.5w,0)}{vc5,vc8}
\fmf{plain,tension=1,right=0.125}{v1,vc1}
\fmf{plain,tension=0.25,right=0.25}{v2,vc2}
\fmf{plain,tension=0.25,left=0.25}{v3,vc2}
\fmf{plain,tension=1,left=0.125}{v4,vc4}
\fmf{plain,tension=1,left=0.125}{v5,vc5}
\fmf{plain,tension=0.25,left=0.25}{v6,vc6}
\fmf{plain,tension=0.25,right=0.25}{v7,vc6}
\fmf{plain,tension=1,right=0.125}{v8,vc8}
\fmf{plain,tension=0.5}{vc1,vc3}
\fmf{plain,tension=0.5}{vc2,vc3}
\fmf{plain,tension=0.5}{vc3,vc4}
\fmf{plain,tension=0.5}{vc5,vc7}
\fmf{plain,tension=0.5}{vc6,vc7}
\fmf{plain,tension=0.5}{vc7,vc8}
\fmf{plain,tension=2}{vc1,vc8}
\fmf{phantom,tension=2}{vc5,vc4}
\fmffreeze
\fmfposition
\plainwrap{vc5}{v5}{v8}{vc4}
\fmf{plain,tension=1,left=0,width=1mm}{v5,v8}
\fmffreeze
\end{fmfchar*}}}
}
\caption{Wrapping diagrams with only chiral interactions}
\label{diagrams-chi}
\end{figure}

\begin{itemize}
\item First of all, the asymptotic dilatation operator depends on a set of coefficients that parameterize the behaviour under similarity transformations. An example is given by the $\epsilon_{2a}$ coefficient at three loops in~\eqref{Duptothree}.  Such coefficients do not affect the spectrum, and depend on the renormalization scheme. Since they are non-physical, they cannot alter the spectrum even in the finite-length case, and hence the contributions proportional to them from the subtraction and the wrapping terms must combine in such a way that the final eigenvalues of the exact dilatation operator do not depend on them. However, the wrapping supergraphs must be calculated in a particular renormalization scheme, and thus their dependence on the similarity coefficients is hidden. This is why the values of the relevant coefficients must be determined explicitly in the chosen scheme, through the computation of a subset of the range-five diagrams.

\item Special care must be dedicated to the listing of all the possible wrapping supergraphs, following~\cite{Sieg:2005kd}. It is natural to organize them starting from the completely chiral ones. There are three of them, shown in Figure~\ref{diagrams-chi}. Their action on the length-four subsector can still be described in terms of the standard chiral functions, after the identification of the first and the fifth lines in the operator, as
\begin{equation}
\label{chiMr4}
\begin{aligned}
W_{\mathrm{chiral}}^{(1)}\quad&\sim\quad\chi(2,4,1,3)\col\\
W_{\mathrm{chiral}}^{(2)}\quad&\sim\quad\chi(4,1,2,3)\col\\
W_{\mathrm{chiral}}^{(3)}\quad&\sim\quad\chi(4,3,1,2)\pnt
\end{aligned}
\end{equation}

We then consider wrapping diagrams with vector interactions. Since we work at the critical order they can all be drawn so that the wrapping line is a vector one. Thus, any supergraph of this kind can be obtained from one of the chiral structures with range up to three, by adding the right number of vectors in all the possible ways. The cancellation result of Section~\ref{superspace} reduces the number of relevant contributions. As an example, we list the relevant supergraphs with chiral structure $\chi(2,1)$ in Figure~\ref{diagrams-21}.
\begin{figure}[t]
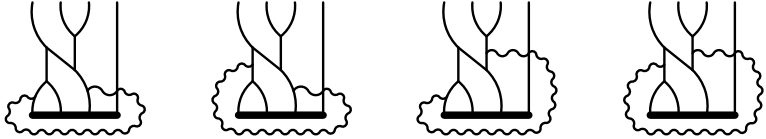

\centering
\renewcommand*{\thesubfigure}{}
\addtolength{\subfigcapskip}{5pt}
\footnotesize
\unitlength=0.75mm
\settoheight{\eqoff}{$\times$}%
\setlength{\eqoff}{0.5\eqoff}%
\addtolength{\eqoff}{-12.5\unitlength}%
\settoheight{\eqofftwo}{$\times$}%
\setlength{\eqofftwo}{0.5\eqofftwo}%
\addtolength{\eqofftwo}{-7.5\unitlength}%
\subfigure[]{
\fmfframe(3,1)(1,4){%
\begin{fmfchar*}(20,20)
\Wthreeplain
\fmfipair{wu[]}
\fmfipair{w[]}
\fmfipair{wd[]}
\svertex{w3}{p3}
\svertex{w6}{p6}
\vvertex{w8}{w6}{p7}
\fmfi{wiggly}{w6..w8}
\wigglywrap{w3}{v5}{v8}{w8}
\end{fmfchar*}}}
\subfigspace
\subfigspace
\subfigure[]{
\fmfframe(3,1)(1,4){%
\begin{fmfchar*}(20,20)
\Wthreeplain
\fmfipair{wu[]}
\fmfipair{w[]}
\fmfipair{wd[]}
\svertex{w2}{p2}
\svertex{w6}{p6}
\vvertex{w8}{w6}{p7}
\fmfi{wiggly}{w6..w8}
\wigglywrap{w2}{v5}{v8}{w8}
\end{fmfchar*}}}
\subfigspace
\subfigspace
\subfigure[]{
\fmfframe(3,1)(1,4){%
\begin{fmfchar*}(20,20)
\Wthreeplain
\fmfipair{wu[]}
\fmfipair{w[]}
\fmfipair{wd[]}
\svertex{w3}{p3}
\svertex{w5}{p5}
\vvertex{w8}{w5}{p7}
\fmfi{wiggly}{w5..w8}
\wigglywrap{w3}{v5}{v8}{w8}
\end{fmfchar*}}}
\subfigspace
\subfigspace
\subfigure[]{
\fmfframe(3,1)(1,4){%
\begin{fmfchar*}(20,20)
\Wthreeplain
\fmfipair{wu[]}
\fmfipair{w[]}
\fmfipair{wd[]}
\svertex{w2}{p2}
\svertex{w5}{p5}
\vvertex{w8}{w5}{p7}
\fmfi{wiggly}{w5..w8}
\wigglywrap{w2}{v5}{v8}{w8}
\end{fmfchar*}}}
\normalsize
\caption{Wrapping diagrams with chiral structure $\chi(2,1)$}
\label{diagrams-21}
\end{figure}

\item Another non-trivial aspect of the procedure concerns the symmetry factors for wrapping diagrams. The best way to draw a wrapping graph is on the surface of a cylinder, with one of the bases representing the composite operator. Such a representation is suggested by the cyclicity of the trace, and is useful to analyze the behaviour under a parity transformation, which reverses the order of the fields in the chain, thus leaving two-impurity states of the $SU(2)$ sector unchanged. When a diagram is not symmetric, we can account for its reflection by simply doubling its contribution.
\end{itemize}

Once we have listed all the possible supergraphs, we perform the D-algebra~\cite{Gates:1983nr} and obtain a set of momentum integrals that can be computed, for example, by means of the Gegenbauer Polynomial $x$-space technique (GPXT)~\cite{Chetyrkin:1980pr}. This method is particularly effective in our case because the insertion of the composite operator, being a vertex from where a large number of lines start, is typically a very good candidate for the root vertex~\cite{Chetyrkin:1980pr,usfive}, whose wise choice can simplify the calculation considerably. As a consequence, trying to extract anomalous dimensions from the two-point function would be much more difficult using the GPXT. 
Moreover, in the calculation of anomalous dimensions we only need the divergent parts of diagrams. This fact result in a further simplification, since it allows us to neglect the exponential factor depending on the external momentum, which does not affect the ultraviolet behaviour of the integral. This is equivalent to putting the external momentum to zero, so a cutoff $R$ on the radial integrations is required to avoid the appearance of infrared divergences~\cite{Chetyrkin:1980pr}. In this way, we reduce by one the number of infinite summations, and we obtain integrands that are simple monomials in the radial variables, free of Bessel functions. In more complicated situations, where more than one power of the same momentum appears in the numerator, an additional regularization must be introduced to correctly consider the dimensional continuation of the appearing traceless products, as discussed in detail in~\cite{uslong}.

Proceeding in this way, we were able to compute the exact correction to the four-loop dilatation operator on the length-four subsector. The linear combinations of the basis~\eqref{Opbasis} that renormalize multiplicatively are found to be
\begin{equation}
\begin{aligned}
&\mathcal{O}_{\mathrm{protected}}=\mathcal{O}_{4,1}+2\mathcal{O}_{4,2}=\frac{1}{2}[3\,\mathrm{tr}(\phi\{Z,\phi\}Z)-\mathrm{tr}(\phi[Z,\phi]Z)] \col \\
&\mathcal{O}_K=\basisop{4}{1}-\basisop{4}{2}=\mathrm{tr}(\phi[Z,\phi]Z) \pnt
\end{aligned}
\end{equation}
The first one is protected, whereas the second one, which is a Konishi descendant, corresponds to the four-loop eigenvalue
\begin{equation}
\gamma_4=-2496+576\zeta(3)-1440\zeta(5) \pnt
\end{equation}
Restoring the lower-order components, we can write the exact expression for the Konishi anomalous dimension up to four loops
\begin{equation}
\label{finalgamma}
\gamma=12\lambda-48\lambda^2+336\lambda^3+\lambda^4(-2496+576\zeta(3)-1440\zeta(5))
\pnt
\end{equation}
The most evident feature of this result is the presence of the $\zeta(5)$ term, which comes entirely from wrapping interactions and which cannot arise in the asymptotic regime, where the only allowed transcendental term is $\zeta(3)$.\footnote{Based on this, in the recent paper \cite{Kataev:2010tm} the absence of 
a $\zeta(5)$ term in the results of perturbative quenched QED
was interpreted as the absence of wrapping interactions.} The appearance of these transcendental functions is expected from the structure of wrapping integrals and the Gegenbauer Polynomial $x$-space technique. Moreover, our exact result ruled out previous conjectures based on analogies with the Hubbard model~\cite{Rej:2005qt} or the BFKL equation~\cite{Lipatov:1976zz,Kuraev:1977fs,Balitsky:1978ic,Kotikov:2007cy}.

The exact value of the anomalous dimension~\eqref{finalgamma} was later confirmed by an independent computation performed on the string side and based on the L\"uscher approach~\cite{Bajnok:2008bm}. This is a highly non-trivial check for both of the procedures and for the AdS/CFT correspondence itself. Afterwards, a further check was obtained from a computer-made perturbative computation based on the component-field formalism~\cite{Velizhanin:2009zz}, which has also been extended to take non-planar contributions into account~\cite{Velizhanin:2009gv}. The comparison between our approach and the one used in~\cite{Velizhanin:2009zz} allows to understand better the power of integrability and superspace techniques: we only had to compute explicitly less than fifty supergraphs, whereas the component-field approach involves more than 130\!~000 diagrams.

Finally, the wrapping correction to the Konishi anomalous dimension was obtained also from the Y-system in~\cite{Gromov:2009tv}, thus representing the first explicit check on the validity of the new proposal.

\subsection{The five-loop case}
The four-loop analysis can be extended to five loops in a straightforward way to compute wrapping effects on the length-five operators
\begin{equation}
\label{Opbasis5}
\basisop{5}{1}=\mathrm{tr}(\phi Z\phi ZZ)\col\qquad\basisop{5}{2}=\mathrm{tr}(\phi\phi ZZZ) \pnt
\end{equation}
With this choice for the length, order five is again the critical perturbative order for finite-size contributions. 
All the details of the computation have been presented in~\cite{usfive}.

Some care must be dedicated to the determination of the most general form of the asymptotic operator $\mathcal{D}_5$, in order to fully parameterize the behaviour under similarity transformations, so that explicit computations in any renormalization scheme are possible. For a detailed description of this procedure and of the derivation of $\mathcal{D}_5$ from the integrability hypothesis, see~\cite{Beisert:2004ry,usfive}.
In the five-loop case, we must subtract the contribution of range-six interactions. This task can be accomplished again through the simple cancellation of the range-six chiral functions in the expression of the asymptotic five-loop dilatation operator.

A difference with respect to the four-loop case is represented by the fact that now not all of the lower-range structures appear in independent
wrapping diagrams. In particular, $\chi(1,2,4)$ and $\chi(2,1,4)$ lead to the same wrapping diagrams. Similarly, $\chi(1,3)$ is equivalent to $\chi(1,4)$.
 Hence, an independent set of structures is made of 
\begin{gather}
\chi(2,4,1,3)\col\  \chi(3,2,1,4)\col\  \chi(1,2,3,4)\col\  \chi(1,4,3,2)\col\  \chi(1,3,2)\col\\
\chi(2,1,3)\col\ \chi(1,2,3)\col\  \chi(2,1,4)\col\  \chi(2,1)\col\  \chi(1,4)\col\  \chi(1) \pnt
\end{gather}
Putting all the contributions together, the total wrapping correction to the five-loop dilatation operator can be assembled as a $2\times2$ matrix on the length-five subsector.
The linear combination of the basis operators~\eqref{Opbasis5}
\begin{equation}
\basisop{5}{1}'=\mathrm{tr}(\phi Z\phi ZZ)+\mathrm{tr}(\phi\phi ZZZ)\col
\end{equation}
is protected. On the contrary, the combination
\begin{equation}
\label{op5mult}
\basisop{5}{2}'=\mathrm{tr}(\phi Z\phi ZZ)-\mathrm{tr}(\phi\phi ZZZ)\col
\end{equation}
is not protected, and its exact anomalous dimension up to five loops is
\begin{multline}
\label{fullgamma5}
\gamma=8\lambda-24\lambda^2+136\lambda^3-8[115+16\zeta(3)]\lambda^4 \\
+[6664+1152\zeta(3)+3840\zeta(5)-2240\zeta(7)]\lambda^5 \pnt
\end{multline}
As in the four-loop case, the maximum-transcendentality term in the final result, which in this case is $\zeta(7)$, is generated entirely by wrapping effects. The $\zeta(5)$ and $\zeta(3)$ terms get contributions also from the dressing phase of the asymptotic regime.

The result~\eqref{fullgamma5} confirms the computation of~\cite{Beccaria:2009eq}, based on a conjecture. Moreover, it agrees~\cite{usfive} with the prediction of the Y-system, thus representing a new independent check for it.

It would be very interesting to be able to compute also the five-loop exact anomalous dimension of the Konishi descendant, which we studied at four loops in the previous section and which has length four. In fact, its value has been obtained in~\cite{Bajnok:2009vm} by means of the L\"uscher approach and in~\cite{Arutyunov:2010gb,Balog:2010xa} using the Thermodynamic Bethe Ansatz, so it would be important to check the result against a direct perturbative computation based on pure field-theoretical techniques. At the moment, however, such a computation seems to be out of reach, even with the help of superspace methods, since when the condition of criticality of the perturbative order no longer holds the number of possibly relevant diagrams greatly increases, and the D-algebra becomes much more complicated. Anyway, we think that such difficulties may be overcome if new and more powerful cancellation identities, similar to those described in Section~\ref{superspace}, will be found.

\section{Wrapping in the $\beta$-deformed $\N=4$ SYM}
\label{betadef}
We now move to the analysis of wrapping effects in the $\beta$-deformed $\N=4$ SYM~\cite{betadef,Fiamberti:2008sn}. We will see that the features of this theory allow to perform computations at much higher orders.

\subsection{General considerations}
The general procedure described in Section~\ref{undeformed} to deal with finite-size effects can be applied in the deformed case, too. In order to do this, we need the expressions for the components of the asymptotic dilatation operator. It is easy to see that this operator can be found directly from its undeformed counterpart, without the need for full computations from scratch. Let us consider in fact the deformed permutations~\cite{Berenstein:2004ys}
\begin{equation}
\dperm_{i,j}=\frac{1}{2}\left[\unitmatrix_{i,j}+\sigma_i^3\sigma_j^3+q^2\,\sigma_i^+\sigma_j^-+\bar{q}^2\,\sigma_i^-\sigma_j^+\right] 
\col\qquad\qquad q\equiv e^{i\pi\beta} \col
\label{defperm}
\end{equation}
where the $\sigma_k^j$ are the Pauli matrices at chain site $k$ and $\sigma_k^\pm=\sigma_k^1\pm i\sigma_k^2$, which reduce to the standard ones in the undeformed limit $\beta\to0$. We can use them to build a set of basis operators similar to~\eqref{permstrucdef}
\begin{equation}
\label{defbasisops}
\dpthree{a_1}{\dots}{a_n}=\sum_{r=0}^{L-1}\dperm_{a_1+r,\;a_1+r+1}\cdots\dperm_{a_n+r,\;a_n+r+1}
\pnt
\end{equation}
This definition is suggested by the fact that the deformation affects only the three-scalar interactions, and is useful because it allows to write the deformed chiral functions $\dchi{\ldots}$, describing the chiral structure of Feynman supergraphs in the deformed theory, as linear combinations of the operators~\eqref{defbasisops} with the same coefficients that appear in~\eqref{chistruc}. The function $\dchi{1}$ is still the building block for all the possible non-trivial chiral structures, and all the dependence of a diagram on $\beta$ is encoded in the corresponding chiral function. Hence, the deformed dilatation operator must be writable in the basis of the $\dchi{\ldots}$ functions with coefficients that do not depend on $\beta$. Since the chiral functions reduce to their undeformed counterparts when $\beta\to0$, we can conclude that such coefficients must be exactly the same as in the undeformed case. 

So we can obtain the deformed dilatation operator simply by replacing every chiral function in the expansion of the standard dilatation operator with its deformed version. In particular, we can deform the result we found in the previous section to compute wrapping effects at four and five loops on two-impurity states of length four and five respectively. The four-loop result has been
confirmed recently using the L\"uscher technique in \cite{Ahn:2010yv}.
The resulting expressions are not very enlightening, being complicated functions of $\beta$, and so we will not present them here. Instead, it is more interesting to focus on the class of single-impurity operators, which were protected by supersymmetry in standard $\N=4$ SYM, but which are not in the $\beta$-deformed case. These states in fact are easier to study than two-impurity ones, and thus we will be able to push perturbative computations up to higher orders. It is useful to summarize the main features of single-impurity operators that give rise to such simplifications:
\begin{itemize}
\item for every value of $L$, only one single-impurity operator $\mathcal{O}_L=\tr(\phi Z^{L-1})$ of length $L$ exists, so that there is no mixing under renormalization and Feynman supergraphs reduce to numbers instead of matrices.
\item We do not need the explicit expression for the asymptotic dilatation operator: the asymptotic contribution to the anomalous dimension can be extracted from the all-loop formula~\cite{Mauri:2005pa} 
\begin{equation}
\label{single-all-orders}
\gamma(\mathcal{O}_\text{as})=-1+\sqrt{1+4\lambda\Big\vert q-\frac{1}{q}\Big\vert^2}=-1+\sqrt{1+16\lambda\sin^2(\pi\beta)} \col
\end{equation}
as
\begin{equation}
\gamma^{as}_L= \alpha_L\,\lambda^L \sin^{2L}(\pi\beta)\col\qquad\alpha_L=-(-8)^L \frac{(2L-3)!!}{L!} \pnt
\end{equation}
\item The form of the possible chiral structures of Feynman graphs is restricted: apart from completely chiral wrapping diagrams, the only allowed functions are those of the form $\dchi{1,2,\ldots,k}$ and their parity reflections. Note that, besides the reduction in the possible number of contributions, with respect to the two-impurity case here the most complicated structures (especially those with two $\chi(1)$ blocks acting directly on the composite operator) do not contribute.
\end{itemize}

\begin{figure}[t]
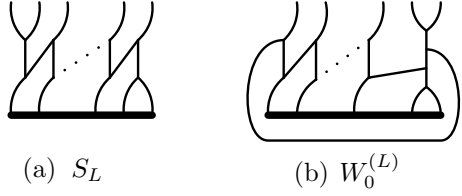

\centering
\unitlength=0.75mm
\addtolength{\subfigcapskip}{5pt}
\settoheight{\eqoff}{$\times$}%
\setlength{\eqoff}{0.5\eqoff}%
\addtolength{\eqoff}{-12.5\unitlength}%
\settoheight{\eqofftwo}{$\times$}%
\setlength{\eqofftwo}{0.5\eqofftwo}%
\addtolength{\eqofftwo}{-7.5\unitlength}%
\subfigure[$\protect\phantom{.^{(L)}}\!\!\!\!\!\!\!\! S_L\quad$]{
\label{SL}
\raisebox{\eqoff}{%
\fmfframe(3,1)(1,4){%
\begin{fmfchar*}(25,20)
\fmftop{v1}
\fmfbottom{v5}
\fmfforce{(0w,h)}{v1}
\fmfforce{(0w,0)}{v5}
\fmffixed{(0.2w,0)}{v1,v2}
\fmffixed{(0.2w,0)}{v2,v3}
\fmffixed{(0.2w,0)}{v3,w1}
\fmffixed{(0.2w,0)}{w1,v4}
\fmffixed{(0.2w,0)}{v4,v9}
\fmffixed{(0.2w,0)}{v5,v6}
\fmffixed{(0.2w,0)}{v6,v7}
\fmffixed{(0.2w,0)}{v7,w2}
\fmffixed{(0.2w,0)}{w2,v8}
\fmffixed{(0.2w,0)}{v8,v10}
\fmffixed{(0,whatever)}{vc1,vc2}
\fmffixed{(0,whatever)}{vc3,vc4}
\fmffixed{(0,whatever)}{vc5,vc6}
\fmffixed{(0,whatever)}{vc7,vc8}
\fmf{plain,tension=0.25,right=0.25}{v1,vc1}
\fmf{plain,tension=0.25,left=0.25}{v2,vc1}
\fmf{plain,left=0.25}{v5,vc2}
\fmf{plain,tension=1,left=0.25}{v3,vc3}
\fmf{phantom,tension=1,left=0.25}{w1,wc1}
\fmf{plain,tension=1,left=0.25}{v4,vc5}
\fmf{plain,tension=1,left=0.25}{v9,vc7}
\fmf{plain,left=0.25}{w2,vc6}
\fmf{plain,tension=0.25,left=0.25}{v8,vc8}
\fmf{plain,tension=0.25,right=0.25}{v10,vc8}
\fmf{plain,left=0.25}{v6,vc4}
\fmf{phantom,left=0.25}{v7,wc2}
\fmf{plain,tension=0.5}{vc1,vc2}
\fmf{phantom,tension=0.5}{wc1,wc2}
\fmf{plain,tension=0.5}{vc2,vc3}
\fmf{phantom,tension=0.5}{wc2,vc5}
\fmf{plain,tension=0.5}{vc3,vc4}
\fmf{phantom,tension=0.5}{vc4,wc1}
\fmf{plain,tension=0.5}{vc5,vc6}
\fmf{plain,tension=0.5}{vc6,vc7}
\fmf{plain,tension=0.5}{vc7,vc8}
\fmf{plain,tension=0.5,right=0,width=1mm}{v5,v10}
\fmffreeze
\fmf{dots,tension=0.5}{vc4,vc5}
\fmfposition
\end{fmfchar*}}}
}
\subfigspace
\subfigspace
\subfigure[$W_0^{(L)}$]{
\label{WL}
\raisebox{\eqoff}{%
\fmfframe(3,1)(1,4){%
\begin{fmfchar*}(30,20)
\fmftop{v1}
\fmfbottom{v5}
\fmfforce{(0w,h)}{v1}
\fmfforce{(0w,0)}{v5}
\fmffixed{(0.17w,0)}{v1,v2}
\fmffixed{(0.17w,0)}{v2,v3}
\fmffixed{(0.17w,0)}{v3,w1}
\fmffixed{(0.17w,0)}{w1,w3}
\fmffixed{(0.17w,0)}{w3,v4}
\fmffixed{(0.17w,0)}{v4,v9}
\fmffixed{(0.17w,0)}{v5,v6}
\fmffixed{(0.17w,0)}{v6,v7}
\fmffixed{(0.17w,0)}{v7,w2}
\fmffixed{(0.17w,0)}{w2,w4}
\fmffixed{(0.17w,0)}{w4,v8}
\fmffixed{(0.17w,0)}{v8,v10}
\fmffixed{(0,whatever)}{vc1,vc2}
\fmffixed{(0,whatever)}{vc3,vc4}
\fmffixed{(0,whatever)}{vc5,vc6}
\fmffixed{(0,whatever)}{vc7,vc8}
\fmf{phantom,tension=0.25,right=0.25}{v1,vc1}
\fmf{plain,tension=0.25,left=0.25}{v2,vc1}
\fmf{plain,left=0.25}{v5,vc2}
\fmf{plain,tension=1,left=0.25}{v3,vc3}
\fmf{phantom,tension=1,left=0.25}{v3,wc3}
\fmf{phantom,tension=1,left=0.25}{w1,wc3}
\fmf{plain,tension=0.5,right=0.25}{v4,vc7}
\fmf{plain,tension=0.5,left=0.25}{v9,vc7}
\fmf{plain,tension=0.5,left=0.25}{v8,vc8}
\fmf{plain,tension=0.5,right=0.25}{v10,vc8}
\fmf{plain,left=0.25}{v6,vc4}
\fmf{phantom,right=0.25}{v7,vc4}
\fmf{phantom}{vc4,wc3}
\fmf{plain,tension=0.5}{vc1,vc2}
\fmf{plain,tension=0.5}{vc2,vc3}
\fmf{plain,tension=0.5}{vc3,vc4}
\fmf{plain,tension=0.5}{vc7,vc8}
\fmf{plain,tension=0.5,right=0,width=1mm}{v5,v10}
\fmf{plain,tension=0.5}{vc5,vc6}
\fmffreeze
\fmf{plain,tension=0.25,left=0.25}{w2,vc6}
\fmf{phantom,tension=0.25,right=0.25}{w4,vc6}
\fmf{phantom,tension=0.25,right=0.25}{w1,wc1}
\fmf{plain,tension=0.25,left=0.25}{w3,wc1}
\fmf{plain,tension=0.5}{vc6,wc1}
\fmffreeze
\fmf{dots}{vc4,wc1}
\fmfposition
\fmfipath{p[]}
\fmfiset{p1}{vpath(__vc8,__vc7)}
\fmfipair{wz[]}
\fmfiequ{wz1}{point length(p1)/3 of p1}
\fmfiequ{wz2}{point 2*length(p1)/3 of p1}
\fmfiequ{wz3}{(xpart(vloc(__vc6)),ypart(vloc(__vc6)))}
\fmfiequ{wz4}{(xpart(vloc(__vc1)),ypart(vloc(__vc1)))}
\fmf{plain}{vc6,v12}
\fmfforce{(xpart(wz1),ypart(wz1))}{v11}
\fmfforce{(xpart(wz2),ypart(wz2))}{v12}
\plainwrap{vc1}{v5}{v10}{v11}
\end{fmfchar*}}}
}
\caption{Range-$(L+1)$ and completely-chiral wrapping diagrams at $L$ loops}
\label{diagram-SL}
\end{figure}
\begin{figure}[t]
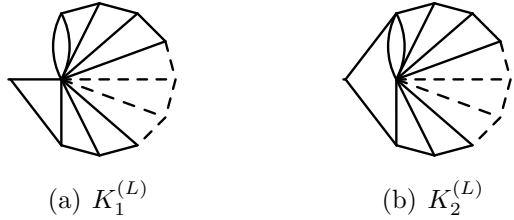

\centering
\addtolength{\subfigcapskip}{5pt}
\unitlength=0.75mm
\settoheight{\eqoff}{$\times$}%
\setlength{\eqoff}{0.5\eqoff}%
\addtolength{\eqoff}{-12.5\unitlength}%
\settoheight{\eqofftwo}{$\times$}%
\setlength{\eqofftwo}{0.5\eqofftwo}%
\addtolength{\eqofftwo}{-7.5\unitlength}%
\centering
\subfigure[$\kint{L}{1}$]{
\label{JL}
\raisebox{\eqoff}{%
\fmfframe(0,0)(0,0){%
\begin{fmfchar*}(30,30)
\fmfleft{in}
\fmfright{out}
\fmf{plain,tension=1}{in,vi}
\fmf{phantom,tension=1}{out,v7}
\fmfpoly{phantom}{v12,v11,v10,v9,v8,v7,v6,v5,v4,v3,v2,v1}
\fmffixed{(0.9w,0)}{v1,v7}
\fmffixed{(0.3w,0)}{vi,v0}
\fmffixed{(0,whatever)}{v0,v3}
\fmf{plain}{v3,v4}
\fmf{plain}{v4,v5}
\fmf{plain}{v5,v6}
\fmf{dashes}{v6,v7}
\fmf{dashes}{v7,v8}
\fmf{dashes}{v8,v9}
\fmf{plain}{v9,v10}
\fmf{plain}{v10,v11}
\fmf{phantom}{v3,vi}
\fmf{plain}{vi,v11}
\fmf{phantom}{v0,v3}
\fmf{plain}{v0,v4}
\fmf{plain}{v0,v5}
\fmf{plain}{v0,v6}
\fmf{dashes}{v0,v7}
\fmf{dashes}{v0,v8}
\fmf{plain}{v0,v9}
\fmf{plain}{v0,v10}
\fmf{plain}{v0,v11}
\fmffreeze
\fmfposition
\fmf{plain}{vi,v0}
\fmf{plain,left=0.25}{v0,v3}
\fmf{plain,left=0.25}{v3,v0}
\fmfipair{w[]}
\fmfiequ{w1}{(xpart(vloc(__v3)),ypart(vloc(__v3)))}
\fmfiequ{w2}{(xpart(vloc(__v4)),ypart(vloc(__v4)))}
\fmfiequ{w3}{(xpart(vloc(__v5)),ypart(vloc(__v5)))}
\fmfiequ{w4}{(xpart(vloc(__v6)),ypart(vloc(__v6)))}
\fmfiequ{w5}{(xpart(vloc(__v7)),ypart(vloc(__v7)))}
\fmfiequ{w6}{(xpart(vloc(__v8)),ypart(vloc(__v8)))}
\fmfiequ{w7}{(xpart(vloc(__v9)),ypart(vloc(__v9)))}
\end{fmfchar*}}
}
}
\qquad\qquad
\subfigure[$\kint{L}{2}$]{
\label{KL}
\raisebox{\eqoff}{%
\fmfframe(0,0)(0,0){%
\begin{fmfchar*}(30,30)
\fmfleft{in}
\fmfright{out}
\fmf{plain,tension=1}{in,vi}
\fmf{phantom,tension=1}{out,v7}
\fmfpoly{phantom}{v12,v11,v10,v9,v8,v7,v6,v5,v4,v3,v2,v1}
\fmffixed{(0.9w,0)}{v1,v7}
\fmffixed{(0.3w,0)}{vi,v0}
\fmffixed{(0,whatever)}{v0,v3}
\fmf{plain}{v3,v4}
\fmf{plain}{v4,v5}
\fmf{plain}{v5,v6}
\fmf{dashes}{v6,v7}
\fmf{dashes}{v7,v8}
\fmf{dashes}{v8,v9}
\fmf{plain}{v9,v10}
\fmf{plain}{v10,v11}
\fmf{plain}{v3,vi}
\fmf{plain}{vi,v11}
\fmf{phantom}{v0,v3}
\fmf{plain}{v0,v4}
\fmf{plain}{v0,v5}
\fmf{plain}{v0,v6}
\fmf{dashes}{v0,v7}
\fmf{dashes}{v0,v8}
\fmf{plain}{v0,v9}
\fmf{plain}{v0,v10}
\fmf{plain}{v0,v11}
\fmffreeze
\fmfposition
\fmf{plain,left=0.25}{v0,v3}
\fmf{plain,left=0.25}{v3,v0}
\fmfipair{w[]}
\fmfiequ{w1}{(xpart(vloc(__v3)),ypart(vloc(__v3)))}
\fmfiequ{w2}{(xpart(vloc(__v4)),ypart(vloc(__v4)))}
\fmfiequ{w3}{(xpart(vloc(__v5)),ypart(vloc(__v5)))}
\fmfiequ{w4}{(xpart(vloc(__v6)),ypart(vloc(__v6)))}
\fmfiequ{w5}{(xpart(vloc(__v7)),ypart(vloc(__v7)))}
\fmfiequ{w6}{(xpart(vloc(__v8)),ypart(vloc(__v8)))}
\fmfiequ{w7}{(xpart(vloc(__v9)),ypart(vloc(__v9)))}
\end{fmfchar*}}
}
}
\caption{$L$-loop integrals from diagrams $\sub{L}$ and $\swrap{L}{0}$}
\label{integrals-SWL}
\end{figure}

\subsection{Single-impurity operators at higher loops}
We are now ready to undertake the general discussion on single-impurity states, following~\cite{betadef,Fiamberti:2008sn}. More precisely, we will give a general formula for the wrapping correction to the $L$-loop anomalous dimension of the length-$L$ single-impurity operator $\mathcal{O}_L$. 

Thanks to the validity of the general supergraph cancellation identities of Section~\ref{superspace}, and given the restrictions on the possible chiral structures for single-impurity states, we conclude that only two diagrams need to be considered for the subtraction of range-$(L+1)$ interactions, the $S_L$ of Figure~\ref{SL} and its reflection. In the deformed theory, the outcome of a supergraph will be complex in general, since every vertex gets a factor of $q$ or $\bar{q}$ depending on the order of the $\phi$, $Z$ and $\psi$ superfields. A parity transformation reverses this order, thus turning each $q$ into a $\bar{q}$ and vice-versa. Hence, the parity reflection of a supergraph is actually the complex conjugate of the original diagram, and their total contribution is real as expected. The structure of $S_L$ is simple enough to allow to complete the D-algebra for any $L$, and we find
\begin{equation}
\begin{aligned}
\sub{L}+\bsub{L}&\rightarrow
(g^2 N)^L \kint{L}{1}\,[\dchi{1,2,\dots,L}+
\dchi{L,\dots,2,1}]\\
&\phantom{{}\rightarrow{}}
=(g^2 N)^L \kint{L}{1}\,
(q-\bar{q})^2\left[q^{2(L-1)}+\bar{q}^{2(L-1)}
\right]\col
\end{aligned}
\end{equation}
where $\kint{L}{1}$ is the $L$-loop integral of Figure~\ref{JL}.

Similarly, we have to consider a single completely chiral wrapping supergraph, the $W_0^{(L)}$ of Figure~\ref{WL}. In this case, the D-algebra gives
\begin{equation}
\begin{aligned}
\swrap{L}{0}+\bswrap{L}{0}&\rightarrow(g^2 N)^L \kint{L}{2}\,[\dchi{L,1,2,\dots,L-1}+\dchi{1,L,L-1,\dots,2}]\\
&\phantom{{}\rightarrow{}}=(g^2 N)^L \kint{L}{2}\,
(q-\bar{q})^2\left[q^{2(L-1)}+\bar{q}^{2(L-1)}
\right]\col
\end{aligned}
\end{equation}
where the $L$-loop integral $\kint{L}{2}$ is shown in Figure~\ref{KL}.

\begin{figure}[t]
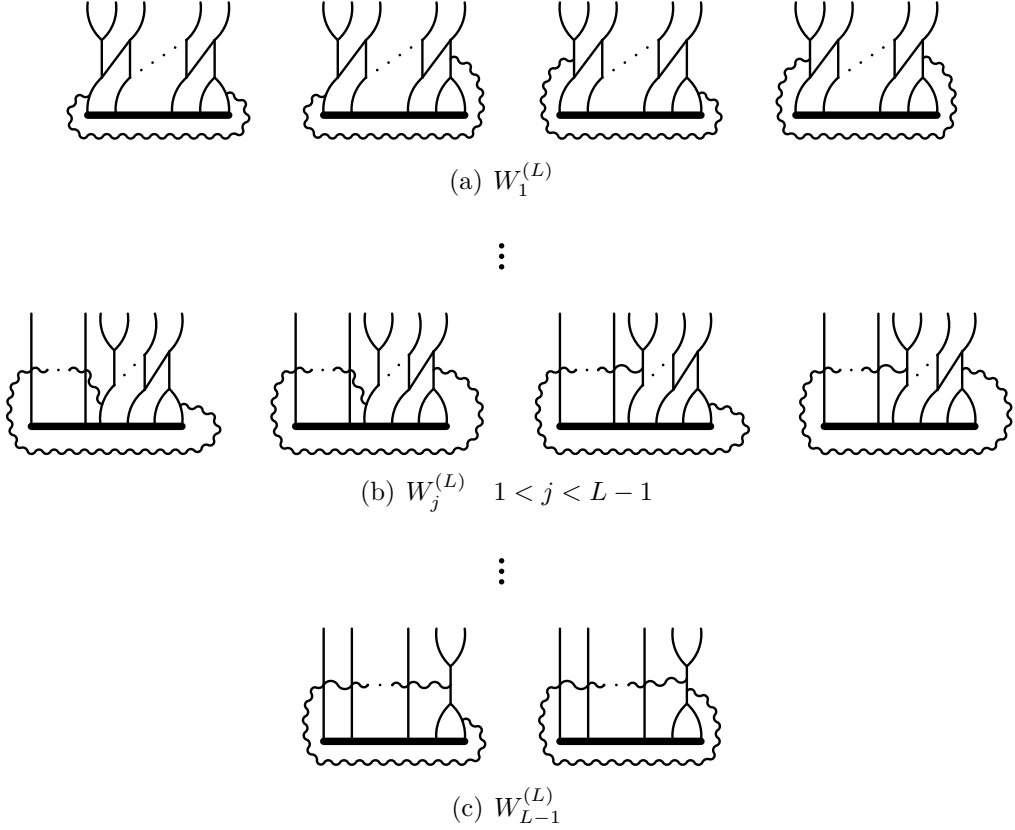

\centering
\addtolength{\subfigcapskip}{5pt}
\unitlength=0.75mm
\settoheight{\eqoff}{$\times$}%
\setlength{\eqoff}{0.5\eqoff}%
\addtolength{\eqoff}{-12.5\unitlength}%
\settoheight{\eqofftwo}{$\times$}%
\setlength{\eqofftwo}{0.5\eqofftwo}%
\addtolength{\eqofftwo}{-7.5\unitlength}%
\centering
\raisebox{\eqoff}{%
\subfigure[$\swrap{L}{1}$]
{
\fmfframe(3,1)(1,4){%
\begin{fmfchar*}(25,20)
\fmftop{v1}
\fmfbottom{v5}
\fmfforce{(0w,h)}{v1}
\fmfforce{(0w,0)}{v5}
\fmffixed{(0.2w,0)}{v1,v2}
\fmffixed{(0.2w,0)}{v2,v3}
\fmffixed{(0.2w,0)}{v3,w1}
\fmffixed{(0.2w,0)}{w1,v4}
\fmffixed{(0.2w,0)}{v4,v9}
\fmffixed{(0.2w,0)}{v5,v6}
\fmffixed{(0.2w,0)}{v6,v7}
\fmffixed{(0.2w,0)}{v7,w2}
\fmffixed{(0.2w,0)}{w2,v8}
\fmffixed{(0.2w,0)}{v8,v10}
\fmffixed{(0,whatever)}{vc1,vc2}
\fmffixed{(0,whatever)}{vc3,vc4}
\fmffixed{(0,whatever)}{vc5,vc6}
\fmffixed{(0,whatever)}{vc7,vc8}
\fmf{plain,tension=0.25,right=0.25}{v1,vc1}
\fmf{plain,tension=0.25,left=0.25}{v2,vc1}
\fmf{plain,left=0.25}{v5,vc2}
\fmf{plain,tension=1,left=0.25}{v3,vc3}
\fmf{phantom,tension=1,left=0.25}{w1,wc1}
\fmf{plain,tension=1,left=0.25}{v4,vc5}
\fmf{plain,tension=1,left=0.25}{v9,vc7}
\fmf{plain,left=0.25}{w2,vc6}
\fmf{plain,tension=0.25,left=0.25}{v8,vc8}
\fmf{plain,tension=0.25,right=0.25}{v10,vc8}
\fmf{plain,left=0.25}{v6,vc4}
\fmf{phantom,left=0.25}{v7,wc2}
\fmf{plain,tension=0.5}{vc1,vc2}
\fmf{phantom,tension=0.5}{wc1,wc2}
\fmf{plain,tension=0.5}{vc2,vc3}
\fmf{phantom,tension=0.5}{wc2,vc5}
\fmf{plain,tension=0.5}{vc3,vc4}
\fmf{phantom,tension=0.5}{vc4,wc1}
\fmf{plain,tension=0.5}{vc5,vc6}
\fmf{plain,tension=0.5}{vc6,vc7}
\fmf{plain,tension=0.5}{vc7,vc8}
\fmf{plain,tension=0.5,right=0,width=1mm}{v5,v10}
\fmffreeze
\fmf{dots,tension=0.5}{vc4,vc5}
\fmfposition
\fmfipath{p[]}
\fmfiset{p1}{vpath(__v5,__vc2)}
\fmfiset{p2}{vpath(__v10,__vc8)}
\fmfipair{wz[]}
\fmfiequ{wz2}{point length(p2)/2 of p2}
\vvertex{wz1}{wz2}{p1}
\svertex{wz2}{p2}
\wigglywrap{wz1}{v5}{v10}{wz2}
\fmfposition
\end{fmfchar*}}
\qquad
\fmfframe(3,1)(1,4){%
\begin{fmfchar*}(25,20)
\fmftop{v1}
\fmfbottom{v5}
\fmfforce{(0w,h)}{v1}
\fmfforce{(0w,0)}{v5}
\fmffixed{(0.2w,0)}{v1,v2}
\fmffixed{(0.2w,0)}{v2,v3}
\fmffixed{(0.2w,0)}{v3,w1}
\fmffixed{(0.2w,0)}{w1,v4}
\fmffixed{(0.2w,0)}{v4,v9}
\fmffixed{(0.2w,0)}{v5,v6}
\fmffixed{(0.2w,0)}{v6,v7}
\fmffixed{(0.2w,0)}{v7,w2}
\fmffixed{(0.2w,0)}{w2,v8}
\fmffixed{(0.2w,0)}{v8,v10}
\fmffixed{(0,whatever)}{vc1,vc2}
\fmffixed{(0,whatever)}{vc3,vc4}
\fmffixed{(0,whatever)}{vc5,vc6}
\fmffixed{(0,whatever)}{vc7,vc8}
\fmf{plain,tension=0.25,right=0.25}{v1,vc1}
\fmf{plain,tension=0.25,left=0.25}{v2,vc1}
\fmf{plain,left=0.25}{v5,vc2}
\fmf{plain,tension=1,left=0.25}{v3,vc3}
\fmf{phantom,tension=1,left=0.25}{w1,wc1}
\fmf{plain,tension=1,left=0.25}{v4,vc5}
\fmf{plain,tension=1,left=0.25}{v9,vc7}
\fmf{plain,left=0.25}{w2,vc6}
\fmf{plain,tension=0.25,left=0.25}{v8,vc8}
\fmf{plain,tension=0.25,right=0.25}{v10,vc8}
\fmf{plain,left=0.25}{v6,vc4}
\fmf{phantom,left=0.25}{v7,wc2}
\fmf{plain,tension=0.5}{vc1,vc2}
\fmf{phantom,tension=0.5}{wc1,wc2}
\fmf{plain,tension=0.5}{vc2,vc3}
\fmf{phantom,tension=0.5}{wc2,vc5}
\fmf{plain,tension=0.5}{vc3,vc4}
\fmf{phantom,tension=0.5}{vc4,wc1}
\fmf{plain,tension=0.5}{vc5,vc6}
\fmf{plain,tension=0.5}{vc6,vc7}
\fmf{plain,tension=0.5}{vc7,vc8}
\fmf{plain,tension=0.5,right=0,width=1mm}{v5,v10}
\fmffreeze
\fmf{dots,tension=0.5}{vc4,vc5}
\fmfposition
\fmfipath{p[]}
\fmfiset{p1}{vpath(__v5,__vc2)}
\fmfiset{p2}{vpath(__vc7,__vc8)}
\fmfipair{wz[]}
\svertex{wz1}{p1}
\svertex{wz2}{p2}
\wigglywrap{wz1}{v5}{v10}{wz2}
\fmfposition
\end{fmfchar*}}
\qquad
\fmfframe(3,1)(1,4){%
\begin{fmfchar*}(25,20)
\fmftop{v1}
\fmfbottom{v5}
\fmfforce{(0w,h)}{v1}
\fmfforce{(0w,0)}{v5}
\fmffixed{(0.2w,0)}{v1,v2}
\fmffixed{(0.2w,0)}{v2,v3}
\fmffixed{(0.2w,0)}{v3,w1}
\fmffixed{(0.2w,0)}{w1,v4}
\fmffixed{(0.2w,0)}{v4,v9}
\fmffixed{(0.2w,0)}{v5,v6}
\fmffixed{(0.2w,0)}{v6,v7}
\fmffixed{(0.2w,0)}{v7,w2}
\fmffixed{(0.2w,0)}{w2,v8}
\fmffixed{(0.2w,0)}{v8,v10}
\fmffixed{(0,whatever)}{vc1,vc2}
\fmffixed{(0,whatever)}{vc3,vc4}
\fmffixed{(0,whatever)}{vc5,vc6}
\fmffixed{(0,whatever)}{vc7,vc8}
\fmf{plain,tension=0.25,right=0.25}{v1,vc1}
\fmf{plain,tension=0.25,left=0.25}{v2,vc1}
\fmf{plain,left=0.25}{v5,vc2}
\fmf{plain,tension=1,left=0.25}{v3,vc3}
\fmf{phantom,tension=1,left=0.25}{w1,wc1}
\fmf{plain,tension=1,left=0.25}{v4,vc5}
\fmf{plain,tension=1,left=0.25}{v9,vc7}
\fmf{plain,left=0.25}{w2,vc6}
\fmf{plain,tension=0.25,left=0.25}{v8,vc8}
\fmf{plain,tension=0.25,right=0.25}{v10,vc8}
\fmf{plain,left=0.25}{v6,vc4}
\fmf{phantom,left=0.25}{v7,wc2}
\fmf{plain,tension=0.5}{vc1,vc2}
\fmf{phantom,tension=0.5}{wc1,wc2}
\fmf{plain,tension=0.5}{vc2,vc3}
\fmf{phantom,tension=0.5}{wc2,vc5}
\fmf{plain,tension=0.5}{vc3,vc4}
\fmf{phantom,tension=0.5}{vc4,wc1}
\fmf{plain,tension=0.5}{vc5,vc6}
\fmf{plain,tension=0.5}{vc6,vc7}
\fmf{plain,tension=0.5}{vc7,vc8}
\fmf{plain,tension=0.5,right=0,width=1mm}{v5,v10}
\fmffreeze
\fmf{dots,tension=0.5}{vc4,vc5}
\fmfposition
\fmfipath{p[]}
\fmfiset{p1}{vpath(__vc1,__vc2)}
\fmfiset{p2}{vpath(__v10,__vc8)}
\fmfipair{wz[]}
\svertex{wz1}{p1}
\svertex{wz2}{p2}
\wigglywrap{wz1}{v5}{v10}{wz2}
\fmfposition
\end{fmfchar*}}
\qquad
\fmfframe(3,1)(1,4){%
\begin{fmfchar*}(25,20)
\fmftop{v1}
\fmfbottom{v5}
\fmfforce{(0w,h)}{v1}
\fmfforce{(0w,0)}{v5}
\fmffixed{(0.2w,0)}{v1,v2}
\fmffixed{(0.2w,0)}{v2,v3}
\fmffixed{(0.2w,0)}{v3,w1}
\fmffixed{(0.2w,0)}{w1,v4}
\fmffixed{(0.2w,0)}{v4,v9}
\fmffixed{(0.2w,0)}{v5,v6}
\fmffixed{(0.2w,0)}{v6,v7}
\fmffixed{(0.2w,0)}{v7,w2}
\fmffixed{(0.2w,0)}{w2,v8}
\fmffixed{(0.2w,0)}{v8,v10}
\fmffixed{(0,whatever)}{vc1,vc2}
\fmffixed{(0,whatever)}{vc3,vc4}
\fmffixed{(0,whatever)}{vc5,vc6}
\fmffixed{(0,whatever)}{vc7,vc8}
\fmf{plain,tension=0.25,right=0.25}{v1,vc1}
\fmf{plain,tension=0.25,left=0.25}{v2,vc1}
\fmf{plain,left=0.25}{v5,vc2}
\fmf{plain,tension=1,left=0.25}{v3,vc3}
\fmf{phantom,tension=1,left=0.25}{w1,wc1}
\fmf{plain,tension=1,left=0.25}{v4,vc5}
\fmf{plain,tension=1,left=0.25}{v9,vc7}
\fmf{plain,left=0.25}{w2,vc6}
\fmf{plain,tension=0.25,left=0.25}{v8,vc8}
\fmf{plain,tension=0.25,right=0.25}{v10,vc8}
\fmf{plain,left=0.25}{v6,vc4}
\fmf{phantom,left=0.25}{v7,wc2}
\fmf{plain,tension=0.5}{vc1,vc2}
\fmf{phantom,tension=0.5}{wc1,wc2}
\fmf{plain,tension=0.5}{vc2,vc3}
\fmf{phantom,tension=0.5}{wc2,vc5}
\fmf{plain,tension=0.5}{vc3,vc4}
\fmf{phantom,tension=0.5}{vc4,wc1}
\fmf{plain,tension=0.5}{vc5,vc6}
\fmf{plain,tension=0.5}{vc6,vc7}
\fmf{plain,tension=0.5}{vc7,vc8}
\fmf{plain,tension=0.5,right=0,width=1mm}{v5,v10}
\fmffreeze
\fmf{dots,tension=0.5}{vc4,vc5}
\fmfposition
\fmfipath{p[]}
\fmfiset{p1}{vpath(__vc1,__vc2)}
\fmfiset{p2}{vpath(__vc7,__vc8)}
\fmfipair{wz[]}
\svertex{wz1}{p1}
\svertex{wz2}{p2}
\wigglywrap{wz1}{v5}{v10}{wz2}
\fmfposition
\end{fmfchar*}}}
}
\\
\vspace{0.2cm}
$\Huge{\vdots}$
\vspace{0.3cm}
\\
\subfigure[$\swrap{L}{j}\quad 1<j<L-1$]{
\fmfframe(3,1)(1,4){%
\begin{fmfchar*}(30,20)
\fmftop{v1}
\fmfbottom{v5}
\fmfforce{(0w,h)}{v1}
\fmfforce{(0w,0)}{v5}
\fmffixed{(0.25w,0)}{v1,v2}
\fmffixed{(0.16w,0)}{v2,v3}
\fmffixed{(0.16w,0)}{v3,w1}
\fmffixed{(0.16w,0)}{w1,v4}
\fmffixed{(0.16w,0)}{v4,v9}
\fmffixed{(0.25w,0)}{v5,v6}
\fmffixed{(0.16w,0)}{v6,v7}
\fmffixed{(0.16w,0)}{v7,w2}
\fmffixed{(0.16w,0)}{w2,v8}
\fmffixed{(0.16w,0)}{v8,v10}
\fmffixed{(0,whatever)}{vc1,vc2}
\fmffixed{(0,whatever)}{vc3,vc4}
\fmffixed{(0,whatever)}{vc5,vc6}
\fmffixed{(0,whatever)}{vc7,vc8}
\fmf{phantom,tension=0.25,right=0.25}{v2,vc1}
\fmf{phantom,tension=0.25,left=0.25}{v3,vc1}
\fmf{phantom,left=0.25}{v6,vc2}
\fmf{plain,tension=1,left=0.25}{w1,vc3}
\fmf{phantom,tension=1,left=0.25}{v4,wc1}
\fmf{plain,tension=1,left=0.25}{v4,vc5}
\fmf{plain,tension=1,left=0.25}{v9,vc7}
\fmf{plain,left=0.25}{w2,vc6}
\fmf{plain,tension=0.25,left=0.25}{v8,vc8}
\fmf{plain,tension=0.25,right=0.25}{v10,vc8}
\fmf{plain,left=0.25}{v7,vc4}
\fmf{phantom,left=0.25}{w2,wc2}
\fmf{phantom,tension=0.5}{vc1,vc2}
\fmf{phantom,tension=0.5}{vc2,vc3}
\fmf{phantom,tension=0.5}{wc2,vc5}
\fmf{plain,tension=0.5}{vc3,vc4}
\fmf{phantom,tension=0.5}{vc4,wc1}
\fmf{plain,tension=0.5}{vc5,vc6}
\fmf{plain,tension=0.5}{vc6,vc7}
\fmf{plain,tension=0.5}{vc7,vc8}
\fmf{plain}{v1,v5}
\fmf{plain,tension=0.5,right=0,width=1mm}{v5,v10}
\fmffreeze
\fmffixed{(0.32w,0)}{v1,v11}
\fmffixed{(0.32w,0)}{v5,v12}
\fmf{plain,tension=0.25,right=0.25}{v3,vc3}
\fmf{plain}{v11,v12}
\fmf{dots,tension=0.5}{vc4,vc5}
\fmfposition
\fmfipath{p[]}
\fmfipair{wz[]}
\fmfiset{p1}{vpath(__v7,__vc4)}
\fmfiset{p2}{vpath(__v1,__v5)}
\fmfiset{p3}{vpath(__v10,__vc8)}
\fmfiset{p4}{vpath(__v11,__v12)}
\svertex{wz1}{p1}
\vvertex{wz2}{wz1}{p2}
\svertex{wz3}{p2}
\svertex{wz4}{p3}
\svertex{wz5}{p4}
\fmfiequ{wz6}{(xpart(wz3)+6,ypart(wz3))}
\fmfiequ{wz7}{(xpart(wz5)-6,ypart(wz5))}
\fmfi{wiggly}{wz5..wz1}
\fmfi{wiggly}{wz6..wz3}
\fmfi{wiggly}{wz7..wz5}
\fmfi{dots}{wz6..wz7}
\wigglywrap{wz3}{v5}{v10}{wz4}
\end{fmfchar*}}
\qquad
\fmfframe(3,1)(1,4){%
\begin{fmfchar*}(30,20)
\fmftop{v1}
\fmfbottom{v5}
\fmfforce{(0w,h)}{v1}
\fmfforce{(0w,0)}{v5}
\fmffixed{(0.25w,0)}{v1,v2}
\fmffixed{(0.16w,0)}{v2,v3}
\fmffixed{(0.16w,0)}{v3,w1}
\fmffixed{(0.16w,0)}{w1,v4}
\fmffixed{(0.16w,0)}{v4,v9}
\fmffixed{(0.25w,0)}{v5,v6}
\fmffixed{(0.16w,0)}{v6,v7}
\fmffixed{(0.16w,0)}{v7,w2}
\fmffixed{(0.16w,0)}{w2,v8}
\fmffixed{(0.16w,0)}{v8,v10}
\fmffixed{(0,whatever)}{vc1,vc2}
\fmffixed{(0,whatever)}{vc3,vc4}
\fmffixed{(0,whatever)}{vc5,vc6}
\fmffixed{(0,whatever)}{vc7,vc8}
\fmf{phantom,tension=0.25,right=0.25}{v2,vc1}
\fmf{phantom,tension=0.25,left=0.25}{v3,vc1}
\fmf{phantom,left=0.25}{v6,vc2}
\fmf{plain,tension=1,left=0.25}{w1,vc3}
\fmf{phantom,tension=1,left=0.25}{v4,wc1}
\fmf{plain,tension=1,left=0.25}{v4,vc5}
\fmf{plain,tension=1,left=0.25}{v9,vc7}
\fmf{plain,left=0.25}{w2,vc6}
\fmf{plain,tension=0.25,left=0.25}{v8,vc8}
\fmf{plain,tension=0.25,right=0.25}{v10,vc8}
\fmf{plain,left=0.25}{v7,vc4}
\fmf{phantom,left=0.25}{w2,wc2}
\fmf{phantom,tension=0.5}{vc1,vc2}
\fmf{phantom,tension=0.5}{vc2,vc3}
\fmf{phantom,tension=0.5}{wc2,vc5}
\fmf{plain,tension=0.5}{vc3,vc4}
\fmf{phantom,tension=0.5}{vc4,wc1}
\fmf{plain,tension=0.5}{vc5,vc6}
\fmf{plain,tension=0.5}{vc6,vc7}
\fmf{plain,tension=0.5}{vc7,vc8}
\fmf{plain}{v1,v5}
\fmf{plain,tension=0.5,right=0,width=1mm}{v5,v10}
\fmffreeze
\fmffixed{(0.32w,0)}{v1,v11}
\fmffixed{(0.32w,0)}{v5,v12}
\fmf{plain,tension=0.25,right=0.25}{v3,vc3}
\fmf{plain}{v11,v12}
\fmf{dots,tension=0.5}{vc4,vc5}
\fmfposition
\fmfipath{p[]}
\fmfipair{wz[]}
\fmfiset{p1}{vpath(__v7,__vc4)}
\fmfiset{p2}{vpath(__v1,__v5)}
\fmfiset{p3}{vpath(__vc7,__vc8)}
\fmfiset{p4}{vpath(__v11,__v12)}
\svertex{wz1}{p1}
\vvertex{wz2}{wz1}{p2}
\svertex{wz3}{p2}
\svertex{wz4}{p3}
\svertex{wz5}{p4}
\fmfiequ{wz6}{(xpart(wz3)+6,ypart(wz3))}
\fmfiequ{wz7}{(xpart(wz5)-6,ypart(wz5))}
\fmfi{wiggly}{wz5..wz1}
\fmfi{wiggly}{wz6..wz3}
\fmfi{wiggly}{wz7..wz5}
\fmfi{dots}{wz6..wz7}
\wigglywrap{wz3}{v5}{v10}{wz4}
\end{fmfchar*}}
\qquad
\fmfframe(3,1)(1,4){%
\begin{fmfchar*}(30,20)
\fmftop{v1}
\fmfbottom{v5}
\fmfforce{(0w,h)}{v1}
\fmfforce{(0w,0)}{v5}
\fmffixed{(0.25w,0)}{v1,v2}
\fmffixed{(0.16w,0)}{v2,v3}
\fmffixed{(0.16w,0)}{v3,w1}
\fmffixed{(0.16w,0)}{w1,v4}
\fmffixed{(0.16w,0)}{v4,v9}
\fmffixed{(0.25w,0)}{v5,v6}
\fmffixed{(0.16w,0)}{v6,v7}
\fmffixed{(0.16w,0)}{v7,w2}
\fmffixed{(0.16w,0)}{w2,v8}
\fmffixed{(0.16w,0)}{v8,v10}
\fmffixed{(0,whatever)}{vc1,vc2}
\fmffixed{(0,whatever)}{vc3,vc4}
\fmffixed{(0,whatever)}{vc5,vc6}
\fmffixed{(0,whatever)}{vc7,vc8}
\fmf{phantom,tension=0.25,right=0.25}{v2,vc1}
\fmf{phantom,tension=0.25,left=0.25}{v3,vc1}
\fmf{phantom,left=0.25}{v6,vc2}
\fmf{plain,tension=1,left=0.25}{w1,vc3}
\fmf{phantom,tension=1,left=0.25}{v4,wc1}
\fmf{plain,tension=1,left=0.25}{v4,vc5}
\fmf{plain,tension=1,left=0.25}{v9,vc7}
\fmf{plain,left=0.25}{w2,vc6}
\fmf{plain,tension=0.25,left=0.25}{v8,vc8}
\fmf{plain,tension=0.25,right=0.25}{v10,vc8}
\fmf{plain,left=0.25}{v7,vc4}
\fmf{phantom,left=0.25}{w2,wc2}
\fmf{phantom,tension=0.5}{vc1,vc2}
\fmf{phantom,tension=0.5}{vc2,vc3}
\fmf{phantom,tension=0.5}{wc2,vc5}
\fmf{plain,tension=0.5}{vc3,vc4}
\fmf{phantom,tension=0.5}{vc4,wc1}
\fmf{plain,tension=0.5}{vc5,vc6}
\fmf{plain,tension=0.5}{vc6,vc7}
\fmf{plain,tension=0.5}{vc7,vc8}
\fmf{plain}{v1,v5}
\fmf{plain,tension=0.5,right=0,width=1mm}{v5,v10}
\fmffreeze
\fmffixed{(0.32w,0)}{v1,v11}
\fmffixed{(0.32w,0)}{v5,v12}
\fmf{plain,tension=0.25,right=0.25}{v3,vc3}
\fmf{plain}{v11,v12}
\fmf{dots,tension=0.5}{vc4,vc5}
\fmfposition
\fmfipath{p[]}
\fmfipair{wz[]}
\fmfiset{p1}{vpath(__vc3,__vc4)}
\fmfiset{p2}{vpath(__v1,__v5)}
\fmfiset{p3}{vpath(__v10,__vc8)}
\fmfiset{p4}{vpath(__v11,__v12)}
\svertex{wz1}{p1}
\vvertex{wz2}{wz1}{p2}
\svertex{wz3}{p2}
\svertex{wz4}{p3}
\svertex{wz5}{p4}
\fmfiequ{wz6}{(xpart(wz3)+6,ypart(wz3))}
\fmfiequ{wz7}{(xpart(wz5)-6,ypart(wz5))}
\fmfi{wiggly}{wz5..wz1}
\fmfi{wiggly}{wz6..wz3}
\fmfi{wiggly}{wz7..wz5}
\fmfi{dots}{wz6..wz7}
\wigglywrap{wz3}{v5}{v10}{wz4}
\end{fmfchar*}}
\qquad
\fmfframe(3,1)(1,4){%
\begin{fmfchar*}(30,20)
\fmftop{v1}
\fmfbottom{v5}
\fmfforce{(0w,h)}{v1}
\fmfforce{(0w,0)}{v5}
\fmffixed{(0.25w,0)}{v1,v2}
\fmffixed{(0.16w,0)}{v2,v3}
\fmffixed{(0.16w,0)}{v3,w1}
\fmffixed{(0.16w,0)}{w1,v4}
\fmffixed{(0.16w,0)}{v4,v9}
\fmffixed{(0.25w,0)}{v5,v6}
\fmffixed{(0.16w,0)}{v6,v7}
\fmffixed{(0.16w,0)}{v7,w2}
\fmffixed{(0.16w,0)}{w2,v8}
\fmffixed{(0.16w,0)}{v8,v10}
\fmffixed{(0,whatever)}{vc1,vc2}
\fmffixed{(0,whatever)}{vc3,vc4}
\fmffixed{(0,whatever)}{vc5,vc6}
\fmffixed{(0,whatever)}{vc7,vc8}
\fmf{phantom,tension=0.25,right=0.25}{v2,vc1}
\fmf{phantom,tension=0.25,left=0.25}{v3,vc1}
\fmf{phantom,left=0.25}{v6,vc2}
\fmf{plain,tension=1,left=0.25}{w1,vc3}
\fmf{phantom,tension=1,left=0.25}{v4,wc1}
\fmf{plain,tension=1,left=0.25}{v4,vc5}
\fmf{plain,tension=1,left=0.25}{v9,vc7}
\fmf{plain,left=0.25}{w2,vc6}
\fmf{plain,tension=0.25,left=0.25}{v8,vc8}
\fmf{plain,tension=0.25,right=0.25}{v10,vc8}
\fmf{plain,left=0.25}{v7,vc4}
\fmf{phantom,left=0.25}{w2,wc2}
\fmf{phantom,tension=0.5}{vc1,vc2}
\fmf{phantom,tension=0.5}{vc2,vc3}
\fmf{phantom,tension=0.5}{wc2,vc5}
\fmf{plain,tension=0.5}{vc3,vc4}
\fmf{phantom,tension=0.5}{vc4,wc1}
\fmf{plain,tension=0.5}{vc5,vc6}
\fmf{plain,tension=0.5}{vc6,vc7}
\fmf{plain,tension=0.5}{vc7,vc8}
\fmf{plain}{v1,v5}
\fmf{plain,tension=0.5,right=0,width=1mm}{v5,v10}
\fmffreeze
\fmffixed{(0.32w,0)}{v1,v11}
\fmffixed{(0.32w,0)}{v5,v12}
\fmf{plain,tension=0.25,right=0.25}{v3,vc3}
\fmf{plain}{v11,v12}
\fmf{dots,tension=0.5}{vc4,vc5}
\fmfposition
\fmfipath{p[]}
\fmfipair{wz[]}
\fmfiset{p1}{vpath(__vc3,__vc4)}
\fmfiset{p2}{vpath(__v1,__v5)}
\fmfiset{p3}{vpath(__vc7,__vc8)}
\fmfiset{p4}{vpath(__v11,__v12)}
\svertex{wz1}{p1}
\vvertex{wz2}{wz1}{p2}
\svertex{wz3}{p2}
\svertex{wz4}{p3}
\svertex{wz5}{p4}
\fmfiequ{wz6}{(xpart(wz3)+6,ypart(wz3))}
\fmfiequ{wz7}{(xpart(wz5)-6,ypart(wz5))}
\fmfi{wiggly}{wz5..wz1}
\fmfi{wiggly}{wz6..wz3}
\fmfi{wiggly}{wz7..wz5}
\fmfi{dots}{wz6..wz7}
\wigglywrap{wz3}{v5}{v10}{wz4}
\end{fmfchar*}}
}
\\
\vspace{0.2cm}
$\Huge{\vdots}$
\vspace{0.3cm}
\\
\subfigure[$\swrap{L}{L-1}$]
{
\fmfframe(3,1)(1,4){%
\begin{fmfchar*}(25,20)
\fmftop{v1}
\fmfbottom{v5}
\fmfforce{(0w,h)}{v1}
\fmfforce{(0w,0)}{v5}
\fmffixed{(0.2w,0)}{v1,v2}
\fmffixed{(0.2w,0)}{v2,v3}
\fmffixed{(0.2w,0)}{v3,w1}
\fmffixed{(0.2w,0)}{w1,v4}
\fmffixed{(0.2w,0)}{v4,v9}
\fmffixed{(0.2w,0)}{v5,v6}
\fmffixed{(0.2w,0)}{v6,v7}
\fmffixed{(0.2w,0)}{v7,w2}
\fmffixed{(0.2w,0)}{w2,v8}
\fmffixed{(0.2w,0)}{v8,v10}
\fmffixed{(0,whatever)}{vc7,vc8}
\fmf{plain,tension=0.25,right=0.25}{v4,vc7}
\fmf{plain,tension=0.25,left=0.25}{v9,vc7}
\fmf{plain,tension=0.25,left=0.25}{v8,vc8}
\fmf{plain,tension=0.25,right=0.25}{v10,vc8}
\fmf{plain}{v2,v6}
\fmf{plain}{w2,w1}
\fmf{plain,tension=0.5}{vc7,vc8}
\fmf{plain}{v1,v5}
\fmf{plain,tension=0.5,right=0,width=1mm}{v5,v10}
\fmffreeze
\fmfposition
\fmfipath{p[]}
\fmfipair{wz[]}
\fmfiset{p1}{vpath(__v2,__v6)}
\fmfiset{p2}{vpath(__w1,__w2)}
\fmfiset{p3}{vpath(__v10,__vc8)}
\fmfiset{p4}{vpath(__v1,__v5)}
\fmfiset{p5}{vpath(__vc8,__vc7)}
\svertex{wz1}{p1}
\svertex{wz2}{p2}
\svertex{wz3}{p3}
\svertex{wz4}{p4}
\svertex{wz5}{p5}
\fmfiequ{wz6}{(xpart(wz1)+6,ypart(wz1))}
\fmfiequ{wz7}{(xpart(wz2)-6,ypart(wz2))}
\fmfi{wiggly}{wz2..wz5}
\fmfi{dots}{wz6..wz7}
\fmfi{wiggly}{wz4..wz1}
\fmfi{wiggly}{wz1..wz6}
\fmfi{wiggly}{wz7..wz2}
\wigglywrap{wz4}{v5}{v10}{wz3}
\end{fmfchar*}}
\qquad
\fmfframe(3,1)(1,4){%
\begin{fmfchar*}(25,20)
\fmftop{v1}
\fmfbottom{v5}
\fmfforce{(0w,h)}{v1}
\fmfforce{(0w,0)}{v5}
\fmffixed{(0.2w,0)}{v1,v2}
\fmffixed{(0.2w,0)}{v2,v3}
\fmffixed{(0.2w,0)}{v3,w1}
\fmffixed{(0.2w,0)}{w1,v4}
\fmffixed{(0.2w,0)}{v4,v9}
\fmffixed{(0.2w,0)}{v5,v6}
\fmffixed{(0.2w,0)}{v6,v7}
\fmffixed{(0.2w,0)}{v7,w2}
\fmffixed{(0.2w,0)}{w2,v8}
\fmffixed{(0.2w,0)}{v8,v10}
\fmffixed{(0,whatever)}{vc7,vc8}
\fmf{plain,tension=0.25,right=0.25}{v4,vc7}
\fmf{plain,tension=0.25,left=0.25}{v9,vc7}
\fmf{plain,tension=0.25,left=0.25}{v8,vc8}
\fmf{plain,tension=0.25,right=0.25}{v10,vc8}
\fmf{plain}{v2,v6}
\fmf{plain}{w2,w1}
\fmf{plain,tension=0.5}{vc7,vc8}
\fmf{plain}{v1,v5}
\fmf{plain,tension=0.5,right=0,width=1mm}{v5,v10}
\fmffreeze
\fmfposition
\fmfipath{p[]}
\fmfipair{wz[]}
\fmfiset{p1}{vpath(__v2,__v6)}
\fmfiset{p2}{vpath(__w1,__w2)}
\fmfiset{p3}{vpath(__v10,__vc8)}
\fmfiset{p4}{vpath(__v1,__v5)}
\fmfiset{p5}{vpath(__vc8,__vc7)}
\svertex{wz1}{p1}
\svertex{wz2}{p2}
\fmfiequ{wz3}{point 2*length(p5)/3 of p5}
\svertex{wz4}{p4}
\fmfiequ{wz5}{point length(p5)/3 of p5}
\fmfiequ{wz6}{(xpart(wz1)+6,ypart(wz1))}
\fmfiequ{wz7}{(xpart(wz2)-6,ypart(wz2))}
\fmfi{wiggly}{wz2..wz5}
\fmfi{dots}{wz6..wz7}
\fmfi{wiggly}{wz4..wz1}
\fmfi{wiggly}{wz1..wz6}
\fmfi{wiggly}{wz7..wz2}
\wigglywrap{wz4}{v5}{v10}{wz3}
\end{fmfchar*}}
}
\caption{Relevant diagrams after cancellations}
\label{defwrapgraphs}
\end{figure}

Coming to wrapping supergraphs with vector interactions, because of the restrictions on the chiral functions, now the number of vector lines uniquely identifies the structure. So we can denote by $\swrap{L}{k}$ the sum of all the diagrams with $k$ vectors, corresponding to the structure $\dchi{1,2,\ldots,L-k}$.
From the cancellation identities of Section~\ref{superspace} we know that vector propagators can be attached to outgoing scalar lines only at double-vector vertices. This observation greatly reduces the number of possibilities, which would otherwise grow with the number of vectors, and in the end we find that only four diagrams can be relevant for each chiral structure, with the exception of $\dchi{1}$, which is associated to two diagrams only. These considerations are summarized in Figure~\ref{defwrapgraphs}.

\begin{figure}[h]
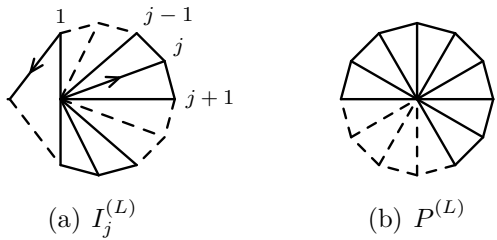

\centering
\addtolength{\subfigcapskip}{5pt}
\unitlength=0.75mm
\settoheight{\eqoff}{$\times$}%
\setlength{\eqoff}{0.5\eqoff}%
\addtolength{\eqoff}{-14.5\unitlength}%
\settoheight{\eqofftwo}{$\times$}%
\setlength{\eqofftwo}{0.5\eqofftwo}%
\addtolength{\eqofftwo}{-7.5\unitlength}%
\centering
\subfigure[$\sint{j}{L}$]{
\label{ILk}
\raisebox{\eqoff}{%
\fmfframe(0,0)(0,0){%
\begin{fmfchar*}(30,30)
\fmfleft{in}
\fmfright{out}
\fmf{plain,tension=1}{in,vi}
\fmf{phantom,tension=1}{out,v7}
\fmfpoly{phantom}{v12,v11,v10,v9,v8,v7,v6,v5,v4,v3,v2,v1}
\fmffixed{(0.9w,0)}{v1,v7}
\fmffixed{(0.3w,0)}{vi,v0}
\fmffixed{(0,whatever)}{v0,v3}
\fmf{dashes}{v3,v4}
\fmf{dashes}{v4,v5}
\fmf{plain}{v5,v6}
\fmf{plain}{v6,v7}
\fmf{dashes}{v7,v8}
\fmf{dashes}{v8,v9}
\fmf{plain}{v9,v10}
\fmf{plain}{v10,v11}
\fmf{derplain}{v3,vi}
\fmf{dashes}{vi,v11}
\fmf{plain}{v0,v3}
\fmf{dashes}{v0,v4}
\fmf{plain}{v0,v5}
\fmf{derplain}{v0,v6}
\fmf{plain}{v0,v7}
\fmf{dashes}{v0,v8}
\fmf{plain}{v0,v9}
\fmf{plain}{v0,v10}
\fmf{plain}{v0,v11}
\fmfposition
\fmfipair{w[]}
\fmfiequ{w1}{(xpart(vloc(__v3)),ypart(vloc(__v3)))}
\fmfiequ{w2}{(xpart(vloc(__v4)),ypart(vloc(__v4)))}
\fmfiequ{w3}{(xpart(vloc(__v5)),ypart(vloc(__v5)))}
\fmfiequ{w4}{(xpart(vloc(__v6)),ypart(vloc(__v6)))}
\fmfiequ{w5}{(xpart(vloc(__v7)),ypart(vloc(__v7)))}
\fmfiequ{w6}{(xpart(vloc(__v8)),ypart(vloc(__v8)))}
\fmfiequ{w7}{(xpart(vloc(__v9)),ypart(vloc(__v9)))}
\fmfiv{l=\scriptsize{$1$},l.a=90,l.d=4}{w1}
\fmfiv{l=\scriptsize{$j-1$},l.a=45,l.d=4}{w3}
\fmfiv{l=\scriptsize{$j$},l.a=30,l.d=4}{w4}
\fmfiv{l=\scriptsize{$j+1$},l.a=0,l.d=4}{w5}
\end{fmfchar*}}}
}
\qquad\qquad
\subfigure[$\pint{L}_{\protect\phantom{j}}$]{
\label{PL}
\raisebox{\eqoff}{%
\fmfframe(0,0)(0,0){%
\begin{fmfchar*}(30,30)
\fmfleft{in}
\fmfright{out}
\fmf{phantom,tension=1}{in,v1}
\fmf{phantom,tension=1}{out,v7}
\fmfpoly{phantom}{v12,v11,v10,v9,v8,v7,v6,v5,v4,v3,v2,v1}
\fmffixed{(0.9w,0)}{v1,v7}
\fmfforce{(0.5w,0.5h)}{v0}
\fmfposition
\fmffreeze
\fmf{plain}{v1,v2}
\fmf{plain}{v2,v3}
\fmf{plain}{v3,v4}
\fmf{plain}{v4,v5}
\fmf{plain}{v5,v6}
\fmf{plain}{v6,v7}
\fmf{plain}{v7,v8}
\fmf{plain}{v8,v9}
\fmf{dashes}{v9,v10}
\fmf{dashes}{v10,v11}
\fmf{dashes}{v11,v12}
\fmf{dashes}{v12,v1}
\fmf{plain}{v1,v0}
\fmf{plain}{v2,v0}
\fmf{plain}{v3,v0}
\fmf{plain}{v4,v0}
\fmf{plain}{v5,v0}
\fmf{plain}{v6,v0}
\fmf{plain}{v7,v0}
\fmf{plain}{v8,v0}
\fmf{plain}{v9,v0}
\fmf{dashes}{v10,v0}
\fmf{dashes}{v11,v0}
\fmf{dashes}{v12,v0}
\end{fmfchar*}}}
}
\caption{$L$-loop momentum integrals}
\label{integrals}
\end{figure}

Once again it is possible to complete the D-algebra for generic $L$. Introducing the deformation factors
\begin{equation}
\label{colfact}
\cfact{L}{j}=(q-\bar{q})^2\left[q^{2(L-j-1)}+\bar{q}^{2(L-j-1)}\right]=-8\sin^2(\pi\beta)\cos[2\pi\beta(L-j-1)] \col
\end{equation}
for $j\in\{0,\ldots,L-1\}$, we can write the contribution from each class as
\begin{equation}
\label{defres}
\begin{aligned}
(\swrap{L}{0}+\bswrap{L}{0})-(\sub{L}+\bsub{L})&\to(g^2 N)^L\ \cfact{L}{0}(\kint{L}{2}-\kint{L}{1})\col\\
\vdots\\
\swrap{L}{j}+\bswrap{L}{j}&\to2(g^2 N)^L\ \cfact{L}{j}\sint{j+1}{L}\col\\
\vdots\\
\swrap{L}{L-1}+\bswrap{L}{L-1}&\to-(g^2 N)^L\ \cfact{L}{L-1}(\kint{L}{2}-\kint{L}{1})\col
\end{aligned}
\end{equation}
where the subtraction of $\sub{L}$ has been combined with the diagram $\swrap{L}{0}$. 
The integrals $\sint{j}{L}$ are shown in Figure~\ref{ILk}, where the pair of arrows indicates that the scalar product of the corresponding momenta appears in the numerator. 
The integrals $I_j^{(L)}$ satisfy the relation
\begin{equation}
\label{ILrel}
\sint{j}{L}=-\sint{L-j+1}{L}
\col
\end{equation}
and thus the total number of integrals that we must compute explicitly is halved.
Moreover, it is useful to rewrite $(\kint{L}{2}-\kint{L}{1})$ in terms of $\sint{1}{L}$ and of the integral $\pint{L}$ of Figure~\ref{PL} as
\begin{equation}
\kint{L}{2}-\kint{L}{1}=\pint{L}-2\sint{1}{L}
\pnt
\end{equation}

We can at last collect all the terms to obtain the correction to the asymptotic anomalous dimension
\begin{equation}
\gamma_L(\mathcal{O}_{L})=\gamma_L^{as}+\delta\gamma_L(\mathcal{O}_{L})
\pnt
\end{equation}
Since $\pint{L}$ and all the $\sint{j}{L}$ are free of subdivergences, we can write our final result as
\begin{equation}
\label{wrapcorr}
\delta\gamma_L(\mathcal{O}_{L})=-2L(g^2 N)^L\lim_{\varepsilon\rightarrow0}\varepsilon\Bigg[(\cfact{L}{0}-\cfact{L}{L-1})\pint{L}(\varepsilon)-2\!\sum_{j=0}^{[\frac{L}{2}]-1}\!(\cfact{L}{j}-\cfact{L}{L-j-1})\sint{j+1}{L}(\varepsilon)\Bigg]
\pnt
\end{equation}

The computation of the wrapping correction has thus been reduced to the calculation of the divergent parts of the $[L/2]$ integrals $\sint{j}{L}$ and of $\pint{L}$. 
For the latter, the result is known as a function of $L$, but we do not have a similar solution for the $\sint{j}{L}$. However, the value of $\sint{j}{L}$ can be computed exactly for any fixed values of $L$ and $j$, by means of a set of recurrence relations obtained from the triangle rule for integration by parts~\cite{Chetyrkin:1981qh,Broadhurst:1985vq,Fiamberti:2008sn}. A different set of relations can be found by applying the GPXT directly in momentum space, as explained in~\cite{Fiamberti:2008sn}. As a particular case of the general result~\eqref{wrapcorr}, we can see that in the three-loop case the correction to the asymptotic result vanishes, in agreement with the explicit computation of~\cite{betadef}.
This is likely to be a consequence of the oversimplified structure of three-loop integrals, since there is no apparent reason why $\mathcal{O}_3$ should be protected against finite-size corrections. In fact, we expect that a non-trivial correction would arise at four loop. The corresponding computation, however, involves a non-critical perturbative order, and would thus be very difficult, for the same reasons explained in Section~\ref{undeformed} for the case of the Konishi operator at five loops.

We performed the explicit calculations of the integrals $\sint{j}{L}$ up to $L=11$~\cite{Fiamberti:2008sn}. In all the cases, the finite-size correction to the asymptotic anomalous dimension contains only transcendental terms, all of the form $\zeta(2L-2k-1)$ with $k\in\{1,2,\ldots,[L/2]\}$. The rational part of the answer is hence protected against wrapping corrections at the critical order! In particular, the cancellation of the lower-transcendentality contributions is a non-trivial feature which was absent in the case of two-impurity states. As an example, the five-loop anomalous dimension of $\mathcal{O}_5$ depends only on $\zeta(7)$ and $\zeta(5)$, whereas the five-loop wrapping correction for two-impurity operators in the undeformed theory contained an additional $\zeta(3)$ term.
About the term with maximum transcendentality, our results suggest that it comes entirely from the integral $P^{(L)}$, 
which in turn is produced only\footnote{The classes 
$\swrap{L}{0}$ and $\swrap{L}{L-1}$ do not contain terms proportional to $P^{(L)}$. In the expression \eqref{wrapcorr} they are cancelled in combination 
with the $j=0$ term of the sum.} 
by the diagrams in the classes $\swrap{L}{1}$ and $\swrap{L}{L-2}$.
We believe that such definite transcendentality pattern will be preserved in general, as far as we restrict to the critical order. 
A hint on the transcendentality properties of quantities that involve computations beyond the critical order is offered by the recent calculation, based on the L\"uscher technique, of the five-loop anomalous dimension of the single-impurity operator $\mathcal{O}_4$~\cite{Bajnok:2009vm}. In the final result, the wrapping correction is still made only of transcendental terms, namely $\zeta(3)$, $\zeta(5)$, $\zeta(7)$ and $\zeta(3)^2$. So one could guess that the wrapping corrections on single-impurity states always involve only transcendental terms. Unluckily, as we have already explained previously, the computation of the results of~\cite{Bajnok:2009vm,Arutyunov:2010gb}, and of other non-critical quantities, by means of direct field-theoretical techniques seems out of reach at the moment.

The anomalous dimensions of the $\mathcal{O}_L$ operators in the case of even $L$ and $\beta=1/2$ have been found also by means of the L\"uscher technique applied to the undeformed theory~\cite{Gunnesson:2009nn,Beccaria:2009hg}, exploiting the correspondence between the actual deformed case and the unphysical single-impurity states with momentum $p=\pi$. The results agree with our calculations. 

A proposal for the general description of wrapping effects in the deformed theory, possibly as an adaptation of the Y-system, has not been formulated yet. If such a solution were available, it might be checked against the whole series of single-impurity anomalous dimensions. We think that, thanks to the simplifications deriving from the possibility to work with single-impurity operators, the deformed theory may be a preferable environment for deep investigations on the validity of the recent proposals for the description of the full 
spectrum.\footnote{After the appearance of the first preprint version of this review, the
paper \cite{Gromov:2010dy} has been presented, in which a Y-system for the
$\beta$-deformed theory was proposed.
Starting from it, the authors were able to reproduce all of our perturbative
results for single-impurity states and to confirm our conjecture about the
transcendentality pattern of the anomalous dimensions.
Moreover, they present a generating function from which one can extract the
general result for any values of $\beta$ and $L$.}

\section{Comments}
\label{comments}
In all the cases that we analyzed, $\N=1$ superspace techniques revealed to be a very powerful tool for the computation of wrapping corrections to the anomalous dimensions of composite operators. First of all, the use of Feynman supergraphs allowed us to find useful cancellation identities coming from supersymmetry. This, together with the fact that every supergraph encodes the information on a large number of component-field diagrams, greatly reduced the number of possible terms. 
In particular, using supergraphs, one never explicitly encounters fermionic matter interactions and their associated $\gamma$-matrix manipulations. 
The standard multi-loop integrals produced after D-algebra can be computed by means of the Gegenbauer Polynomial $x$-space technique, whose application is greatly simplified by the fact that we are only interested in divergent contributions. This allows to find analytic results to very high loop orders.

Direct perturbative computations based on field-theoretical techniques are important because the results can serve as tests for the proposals for the general description of finite-size effects by means of integrable systems. In fact, the recently found Y-system exactly reproduces our results at four and five loops. For the same reason, it would be interesting to extend such perturbative analyses to deal with wrapping effects beyond the critical order. The only known prediction for a non-critical quantity is the five-loop component of the Konishi anomalous dimension, found in computations based on the L\"uscher approach and on the Thermodynamic Bethe Ansatz. The corresponding field-theoretical calculation appears to be very difficult at the moment, but the discovery of new and more powerful identities for supergraph cancellations may make it feasible. 
Thus, we expect that superspace techniques may still reveal to be useful in the future for the analysis of wrapping effects. 

In the case of the $\beta$-deformed theory, thanks to the existence of non-protected single-impurity operators, many more perturbative results are known, so that the tests of the new attempts to the general solution to the wrapping problem could be much deeper. This would require an extension to the $\beta$-deformed case of the recent proposals based on the thermodynamic Bethe ansatz approach and on the Y-system.

\section*{Acknowledgements}
\noindent This work has been supported in part by INFN and by the
Italian MIUR-PRIN contract 20075ATT78. 

\end{fmffile}

\clearpage

\footnotesize
\bibliographystyle{JHEP}
\bibliography{references}

\end{document}
